\documentclass[]{raa}         

\usepackage{graphicx,times}             
\usepackage{natbib}
\usepackage{amssymb,amsmath}
\bibpunct{(}{)}{;}{a}{}{,}

\usepackage[pagebackref=true]{hyperref}

\usepackage{hyperref}
\usepackage{breakurl}

\newcommand{\sunlum}{~L$_{\odot}$~}
\newcommand{\kms}{~km s$^{-1}$~}

\newcommand{\dotMyr}{~M$_{\odot}$~yr$^{-1}$~}

\newcommand{\Tstar}{T$_*$~}
\newcommand{\TstarE}{T$_*$}
\newcommand{\logL}{$\log$~L~}
\newcommand{\logLE}{$\log$~L}
\newcommand{\dotM}{$\log~\dot{M}$~}
\newcommand{\dotME}{$\log~\dot{M}$}
\newcommand{\Vwind}{$v_{\infty}$~}
\newcommand{\VwindE}{$v_{\infty}$}
\usepackage{xspace}

\newcommand{\Teff}{$T_*$\xspace}
\newcommand{\cmfgen}{{\sc cmfgen}\xspace}

\newcommand{\Mdot}{\ensuremath{\dot{M}}\xspace}

\newcommand{\Lstar}{$L_\star$\xspace}

\newcommand{\Vinf}{$v_\infty$\xspace}

\usepackage{xcolor}

\newcommand{\corr}[1]{#1}

\begin{document}


   \title{4D Grid-fitting of UV-optical spectra  
   	of massive stars.\\
	 I. Numerical technique and its associated uncertainties}

\volnopage{Vol.0 (20xx) No.0, 000--000}  
\setcounter{page}{1}   


\author{Blagovest V. Petrov \and Svetozar A. Zhekov}

\institute{Institute of Astronomy and NAO, Bulgarian Academy of Sciences,
72 Tsarigradsko Chaussee Blvd., 1784 Sofia, Bulgaria; 
{\it bpetrov@astro.bas.bg, szhekov@astro.bas.bg} \\
\vs\no
   {\small Received 20xx month day; accepted 20xx month day}
}

\abstract{
The best way to check the validity of our theories (models) is by
direct comparison with the experiment (observations). 
However, this process suffers from numerical inaccuracies, 
which are not frequently studied and often remain mostly unknown.
 In this study, we
focus on addressing the numerical inaccuracies intrinsic to the process
of comparing theory and observations. To achieve this goal, we built four-dimensional (4D)
spectra grids  for  Wolf-Rayet stars (WC and WN spectral classes) and
Blue Supergiants (BSGs) characterized by low metallicity similar to that
of the Small Magellanic Cloud (SMC). 
\corr{In contrast to lighter (three-dimensional) grids, which rely on a priori assumptions about certain stellar parameters (e.g., wind velocity) and thus have limited applicability, our 4D grids vary four independent parameters, enabling more flexible and broadly applicable spectral fitting.}
Utilizing these 4D  grids, we
developed and validated a fitting approach facilitating direct fits to
observed spectra. Through rigorous testing on designated `test' models,
we demonstrated that the numerical precision of derived essential
stellar parameters, including effective temperature, mass-loss rate,
luminosity, and wind velocity, is better than 0.05 dex. 
%
Furthermore, we explored the influence of unaccounted factors, including 
variations in the metal abundances, wind acceleration laws, and
clumping, on the precision of the derived parameters. 
The results indicate that the first two factors have the strongest influence on the
numerical accuracy of the derived stellar parameters.
Variations in
abundances predominantly influenced the mass-loss rate for weak-wind
scenarios, while effective temperature and luminosity remained robust.
We found that the wind acceleration law influence the numerical
uncertainty of the derived wind parameters mostly for models with weak
winds. Interestingly, different degrees of clumping demonstrated good
precision for spectra with strong winds, contrasting with a decrease in
the precision for weak-wind cases. We found also that the accuracy of
our approach  depends on spectral range and the inclusion of ultraviolet
spectral range improves the precision of derived parameters, especially
for object with weak winds.  
}

\authorrunning{B.~V. Petrov \& S.~A.~ Zhekov}

\titlerunning{4D Grid-fitting of UV-optical spectra}

\maketitle

\section{Introduction}
The extreme temperature and luminosities of the massive stars provide a 
large amount 
of ionizing photons,
leading to extreme non-local thermal equilibrium conditions  (non-LTE) in 
their  
expanding atmospheres. 
Additionally, mass-loss rate and velocity structure of the wind contribute to 
altering  the 
density profile of the massive star atmospheres.
This makes  the spectral analysis of massive stars complex problem, 
achievable only after  
development of  sophisticated model atmosphere codes.   

By utilizing stellar atmosphere modelling, it becomes feasible to derive 
essential 
physical properties of massive stars. These include fundamental 
parameters such as 
effective temperature (\Teff), luminosity (\Lstar), mass-loss rate (\Mdot), 
and terminal wind
velocity (\Vinf). These essential parameters play a crucial role as inputs 
in various 
models related to stellar structure and evolution, necessitating accurate 
knowledge of 
their values.

Typically, the interpretation of spectra from massive stars strongly relies on 
model 
atmospheres and a precise spectroscopic analysis should incorporate 
multi-wavelength 
comparison between observations and models.
Currently, there is no established standard approach for deriving stellar 
parameters 
using the complete spectra information and different studies  often have 
significantly 
different results. In general, such  discrepancies may be attributed to 
disparities in 
atomic data, model atmosphere codes, numerical methods and/or analysis 
techniques.  
Nonetheless,  adopting a consistent approach that incorporates the full 
spectral 
information in parameter derivation has the potential to produce more 
comparable 
results and enhance our understanding of the evolution of massive stars.

Over the past decades, there has been significant progress in computer 
hardware and 
astronomical software. This allowed for the inclusion of iron line blanketing 
and 
microclumping in the models \citep{hillier01, grafener02} leading 
to producing more  realistic stellar spectra.
The realism of these spectra can be revealed solely through direct 
comparison of the theory with observed data. Given the complexity of the
stellar atmosphere models, we must have {\it a tool (e.g., numerical
technique)} to carry out such a comparison.
Three primary sources of inaccuracy arise during this comparison: 
a) inaccuracies inherent in the theory; 
b) inaccuracies introduced by the experiment (observations); 
c) inaccuracies of the numerical technique for comparing theory and observations. 
For the latter, the terms numerical uncertainty, numerical precision and 
numerical accuracy will be used throughout the text.
Evaluating the overall accuracy of the model lies beyond the scope of our research.  
In this study, our emphasis is placed on addressing the numerical inaccuracies
resulting from the process of comparing theory and observations: item c)
above. 
This task is fundamentally significant because it serves as a bridge connecting theory with observations.
 If our method of comparing theory and observations bears significant uncertainties, 
 it could  obscure the interpretation of the physical parameters we derive, 
 regardless  how accurate our theory and observations are.

To achieve this goal (evaluation of uncertainties of the numerical
technique for comparing theory and observations), we need a `work 
field'. So, we have 
calculated four-dimensional grids of stellar atmosphere models for massive stars,
aiming to develop and access a rigorous method for deriving stellar parameters through the direct fitting of observed spectra. 
The four `axes' of these grids are effective temperature (\Teff), 
luminosity (\Lstar), mass-loss rate (\Mdot), and terminal wind velocity
(\Vinf).
Although it is well known, we recall that such a choice of basic 
parameters is due to the fact that the spectrum of a massive star is 
determined by the ionization structure of its wind. At a given point in
the stellar wind, the latter basically depends on the ratio of the 
ionizing photons to the gas number density. The stellar luminosity and 
the stellar temperature (controlling the shape of the underlying 
continuum) are responsible for the available ionizing photons.
The mass-loss rate and the wind velocity of the massive star are the 
factors that set the gas number density, as the latter determines the 
shape of the line profiles in the spectrum.

The spectral grid-modelling may adopt `lighter' grids (e.g., 3D grids;
three-dimensional), provided we have a priori information on some of
the basic stellar parameters. It is our understanding that this is
possible only for the terminal wind velocity of a massive star and in
such a case all the model spectra in the 3D grid will have the same 
value of the stellar wind. However, such a 3D grid will have limitted
applications: it is suitable only for analysis of the specific object 
at hand or of spectra of massive stars with stellar wind velocities 
not very different from the value adopted in the corresponding 3D 
spectral grid.
An illustrative application of a three-dimensional grid and its use 
for determining fundamental physical parameters is found in the study 
by \cite{zhekov20}.

We note that `lighter' spectral grids have been actively used in the
last two-three decades or so, as the case of the Potsdam Wolf-Rayet 
Models\footnote{\url{https://www.astro.physik.uni-potsdam.de/~wrh/PoWR/powrgrid1.php}}
is probably the best example of using the stellar atmosphere models for
deriving stellar parameters of an appreciable number
of massive stars in a consistent manner
(\citealt{hamann_19}; \citealt{sander_19} and references therein).
However, these are two-dimentional (2D) grids and their use relies on 
analysis of spectra, normalized to continuum. On the one hand, this 
makes the comparison between theory and observations more 
efficient in sense of technical (computer) resources. On the other hand, 
it adopts some approximate relations (scaling laws) for the stellar 
parameters, that is it is done at the expense of not using all the 
available information about the physical conditions in the 
spectral-formation regions (stellar winds).
Thus, our choice for a `work field' in this study is to make use of 
4D grids of spectral models of massive stars.

This paper is organised as follows: In Section\,\ref{sec:modelling}, we 
describe
 the employed model of stellar atmospheres 
and outline  the approach developed for directly fitting the stellar spectra.
In Section\,\ref{sec:tests},
we evaluate the expected accuracy of the derived stellar parameters by 
fitting 
`test' spectra  of models with randomly chosen values of stellar 
parameters.
In Section~\ref{sec:discussion} we discuss the possible implication of our 
results.
We present our conclusions in Section~\ref{sec:conclusions}.


\section{Spectral modelling} 
\label{sec:modelling}
\subsection{The model of stellar atmosphere}

For the purposes of our investigation, we use the non-LTE radiative
transfer code \cmfgen \citep{hillier98,hillier01}.\footnote{\url{https://sites.pitt.edu/~hillier/web/CMFGEN.htm}}
 This software is a comprehensive 
atmosphere code
that incorporates full line-blanketing, specifically engineered to
address the challenges posed by solving the statistical equilibrium and
radiative transfer equations in spherical geometry.

In  \cmfgen, the determination of a stellar effective temperature follows
the Stefan-Boltzmann law with a reference radius specified at a
particular value for the Rosseland optical depth. For OB stellar models,
we have opted for a reference optical depth of 2/3, whereas for WN and
WC stars, given their strong winds, we have chosen a reference optical
depth of 20. This choice ensures that the stellar radii do not extend
into the wind region. The formula for computing the effective
temperature is as follows:
\begin{equation}
T_*=\left(\frac{L}{4 \pi\sigma R^2_*}\right)^{1/4}
\label{eqn:tstar}
\end{equation}
Here, $L$ denotes the
stellar luminosity, $\sigma$ is the Stefan–Boltzmann constant, and $R_*$ is the specified  radius for the
reference Rosseland opacity.

From observations, we know that the winds of massive  stars exhibit
non-uniform characteristics, characterized by the presence of
inhomogeneities or `clumps'. These clumps have a notable impact on the
appearance of stellar spectra, making it imperative to model non-uniform
stellar atmospheres.

\cmfgen provides a mechanism to account for optically thin clumping,
often referred to as microclumping. This approach is grounded in the
concept that the stellar wind is composed of clumps characterized by
heightened density and dimensions smaller than the mean free path of
photons. These clumps possess an increased density, expressed as a
clumping factor denoted as $D$, in comparison to the average wind
density \corr{\citep[see][for detailed description]{hamann98}.} 
The models are calculated with the assumption that the clumps
are created by a volume filling factor represented by the reciprocal of
$D$, which is designated as $f$, assuming that the regions between
clumps contain minimal matter. This factor is related to the volume
filling factor, and it follows the relationship $f = 1/D$, assuming that
the region between the clumps is essentially devoid of matter.

In our models, $f_{\infty}=  0.1$, described by the following clumping
law:
\begin{equation}
f(r)=f_{\infty} +(1-f_{\infty})e^{-v(r)/v_{cl}}
\label{eqn:clump}
\end{equation}
\corr{introduced in \cite{martins09}}.
Here, $r$ is the distance from the star, $v(r)$ is the wind velocity and 
$v_{cl}$ is its value from  which clumping is taken 
into
account. We have chosen the clumping to start at $v_{cl} = 30$\kms.

It is important to note that \cmfgen does not solve the wind
momentum equation. Consequently, the structure of the wind velocity must
be predefined. In our models, we describe the wind velocity structure
using a standard $\beta$-type velocity law with a specific exponent,
which is set to $\beta$ = 1. This velocity law is  connected to the
hydrostatic section of the wind, situated just below the sonic point,
where the wind velocity attains the local speed of sound.
We note that the turbulent gas velocity ($v_{turb}$) contributes to the 
`micro-structure' of the wind, and a specific value must be adopted.

Selecting uniform values for the volume filling factor, 
velocity-law exponent and  turbulent velocity is done with the intention 
of reducing the number of independent parameters within the models.
Adopting uniform 
parameter values of $\beta$, $f_{\infty}$ and $v_{turb}$ can be regarded as
conventional in the field of spectral modelling for massive Wolf-Rayet
stars (e.g., \citealt{hamann_19} \citealt{sander_19} and references
therein). 
It is important to acknowledge that
there may be valid reasons 
to explore alternative values for these parameters in specific situations. 
This approach 
could be justified when seeking a closer alignment with actual physical 
conditions and 
observational data.

\subsection{Description of WN, WC and BSG grids}
\label{subsec:grids}

\begin{table}
\caption{Grid parameters.}

\label{tab:grids}
\bc
\begin{tabular}{ccc}
\hline\hline
\multicolumn{1}{c}{Parameter} &
\multicolumn{1}{c}{Values} &
\multicolumn{1}{c}{Number of spectra} \\
\hline
\multicolumn{2}{c}{WN grid} & \multicolumn{1}{c}{432}\\
 \dotM  & -5.0, -4.8, -4.5, -4.3 \\
 \logL  &  5.0, 5.2, 5.4, 5.6, 5.8, 6.0 \\
 \Tstar &  40\,000, 45\,000, 50\,000, 55\,000, 60\,000, 65\,000\\
 \Vwind &  1\,000, 1\,500, 2\, 250 \\
\hline
\multicolumn{2}{c}{WC grid} & \multicolumn{1}{c}{792}\\
 \dotM  & -5.0, -4.8, -4.5, -4.3 \\
 \logL  &  4.8, 5.2, 5.4, 5.6, 5.8 \\
 \Tstar &  45\,000, 50\,000, 55\,000, 60\,000, 65\,000, 70\,000 \\ 
	&  80\,000, 90\,000, 100\,000, 110\,000, 120\, 000 \\
 \Vwind &   1\,500, 2\,000, 3\,000 \\
\hline
\multicolumn{2}{c}{SMC grid} & \multicolumn{1}{c}{600}\\
 \dotM  &  -7.00, -6.75, -6.50, -6.25, -6.00\\
 \logL  &   5.00, 5.25, 5.50,  5.75 \\
 \Tstar &   15\,000, 17\,500, 20\,000, 22\,500, 25\,000, 27\,500 \\
 \Vwind &   500, 1\,000, 1\,500, 2\, 000, 2\,500 \\
\hline
\end{tabular}
\ec
\tablecomments{0.86\textwidth}{
Mass-loss rates (\Mdot) are in units of in\dotMyr.
Luminocities (\Lstar) are in units of $L_{\odot}$.
Stellar temperatures (\Teff) are in Kelvin.
Wind velocities (\Vinf) are in units of \kms.
}
\end{table}

The spectral characteristics of a massive star are intricately tied to the 
ionization structure of its stellar wind. Broadly 
speaking, the ionization state within a specific region of the stellar wind 
depends on the equilibrium between the 
quantity of ionizing photons and the density of gas in that region. 
While, the primary sources of ionizing photons in the stellar wind are 
generally the stellar luminosity and temperature, 
density structure of the wind is predominantly determined by the mass-loss rate and wind velocity.
Consequently, these four parameters (\Lstar, \Teff, \Mdot and \Vinf) serve 
as the fundamental physical parameters 
defining the features of stellar spectra. Therefore, we have chosen these 
parameters as the primary inputs for our 
model grids. 

Considering all of this and the needs of our future studies, we have 
calculated  model grids for nitrogen-rich (WN) and 
carbon-rich (WC) Wolf-Rayet stars, as well as Blue Supergiants (BSGs) 
characterized by low metallicity similar to the 
Small Magellanic Cloud (SMC).
Throughout the text, we will denote them WC grid, WN grid and SMC grid.
Each \cmfgen model in our grids incorporate the of elements of H, He, C, N, O, Ne, 
Mg, Al, Si, P, S, Ar, Ca, and Fe.

\corr{The chosen parameter ranges for varying \Lstar, \Mdot, \Vinf and \Teff are
appropriate for these objects \citep[see e.q.][and references
therein]{sander_19,hamann_19} and the corresponding selected values are
given in Table~\ref{tab:grids}.
For the WN and WC grids, the adopted abundances are correspondingly from 
\cite{hucht86}.
For the SMC grid, the adopted He abundance is 25\,\% by mass, while the
other elements have metallicity of  [Fe/H] =
-0.95 dex \citep{choudhury18}.}

For the turbulent velocity,
a value of $v_{turb} = 20$\kms throughout the wind is used for the Wolf-Rayet 
grids, while a value of $v_{turb} = 10$\kms is used for the SMC grid.

 While we are actively working to expand the BSG grids to higher metallicities 
and the WR grids to a broader parameter space, it is important to emphasize that
 building  grids is not the main focus of this study. Instead, these grids are provisional 
 and even temporary,  as key assumptions—such as metallicity, clumping, velocity law etc — remain uncertain

\subsection{4D grid modelling
- a new approach to direct non-linear fitting of stellar spectra} 
\label{subsec: gridmodelling}
	The objective of this study is to develop a numerical methodology for grid-fitting stellar spectra of massive stars and estimate its corresponding numerical uncertainties.
	To accomplish this objective, it is crucial to compare theoretical and fitted spectra with established stellar parameters. However, in reality this is not possible, and associated uncertainties 
	from numerical fitting techniques often remain ambiguous. 
	It is only through a clear understanding of these associated numerical uncertainties that we can gauge the reliability of any parameters derived from the comparison of theoretical spectra with real observations.    

 As  demonstrated by 
\cite{zhekov20}, our choice is to carry out {\it direct} fitting of 
an observed spectrum with a theoretical spectrum `extracted' from a grid 
of theoretical spectra (Section~\ref{subsec:grids}). In that study, we 
used 3D spectral grids while we deal here with technically more complex 
(but physically more sophisticated) case of 4D grids. Thus, our fitting
procedure has two parts: (a) how we calculate a model spectrum, using a 
4D spectral grid; (b) how we estimate the `similarity (correspondence)' 
between the observed and theoretical spectra.
It is important to emphasise that for fitting procedures of this nature,
 whether working with absolutely calibrated observed spectra or observed 
 magnitudes (i.e. using spectral energy distribution), a crucial requirement
 is having knowledge of the distance to the studied object.

It's important to mention that our fitting procedure operates in log-log space. 
In other words, each applied interpolation works with the logarithm of the dependent function (specifically, the spectrum, which represents the spectral density) and the logarithm of the independent variable.
By such a choice, all the
approximations result in physically meaningful (i.e., positive) values
for the function (i.e., spectral density).

\underline{Calculating a model spectrum.}
In our model grids, we provide {\it precise} spectra exclusively at the 
grid nodes. Nevertheless, to generate spectra 
for parameter values that fall between these nodes, we must utilize an 
approximation method. 

\begin{figure}
\begin{center}
\includegraphics[width=2.8in, height=2.0in]{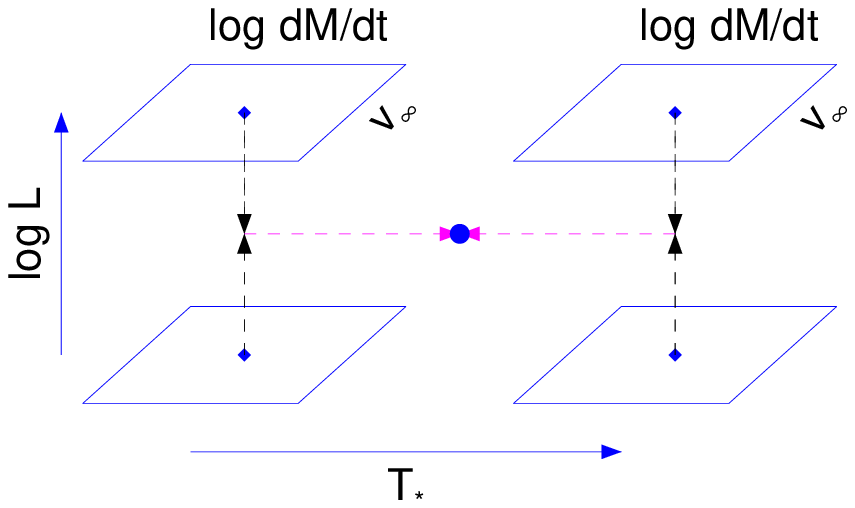}

\end{center}
\caption{Schematic diagram of the {\it quatro}-linear interpolation. To
calculate a model spectrum defined by a set of stellar parameters
(\dotME, \VwindE, \logLE, \TstarE)$_0$, marked by a 
filled-in circle in blue colour, we make the following steps. First, we
run linear interpolation along the \dotM and \Vwind axes to calculate 
the spectra at the (\dotME, \VwindE)$_0$ point in the four 
squares in blue colour. Then, we perform two linear interpolations along 
the \logL axis to get the spectra at the \logLE$_0$ points (marked by 
dashed lines and arrows in black). Finally, we derive the spectrum at 
the point of interest by a linear interpolation along the \Tstar axis 
(marked by dashed lines and arrows in magenta colour).
}
\label{fig:cratoon}

\end{figure}

We tried various approximations and found that the most reliable, that
is the most stable and with acceptable accuracy, is linear interpolation 
in the log-log space. Since we deal with 4D grids, by analogy with the
2D case, for which the standard procedure could be the bi-linear
interpolation we perform {\it quatro}-linear interpolation. So, if we
want to calculate a spectrum for a set of stellar parameters (\TstarE,
\dotME, \logLE, \VwindE) that fell in a specific 4D `cube', we perform
linear interpolation consecutively from 1D through 2D and 3D up to 4D,
that is along all the `axes' of the 4D `cube' (Fig.~\ref{fig:cratoon}
presents a schematic diagram of the {\it quatro}-linear interpolation). 
We recall that the
bi-linear interpolation is of higher order (higher accuracy) than the
linear approximation (e.g., see sec. 3.6 in \citealt{press_92}), thus, 
the adopted {\it quatro}-linear interpolation is of higher order (higher 
accuracy) than the bi-linear interpolation. An important detail is the 
choice of the independent variable for the interpolation process.
 
Understandably, the basic case is to choose 
the logarithm of 
 actual physical
parameters as independent
variables for the quatro-linear interpolation (linear interpolation
along each dimension of the 4D grid). Let us denote it `basic
approximation' or B-approximation.

On the other hand, since the line emission is a very important feature 
in the spectra of massive stars of early types, we consider two other
cases of the independent variable: the emission measure scale and the
transformed radius. Those two are somehow indicative of the stellar 
wind emission. 
We denote these cases EM-approximation and RT-approximation,
respectively.

We note that the emission processes that take place
in formation of the stellar wind emission are two-particle collisions.
So, it is natural to expect that line and continuum emission of
various ionic species to be proportional to the emission measure scale:
\begin{equation}
EM_S = n^2 V = \left(\frac{\dot{M}}{v_{\infty}}\right)^2 R_*^3 =
\left(\frac{\dot{M}}{v_{\infty}}\right)^2 \frac{L^{3/2}}{T_*^6}
\label{eqn:em}
\end{equation}
where $n$ is the gas number density and $V$ is the corresponding 
volume
for a specific ion, and each line-emission volume is proportional to the
third power of the reference radius ($R_*$, see eq.~\ref{eqn:tstar}), 
because the latter sets the scale of the emission region in the stellar 
wind of a massive star. 

The term transformed radius ($R_t$) was introduced by 
\citet{schmutz_89} 
and has been used in modelling of observed spectra of massive stars (see
\citealt{hamann_19}, \citealt{sander_19} and references therein). And,
in terms of stellar parameters it reads:
\begin{equation}
\label{eqn:RT}
R_t = R_* \left(\frac{v_{\infty}}{\dot{M}}\right)^{\frac{2}{3}} =
\frac{L^{1/2}}{T_*^2} \left(\frac{v_{\infty}}{\dot{M}}\right)^{\frac{2}{3}} 
\end{equation}
where again we made use of eq.~\ref{eqn:tstar}. 

As seen from eqs.~\ref{eqn:em} and \ref{eqn:RT}, both interpolation
variables depend on all the basic stellar parameters considered in this
study. Therefore, when performing grid interpolation, we can use any of 
these variables only for one of the `axes' of our 4D grid that is
directly responsible for the stellar wind emission (i.e., \dotM and
\VwindE). So, for the EM-approximation and RT-approximation we use the
corresponding interpolation variable on one of these two axes and we
adopt linear interpolation for the rest three axes (as in the
B-approximation).

As result, our fitting procedure has FIVE approximation cases to 
calculate a model spectrum based on a 4D grid of stellar atmosphere
model spectra: one B-approximation, 
two EM-approximations and two RT-approximations.

\underline{\corr{Match} between observed and theoretical spectra.}
A standard way in spectral fitting is to adopt some likelihood
estimator ($LE$) in order to check how well a theoretical spectrum 
matches an observed one. As a rule, the minimum value of this likelihood 
estimator defines a set of fitting parameters reproducing the observed 
spectrum `best'.
We explored spectral fitting with different likelihood estimators 
($\chi^2$-minimization among others) and we achieved best results with 
estimators based on absolute deviations between the observed and 
theoretical spectra: 
$ z_i = Sp_{obs}(\lambda_i) - Sp_{mod}(\lambda_i) $.
Since we fit an observed spectrum, the model spectrum is subject to the
interstellar extinction (reddening), that is we fit for five parameters:
four stellar parameters \TstarE, \dotME, \logLE, \Vwind and E(B-V).

Our fitting procedure adopts the following robust likelihood estimators:

\begin{eqnarray}
\label{eq:LE}
LE_1 = \sum_{i} |z_i| &\,& \hspace{0.10in} LE_2 = \sum_{i} \ln (1 +
|z_i|) \nonumber \\
                      &\,&   \\
LE_3 = \sum_{i} \frac{|z_i|}{1 + |z_i|} &\,& \hspace{0.10in} LE_4 =
\sum_{i} \tanh(|z_i|) \nonumber 
\end{eqnarray}

$LE_1$ is the sum of the absolute deviations between the observed
and model spectra, $LE_2$ is defined by analogy with the Cauchy
distribution (e.g., see sec. 15.7 in \citealt{press_92}), $LE_3$ and
$LE_4$ are experimental estimators that have the property to decrease
the weight of the deviant points in the fit while searching for the 
minimum of a $LE$-function. We note that the $LE_2$-, $LE_3$- and 
$LE_4$-function behaviour is similar to that of the standard
$LE_1$-function for small values of $z_i$, but $LE_2$-, $LE_3$- and 
$LE_4$-functions are more robust for the deviant points. Such points
might be a result both from the observational uncertainties as well as 
from uncertainties in numerical approximations.

On the technical side, we adopted the downhill simplex method to search
for the minimum of the $LE$-functions as its algorithm is exemplified in
the {\it amoeba} procedure (see sec. 10.4 in
\citealt{press_92})\footnote{For each spectral approximation, we 
performed tests to compare results from {\it amoeba} procedure and those
from a gradient method for searching the minimum of a function (e.g., 
Davidson-Fletcher-Powell algorithm, see sec. 10.7 in \citealt{press_92}) 
and the derived `best' sets of parameters were identical. So, we adopted 
the downhill simplex method in this study, because it does not require 
calculations of derivatives, thus, it introduces less numerical noise 
(uncertainties).}.

In summary, our 4D-grid fitting process consists of the following steps.
First, an individual fit obtains the 'best' set of model parameters by
making use of the five approximations previously described: one 
B-approximation, two EM-approximations and two TR-approximations.
Namely, it picks up the result from that approximation which provides
the smallest minimum of the  $LE$-function at hand.
Second, we carry out spectral fit for each of the four $LE$-functions 
defined above (see eq.~\ref{eq:LE}). As a result, we derive four `best'
sets of stellar parameters that define a model spectrum matching the
observed one. And, the mean of these four `best' sets constitutes the
solution from our fitting process.

\section{Test results}
\label{sec:tests}
The accuracy of the derived stellar parameters from 4D-grid fitting of
observed spectra relies on the quality of the observed spectra and used
fitting approach. An illustration of how 4D-grid fitting can be used to
estimate stellar parameters from fitting an observed spectrum is
provided in Appendix~\ref{app:wr23}. However, before estimating a
reliable range of derived stellar parameters, it is imperative to
investigate and establish the level of uncertainties arising from
applying the fitting procedure.

In order to investigate this, we carried out tests for each of the 4D grids of
model spectra (WC, WN and SMC grid). Specifically, for each grid we calculated over 20 synthetic spectra using the \cmfgen code. These were used as 'perfect' observed spectra in the UV-optical domain.
For each `perfect' observed spectrum, we used a set of stellar
parameters that were randomly chosen within the boundaries of the
corresponding 4D grid. Each `perfect' observed spectrum was subject to
interstellar reddening, the value of which was randomly  chosen as well.
And, some fiducial value for the distance (e.g., 2000 pc) was associated with each object.
We then fitted these test spectra to derive the
corresponding stellar parameters. Differences between the {\it input}
and {\it derived} values of the stellar parameters provide estimates
of the expected accuracy from using our fitting procedure.
Also, we performed additional tests by adding Gaussian noise to the
`perfect' observed spectra (`noisy' observed spectra), which could be
considered more realistic representation of the observed spectra.

\corr{We note that stars with strong winds, such as those represented by WN and WC grids, 
typically exhibit UV-optical spectra that are richer in line emission
 than those from the SMC grid, which correspond to objects with weaker winds. 
 As a result, }we would expect that the stellar wind
parameters for objects with strong winds would be derived with a better
accuracy by our 4D-grid fitting procedure.

\begin{table}
\caption{Test results (absolute difference, Model - Fit),
spectral range 1150 - 11000 \AA.}

\label{tab:test}
\bc
\begin{tabular}{ccccccc}
\hline\hline
\multicolumn{1}{c}{Parameter} &
\multicolumn{2}{c}{WC}  &
\multicolumn{2}{c}{WN}  &
\multicolumn{2}{c}{SMC} \\
\hline
\multicolumn{1}{c}{} &
\multicolumn{1}{c}{mean} & \multicolumn{1}{c}{max} &
\multicolumn{1}{c}{mean} & \multicolumn{1}{c}{max} &
\multicolumn{1}{c}{mean} & \multicolumn{1}{c}{max} \\
\multicolumn{1}{c}{} &
\multicolumn{1}{c}{(dex)} & \multicolumn{1}{c}{(dex)} &
\multicolumn{1}{c}{(dex)} & \multicolumn{1}{c}{(dex)} &
\multicolumn{1}{c}{(dex)} & \multicolumn{1}{c}{(dex)} \\
\hline
\multicolumn{7}{c}{`Perfect' spectra}        \\
 \Tstar  & 0.006 & 0.021   &   0.003 & 0.010   &   0.001 & 0.002   \\
 \dotM   & 0.008 & 0.019   &   0.006 & 0.020   &   0.016 & 0.047   \\
 \logL   & 0.006 & 0.029   &   0.004 & 0.012   &   0.002 & 0.007   \\
 \Vwind  & 0.006 & 0.019   &   0.006 & 0.017   &   0.012 & {\bf 0.052} \\
\multicolumn{7}{c}{SNR $\approx$ 100}        \\
 \Tstar  & 0.006 & 0.021   &   0.004 & 0.012   &   0.001 & 0.002   \\
 \dotM   & 0.008 & 0.019   &   0.007 & 0.020   &   0.014 & {\bf 0.058} \\
 \logL   & 0.006 & 0.029   &   0.005 & 0.013   &   0.002 & 0.005   \\
 \Vwind  & 0.006 & 0.019   &   0.008 & 0.019   &   0.009 & 0.048   \\
\multicolumn{7}{c}{SNR $\approx$ 50}        \\
 \Tstar  & 0.006 & 0.021   &   0.004 & 0.013   &   0.001 & 0.002   \\
 \dotM   & 0.008 & 0.019   &   0.006 & 0.020   &   0.014 & {\bf 0.055} \\
 \logL   & 0.006 & 0.029   &   0.005 & 0.013   &   0.002 & 0.006   \\
 \Vwind  & 0.006 & 0.019   &   0.008 & 0.019   &   0.009 & 0.042   \\
\multicolumn{7}{c}{SNR $\approx$ 25}        \\
 \Tstar  & 0.007 & 0.034   &   0.005 & 0.013   &   0.001 & 0.002   \\
 \dotM   & 0.008 & 0.020   &   0.007 & 0.020   &   0.016 & {\bf 0.077} \\
 \logL   & 0.007 & 0.028   &   0.006 & 0.014   &   0.003 & 0.007   \\
 \Vwind  & 0.006 & 0.020   &   0.009 & 0.020   &   0.010 & 0.038   \\
\hline
\end{tabular}
\ec
\tablecomments{0.86\textwidth}{The labels WC, WN and SMC denote the tests for the WC grid, WN grid 
and 
SMC grid, correspondingly. The `mean' and `max' columns give the mean
and maximum absolute difference between the input model parameter and 
its value derived from the fits.
The SNR label denotes the mean signal-to-noise ratio of the model
spectra with added Gaussian noise.
The values given in bold are those that are beyond the boundary accuracy
(0.05 dex) for the derived parameter. 
In each of those cases, there is only {\it one} value that is beyond the 
mentioned boundary, i.e. in $\sim 1$\% of the SMC cases.}
\end{table}

\begin{figure*}
\centering
\includegraphics[width=2.8in, height=2.0in]{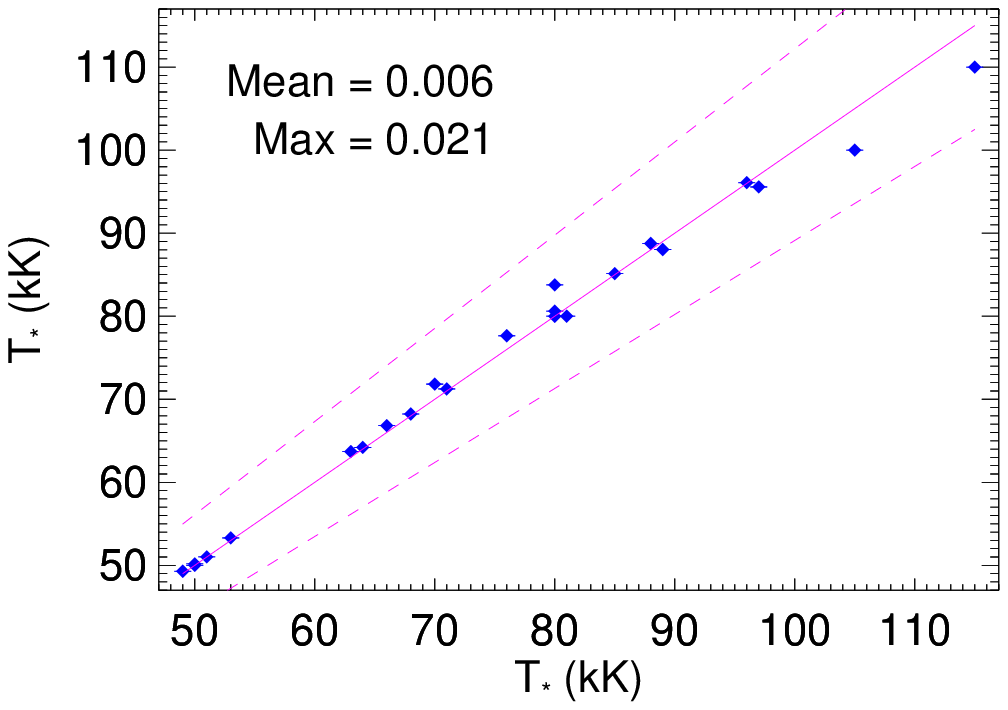}
\includegraphics[width=2.8in, height=2.0in]{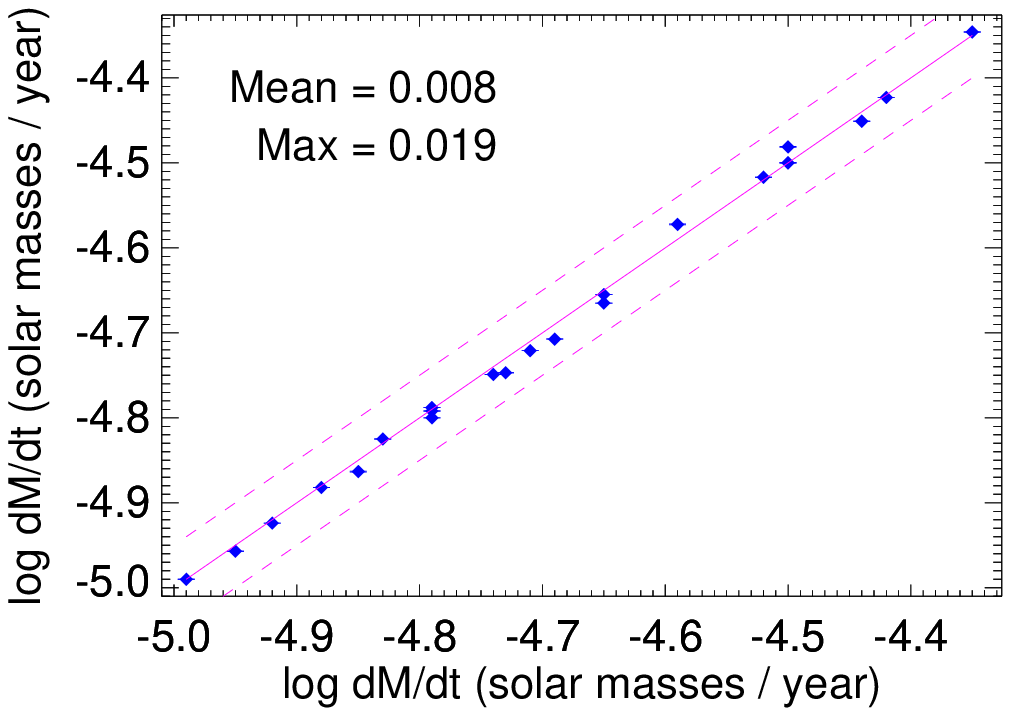}
\includegraphics[width=2.8in, height=2.0in]{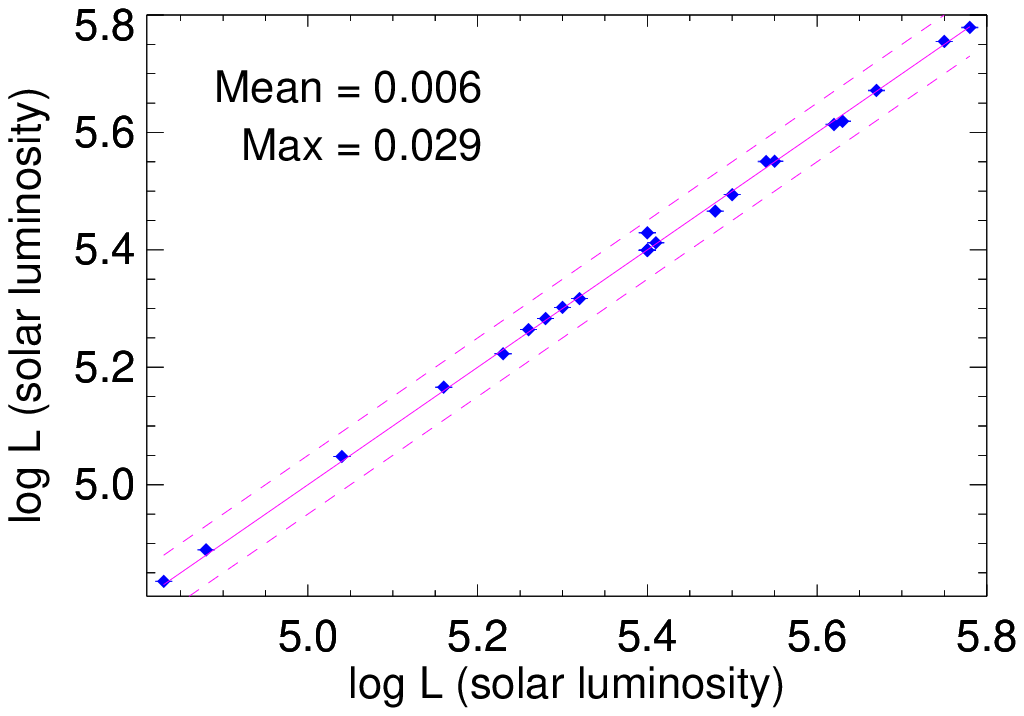}
\includegraphics[width=2.8in, height=2.0in]{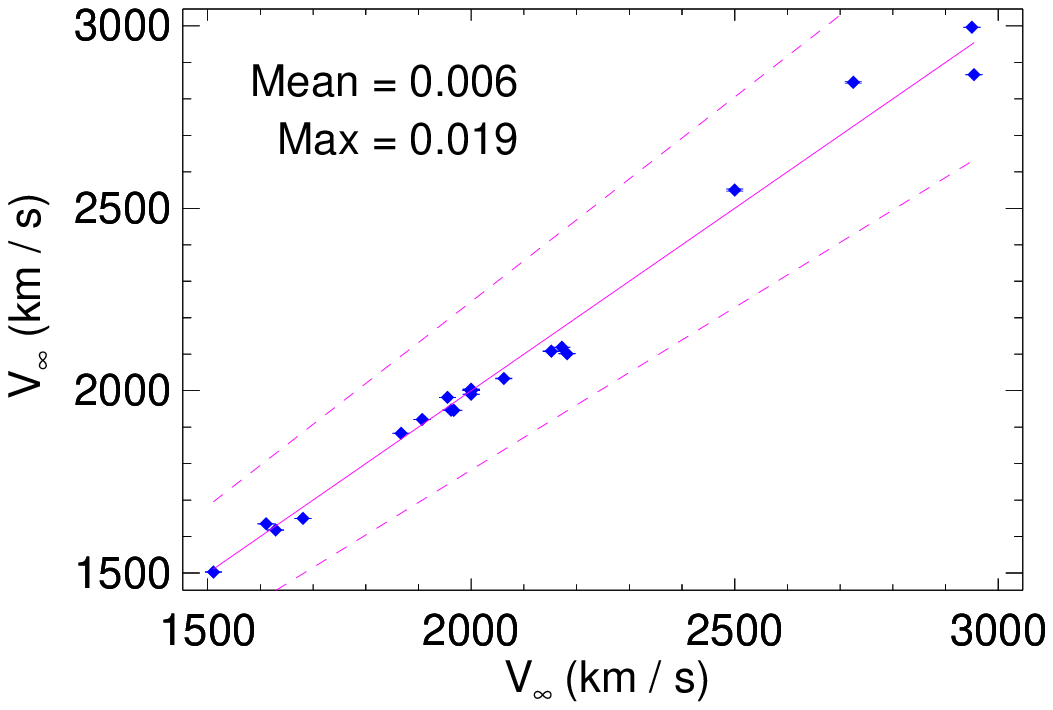}
\caption{Test results for the WC grid (`perfect' observed spectra): 
effective temperature, mass-loss
rate, luminosity and terminal wind velocity. Lines in magenta colour:
(a) the solid line represents the perfect correspondence between the 
input model parameters and their values derived from the grid-modelling; 
(b) the two dashed lines represent the boundary for the $\pm 0.05$~dex 
absolute deviation from the expected perfect correspondence. Mean and
Max labels denote the mean absolute deviation for the test sample and
its maximum value.
Error bars are the standard deviation of the four minimizations adopted.
}
\label{fig:WC_grid}
\end{figure*}

\begin{figure*}
\centering
\includegraphics[width=2.8in, height=2.0in]{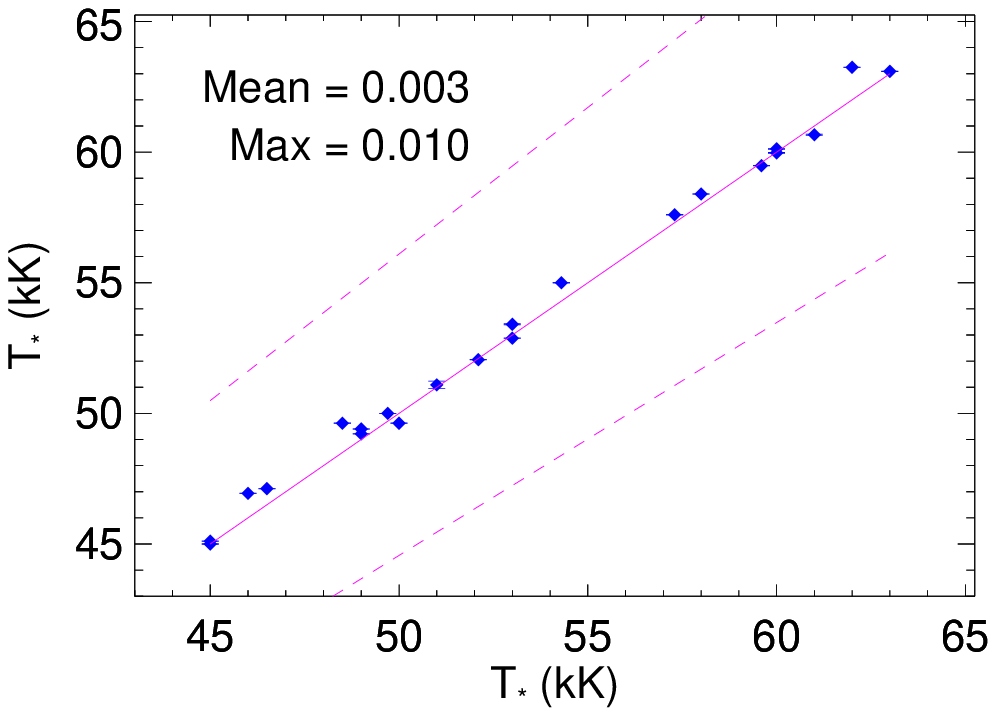}
\includegraphics[width=2.8in, height=2.0in]{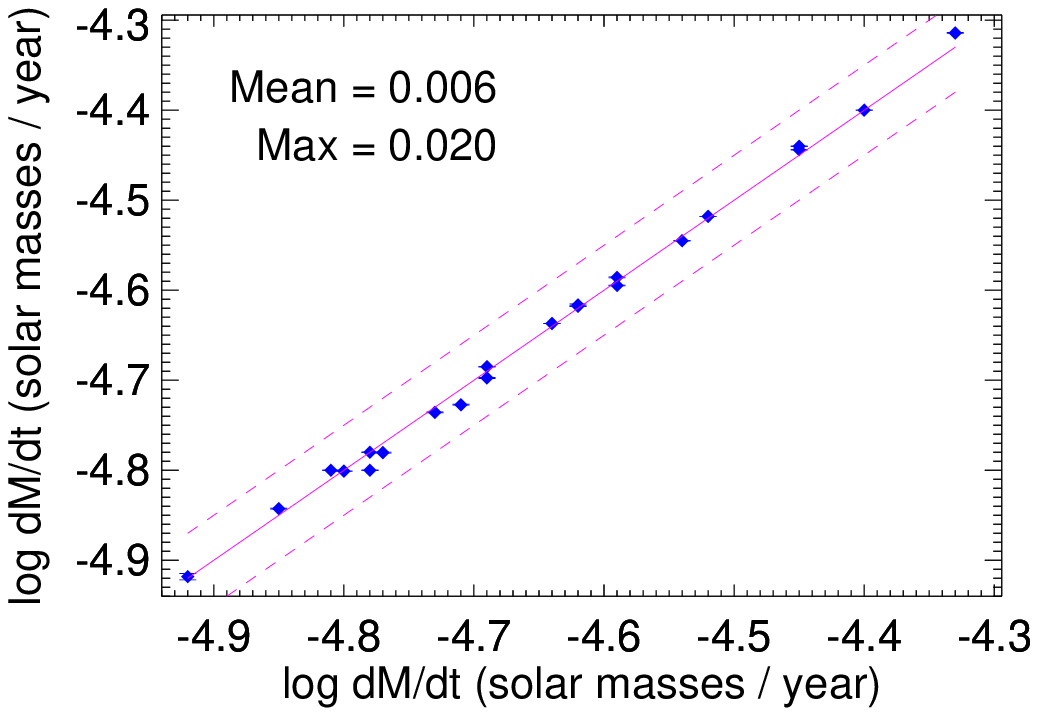}
\includegraphics[width=2.8in, height=2.0in]{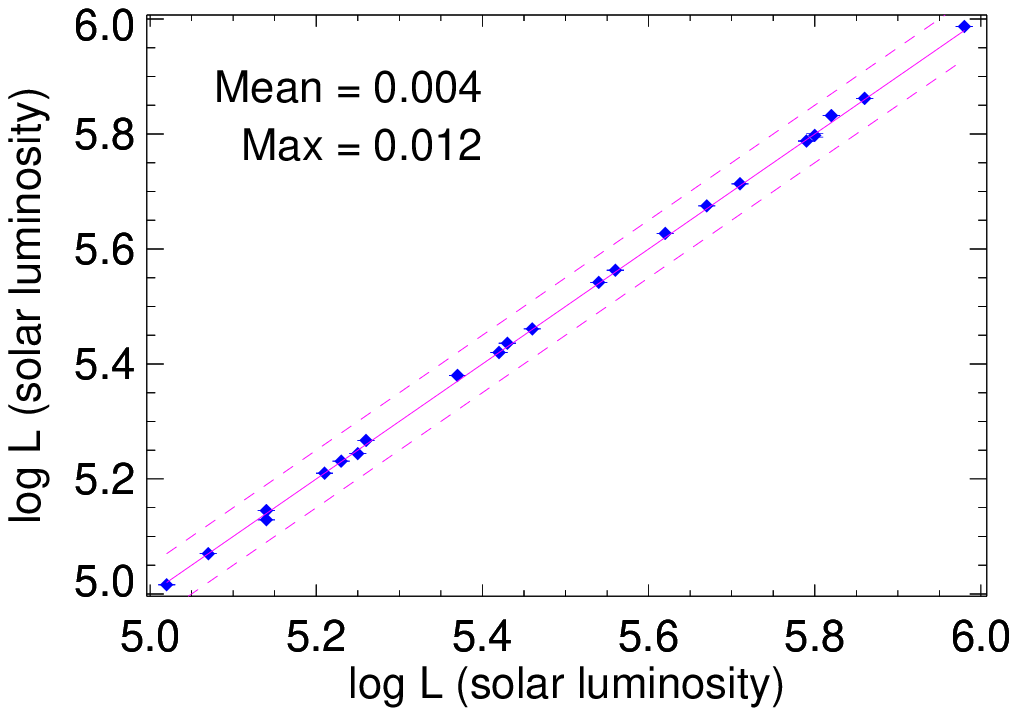}
\includegraphics[width=2.8in, height=2.0in]{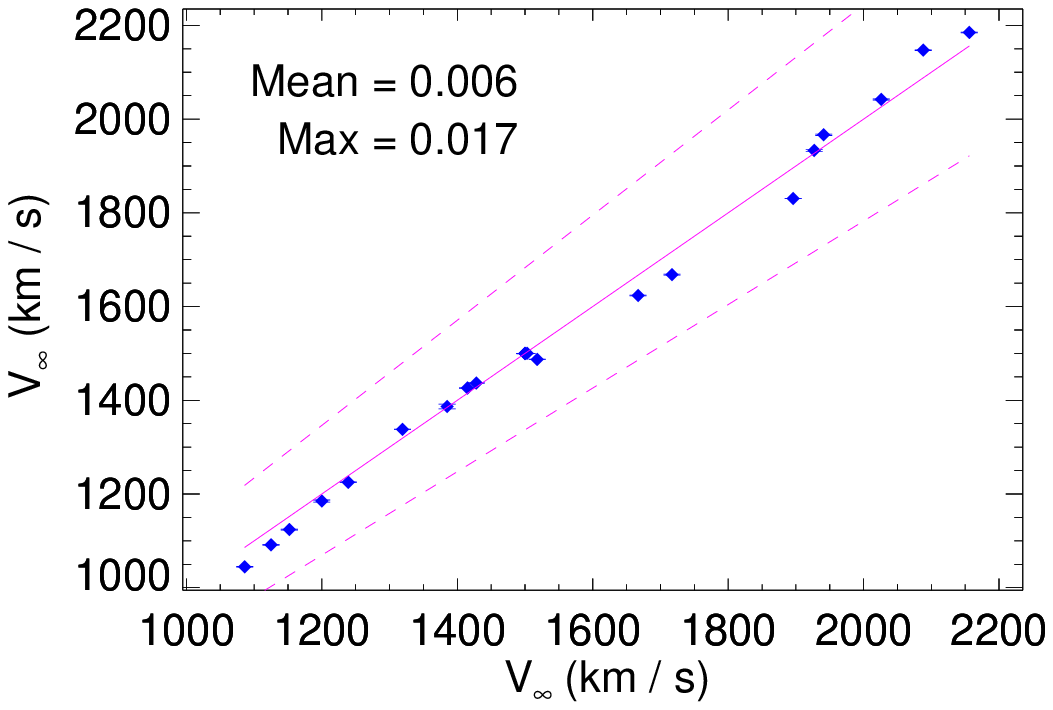}
\caption{The same as Fig.~\ref{fig:WC_grid} but for the WN grid
(`perfect' observed spectra).
}
\label{fig:WN_grid}
\end{figure*}

\underline{Objects with strong winds.}
Results from the tests for the `perfect' observed spectra of WC and WN
stars are shown in Figs.~\ref{fig:WC_grid} and \ref{fig:WN_grid},
respectively. 
Detailed results from these tests, along with those for the 'noisy' 
observed spectra, 
are presented in Table~\ref{tab:test}. 
It is evident
that all the {\it absolute} deviations of the derived stellar parameters
from their respected {\it input} values are well within the limits of
(smaller than) 0.05 dex. 
 Moreover, the mean absolute deviation for the WC and WN samples 
is significantly below 0.05 dex. However, we 
opt for a more conservative approach by assuming that the numerical uncertainty
is determined by the maximum deviation 
observed in the tests
(shown in the `max' column of Table~\ref{tab:test}). Interestingly, this 
uncertainty limit (0.05 dex)
also holds for the case of `noisy' test spectra (Table~\ref{tab:test}).

\begin{figure*}
\centering
\includegraphics[width=2.8in, height=2.0in]{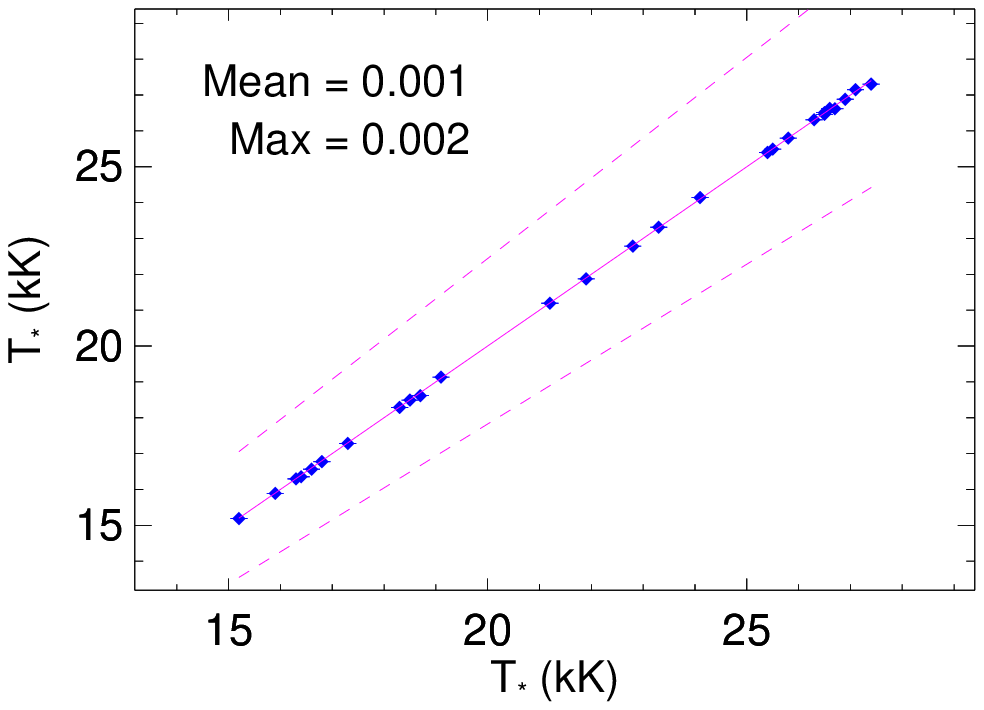}
\includegraphics[width=2.8in, height=2.0in]{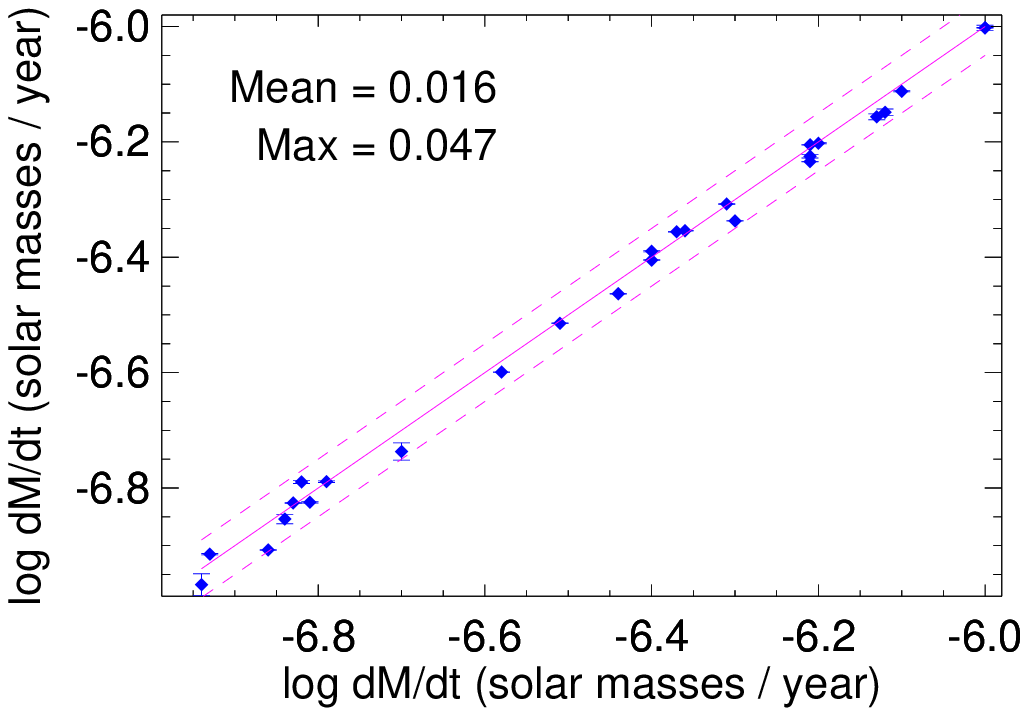}
\includegraphics[width=2.8in, height=2.0in]{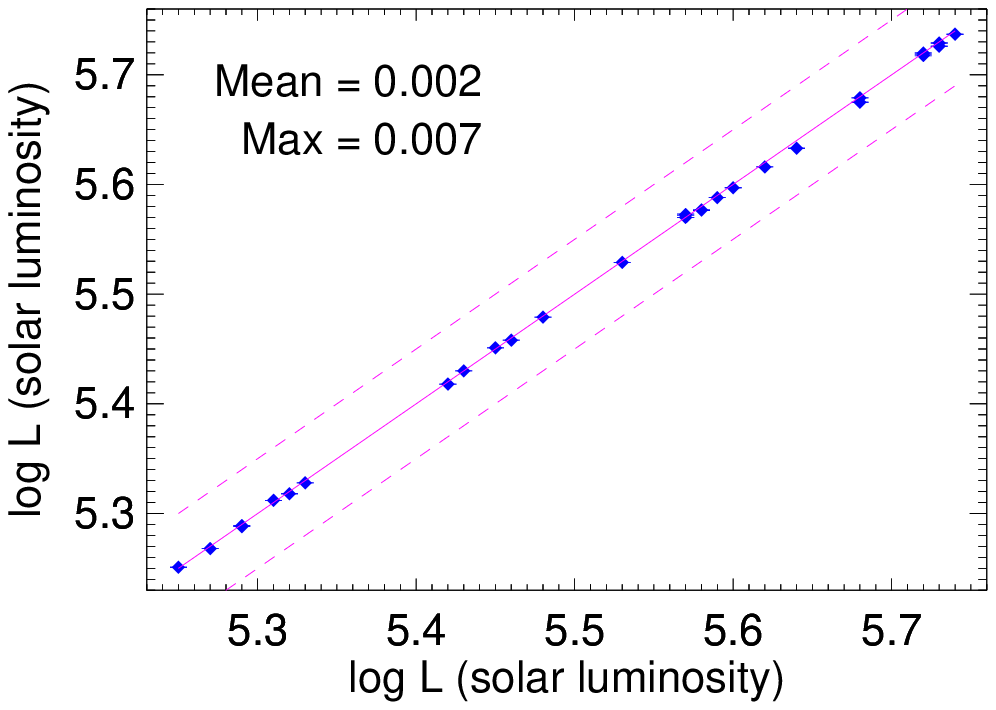}
\includegraphics[width=2.8in, height=2.0in]{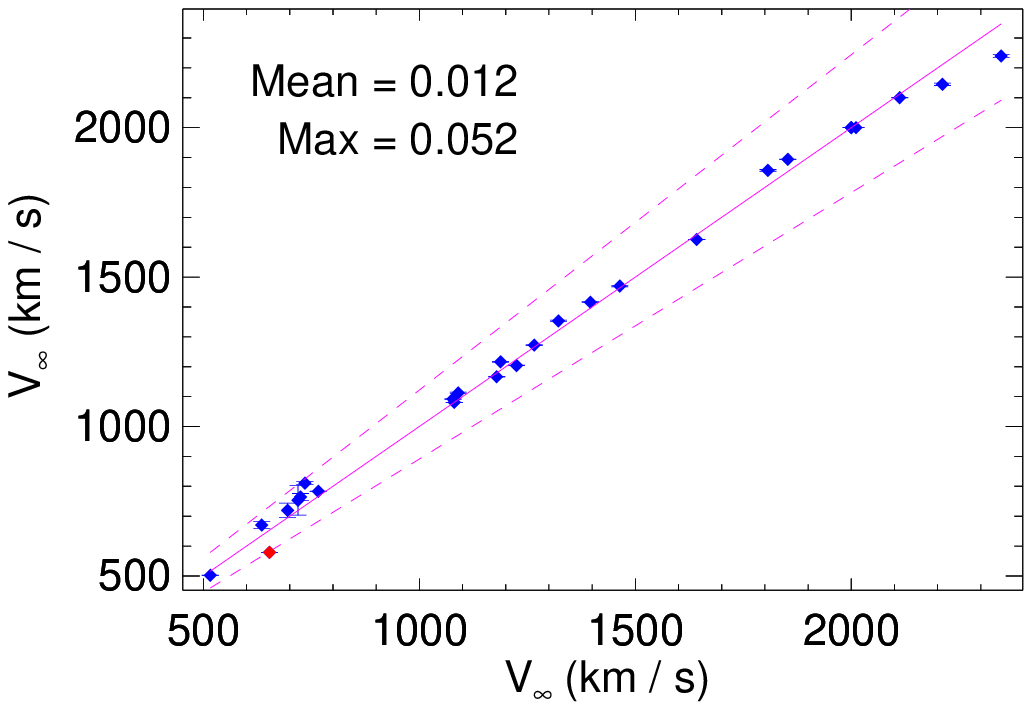}
\caption{The same as Fig.~\ref{fig:WC_grid} but for the SMC
grid (`perfect' observed spectra). The derived parameters that have 
absolute deviation $> 0.05$~dex are given in red colour.
}
\label{fig:SMC_grid}
\end{figure*}

\underline{Objects with weak winds.}
Figure~\ref{fig:SMC_grid} (`perfect' observed spectra) and 
Table~\ref{tab:test} (`perfect' and `noisy' observed spectra) present 
results from the tests for the SMC objects. As in the case of objects
with strong winds, the mean absolute deviation of the derived stellar
parameters is appreciably less than the accuracy limit of 0.05 dex.
However, we see that occasionally larger deviations occur (in 
$\sim 1$\% of the cases).

In general, we may conclude that in both cases (strong and weak winds)
numerical uncertainty of the derived stellar parameters is {\it not worse} than 0.05
dex. But, it is interesting to check whether the spectral range of the 
observed spectra may have some effect on the parameter accuracy as 
well.

\underline{Spectral range.}
Results from the tests for the `perfect' observed spectra of WC, WN and
SMC objects in the optical (3150 - 11000 \AA) are given in
Table~\ref{tab:test_opt}. We see again that the numerical accuracy of the derived
parameters is {\it not worse} than 0.05 dex for the objects with strong
winds (WC and WN spectra). However, the quality of the derived 
parameters
slightly deteriorates for the objects with weak winds (SMC spectra): the
fraction of cases with parameter accuracy greater than 0.05 dex 
increases from $\sim 1$ \% to $\sim 8$ \%.

Thus, it is reasonable to conclude that the larger the spectral range
and the better the quality of the observed spectra of massive stars is, 
the better constrained are the derived stellar parameters from the 
4D-grid fitting.

\begin{table}
\caption{Test results (absolute difference, Model - Fit);
spectral range 3150 - 11000 \AA.}
\bc
\begin{tabular}{ccccccc}
\hline\hline
\multicolumn{1}{c}{Parameter} &
\multicolumn{2}{c}{WC}  &
\multicolumn{2}{c}{WN}  &
\multicolumn{2}{c}{SMC} \\
\multicolumn{1}{c}{} &
\multicolumn{1}{c}{mean} & \multicolumn{1}{c}{max} &
\multicolumn{1}{c}{mean} & \multicolumn{1}{c}{max} &
\multicolumn{1}{c}{mean} & \multicolumn{1}{c}{max} \\
\multicolumn{1}{c}{} &
\multicolumn{1}{c}{(dex)} & \multicolumn{1}{c}{(dex)} &
\multicolumn{1}{c}{(dex)} & \multicolumn{1}{c}{(dex)} &
\multicolumn{1}{c}{(dex)} & \multicolumn{1}{c}{(dex)} \\
\hline
\multicolumn{7}{c}{`Perfect' spectra}        \\
 \Tstar & 0.006 &  0.021 &  0.003 &  0.012 &  0.001 &   0.002 \\
 \dotM  & 0.007 &  0.019 &  0.006 &  0.016 &  0.033 &   {\bf 0.126} \\
 \logL  & 0.006 &  0.026 &  0.004 &  0.020 &  0.003 &   0.006 \\
 \Vwind & 0.006 &  0.017 &  0.006 &  0.017 &  0.035 &   {\bf 0.192} \\
\hline
\end{tabular}
\ec
\tablecomments{0.86\textwidth}{All the columns and labels are as in Table~\ref{tab:test}.
The values given in bold are those that are beyond the boundary accuracy
(0.05 dex) for the derived parameter. 
There are 9 (nine) values in total that are beyond the mentioned 
boundary (5 for \dotM and 4 for \VwindE), i.e. in $\sim 8$\% of the 
SMC cases.
\label{tab:test_opt}}
\end{table}

\section{Discussion}
\label{sec:discussion}

It is essential to acknowledge that our fitting technique works within
 the framework of stellar atmosphere models,  
relying on certain assumptions about key properties such as chemical composition, 
wind velocity profile ($\beta$-law)\footnote{In general, stellar rotation may 
influence the velocity profile of the wind and its structure (e.g.,
corotation interaction regions could also form), which will have an impact on the 
emergent spectrum. However, considering this case will require much 
more complex modelling: i.e., at least 2D stellar atmosphere models.}, 
and  volume filling factor (clumping). However, this limitation applies to any study based on stellar atmosphere models.
What if the actual stellar properties do not correspond to those adopted in this study?  It is indeed reasonable to expect some variation within a specific spectral class.
Factors such as abundances, wind acceleration law, and clumping might display a degree of variability around assumed 'typical' values, though generally within a certain range of expected limits.

Thus, we will next try to address in some detail this issue.
We aim to explore the effect of conducting a 4D-grid fitting on observed spectra of objects whose abundances, wind acceleration laws, and clumping deviate from the values of these parameters adopted in our 4D grids. Also we will briefly address the case of 4D grid-fitting of `composite' spectra of massive stars (e.g., of massive binary). 

\subsection{Abundances} 
\label{sec:abund}

For each of the 4D grids considered in our study (WC, WN and SMC), we
defined an additional abundance set labeled `Add.' (Table~\ref{tab:abu}).

The additional WC set is based on the WC abundances for galactic
metallicity as defined in \cite{sander12} and
the additional WN set represents the WN abundance set for galactic
metallicity as defined in \cite{ham_graf_04} with helium mass fraction of
0.78.\footnote{See WC MW abundances and MW WNL-H20 abundances provided in the website of the Potsdam Wolf-Rayet models, \url{https://www.astro.physik.uni-potsdam.de/~wrh/PoWR/powrgrid1.php}}.
The additional SMC abundance  set exhibits metal abundances that have
 approximately doubled (see Table~\ref{tab:abu} for details).
\begin{table*} 
\caption{Mass fraction of the most abundant chemical elements adopted in the spectral grids and test models in this study.}
\label{tab:abu}
\bc
\begin{tabular}{ccccccc}
\hline\hline
\multicolumn{1}{c}{Element} &
\multicolumn{2}{c}{WC}  &
\multicolumn{2}{c}{WN}  &
\multicolumn{2}{c}{SMC} \\
\multicolumn{1}{c}{} &
\multicolumn{1}{c}{Main} & \multicolumn{1}{c}{Add.} &
\multicolumn{1}{c}{Main} & \multicolumn{1}{c}{Add.} &
\multicolumn{1}{c}{Main} & \multicolumn{1}{c}{Add.} \\
\hline
 X$_H$       & 3.84E-4        &  1.99E-3       & 1.60E-2      & 2.01E-1    &  7.48E-1   &  7.47E-1 \\
 X$_{He}$   & 3.22E-1        &  5.50E-1       & 9.53E-1      & 7.80E-1    &  2.50E-1   &  2.50E-1 \\
 X$_C$       & 3.88E-1        &  3.86E-1       & 3.66E-4      & 1.00E-4    &  2.67E-4   &  4.74E-4 \\
 X$_N$       & 1.83E-5        &  1.83E-5       & 2.10E-2      & 1.50E-2    &  7.86E-5   &  1.39E-4 \\
 X$_O$       & 2.50E-1        &  5.00E-2       & 1.12E-3      & 1.00E-5    &  6.51E-4   &  1.15E-3 \\
 X$_{Ne}$   & 3.02E-2        &  3.02E-3       & 3.16E-3      & 1.00E-5    &  1.43E-4   &  2.52E-4 \\
 X$_{Fe}$   & 1.72E-3        &  1.60E-3       & 1.70E-3      & 1.40E-3    &  1.47E-4   &  2.59E-4 \\
\hline
\end{tabular}
\ec
\tablecomments{0.86\textwidth}{The labels WC, WN and SMC denote the basic cases in this study. 
The `Main' column gives the basic set of abundances used in
the corresponding 4D grids and test models, while the `Add.' column
gives the abundances in the additional test models
(see Sections~\ref{subsec:grids} and \ref{sec:abund}).}
\end{table*}

For each 4D grid, we prepared `perfect' observed spectra having the same 
stellar parameters as described in Section~\ref{sec:tests}, but their
abundances were those from the `Add.' set for WC, WN and SMC
objects. We used our 4D-grid fitting procedure and the corresponding
results are shown in Figs.~\ref{fig:WC_abu}, ~\ref{fig:WN_abu} and 
~\ref{fig:SMC_abu}.

In general, changes in abundances affects the derived parameters and the most important change is in the derived values of  the mass-loss rate.
In the case of strong winds, i.e in the WN and WC grids, alterations in \dotM also result in corresponding changes in \logL.
This is understood as the density profile, specific to a given chemical element in the stellar wind, undergoes changes solely due to variations in the mass-loss rate, the ionization structure must adapt in order to align with the observed spectrum. Consequently, the ionization agent (luminosity) is affected.
In the case of weak winds, i.e SMC grid, 
there are practically no changes in the derived values
of the effective temperature and luminosity (\Tstar and \logLE).
This is so since the stellar wind emission is not the dominant 
part of the spectrum, therefore, its changes does not lead to changes in the other stellar parameters.

\subsection{Wind acceleration law}
\label{sec:beta}
We recall that the 4D grids in this study were built adopting wind
acceleration
law, with $\beta = 1$. So for each 4D grid, we
prepared two additional sets of `perfect' observed spectra having the 
same stellar parameters as described in Section~\ref{sec:tests} and
values of $\beta = 0.5$ and $\beta = 2$. This means that we explore the
impact on the derived parameters if a massive star has a faster
accelerating ($\beta = 0.5$) or a slower accelerating ($\beta = 2$) wind
compared to that adopted in our 4D grids ($\beta = 1$).

The results from
fits of these spectra with our 4D-grid fitting procedure are shown in
Figs.~\ref{fig:WC_beta}, ~\ref{fig:WN_beta} and ~\ref{fig:SMC_beta}.
It is evident that there are appreciable {\it systematic} changes in the
derived stellar parameters. In general, their accuracy worsens.
For strong winds, it could be assumed {\it not worse} than 0.1 dex
for all the stellar parameters. However for weak winds, the
stellar wind parameters may have larger deviations (up to 0.2 dex) while
the effective temperature and luminosity are recovered very well.

We think that the reason for such {\it systematic} changes is due to the
fact that different $\beta$-laws result in considerable change of the
velocity and density profiles of the stellar-wind plasma. Therefore,
there are large changes in the ionization structure of the emission 
region of various ionic species as well. And, the 4D-fitting procedure
tries to handle this within the available 4D grid. For example, if the
ionization structure `moved in' closer to the star, it chooses smaller
stellar wind velocities and the opposite is valid if the ionic structure
`moved out' further from the star. Next, it adjusts the amount of line
emission by choosing the value of the mass-loss rate. In turn, different
values of the latter may cause some changes in the derived values for
luminosity. And, the effective temperature is no stranger here since its
value also has some saying when it comes to ionization structure: it
defines the `typical' energy of the ionizing photons. In other words,
the complex behaviour of the derived stellar parameters follows from the
strong {\it non-linearity} of the physics involved in formation of the
emission from massive stars of early spectral types.

Considering the significant impact of wind acceleration on the determined 
stellar parameters, we believe that the $\beta$-law could serve as a 
promising candidate for future expanding  the grid-fitting procedure by 
adding another `dimension'.

It is worth recalling that the $\beta$-law is an approximation
to the velocity profile in the winds of massive stars. A more realistic
velocity profile can be modelled only if radiative hydrodynamics models
are adopted. However, given their complexity and need of computer
resources\footnote{When coupling radiative transfer and hydrodynamics,
even the {\sc fastwind} code (widely used to model the spectra of massive OB 
stars) is already not that fast according to its authors. The method is
described in \citet{sandqvist_19} and general description is found in the
{\sc fastwind} website
\url{https://fys.kuleuven.be/ster/research-projects/equation-folder/codes-folder/fastwind}.}, 
using such models to confront theory and observations 
by direct fitting of observed spectra of
numerous specific objects does not seem feasible. Thus, we feel
confident that expanding our fitting procedure by adding the $\beta$-law
`dimension' is a reasonable step to take. 

\subsection{Clumping}
\label{sec:clumping}
It is generally assumed that massive stars with different mass-loss rates that scale
with their corresponding volume filling factors ( 
$\dot{M}_1 / \sqrt{f_{\infty,1}} = \dot{M}_2 / \sqrt{f_{\infty,2}}$)
have very similar spectra, provided other stellar parameters are kept fixed.
Our 4D fitting procedure relies on approximations of the model spectrum, leading to the introduction of numerical noise. To check the effect of the latter on this scaling law, 
we prepared an additional set of `perfect' observed spectra having the
same stellar parameters as described in Section~\ref{sec:tests} and a
value of $f_\infty = 0.25$, however the mass-loss rates were scaled in accord with the scaling law: 
$\dot{M}_{new} / \sqrt{0.25} = \dot{M}_{old} / \sqrt{0.1}$ or 
\dotM$_{new} =$ \dotM$_{old} + 0.2$. This means
that the stellar parameters derived from applying our 4D-grid fitting 
procedure should remain the same as before. The results from
fits of these spectra are shown in
Figs.~\ref{fig:WC_clump}, ~\ref{fig:WN_clump} and ~\ref{fig:SMC_clump}.

We see that for strong winds although there are some small 
{\it systematic} changes in the derived parameters, the clumping scaling 
law seems to be acceptably working: the derived stellar parameters are 
still within the numerical accuracy of 0.05 dex as before ($f_{\infty} = 0.1$).
In contrast, when it comes to weak winds, the alteration in the mass-loss rate values is not insignificantly small. We attribute this behaviour to the imperfect nature of the clumping scaling law, where minor residuals are compounded with uncertainties arising from the numerical approximations to the model spectra adopted in  our 4D-grid fitting procedure.

Note that stellar atmosphere models of massive stars typically assume a constant filling factor. However,
a physically more realistic approach would involve clumping that varies with distance from the star. 
Considering such variability becomes feasible when analyzing individual objects 
 across diverse spectral domains like UV, optical and radio.  An illustration of such approach can be found in 
\citet{zhekov20} (see also appendix A in that source). 

Another fundamental assumption in these stellar atmosphere models is that the interclump 
medium has negligible contribution to the stellar spectrum, in other words, the intercloud space is `void of matter'
However, it seems improbable that such `empty' space exists in the winds of massive stars. It is more likely
 that the winds of massive stars consist of a {\it two-component} flow: a massive component composed of clumps and a low-density component that fills the interclump space. Evidence supporting this two-component picture also arises from the analysis of X-ray emission from colliding-wind binaries, which are binary systems consisting of two massive stars (see discussion in section 4.3 of \citet{zhekov20}).
We thus believe that addressing the possible emission from the
low-density wind component (interclump matter) is needed and we plan to
try carrying out even simple estimates in a future study.

\subsection{4D grid-fitting of `composite' spectra}
\label{sec:composite}
A `composite' spectrum  may arise, for example, from a binary system 
in which both components are massive stars. If the studied
object is a wide binary (orbital period of hundreds of days, years), 
then the wind-absorption effects for either of the massive-star spectra 
by its binary companion might be neglected. So, the total (`composite') 
spectrum would be a sum of two spectra subject to common  reddening 
(interstellar absorption).
An interesting (and simple) case to consider is when the stellar
components in a massive binary are of quite different spectral
class (type). In such a case, we can adopt the 4D grid-fitting (using 
two different grids), because the spectral components are in fact 
independent vectors (none of the spectra could mimic the other one).
Again, it is interesting to evaluate the level of uncertainties of the
derived stellar parameters, when adopting the 4D grid-fitting to a
`composite' spectrum.

To address this, we made used of the test spectra of WC and WN stars
(`perfect observed spectra') as described in Section~\ref{sec:tests}. 
We constructed more than 20 `composite' WC$+$WN spectra \corr{by combining
individual WC and WN model spectra and applying a common reddening to the resulting summed spectrum}. These spectra
were fitted with our 4D grid-procedure as at each step of the fitting
process a spectrum is calculated from each of the grids (WC and WN) and
the total model spectrum is the sum of both spectra.

Results from the 4D grid-fitting to the `composite' WC$+$WN test spectra
are given in Table~\ref{tab:test_bin_spectrum}. As in the tests of
fitting spectra of single massive stars, we explored different spectral 
ranges: UV-optical (1150 - 11000 \AA) and optical (3150 - 11000 \AA).
Interestingly, the numerical accuracy of the derived parameters is 
{\it not worse} than 0.05 dex in the case of UV-optical spectra and in
just one case (for one stellar parameter, i.e., in less than 1\%) it
exceeds that boundary. 

These results are quite encouraging for using the 4D grid-fitting of 
observed `composite' spectra of massive stars (e.g., wide massive
binaries).  We note that this procedure can be easily `upgraded' for 
a more complex case, e.g., of `composite' spectra with three spectral 
components. Results from applying the 4D grid-fiting in such a case 
are given in \citet{zhpe_25}.

\begin{table}
\caption{Test results (absolute difference, Model - Fit)
for `composite' WC$+$WN spectra.
\label{tab:test_bin_spectrum}}
\bc
\begin{tabular}{ccccccccc}
\hline\hline
\multicolumn{1}{c}{Parameter} &
\multicolumn{2}{c}{WC}  &
\multicolumn{2}{c}{WN}  &
\multicolumn{2}{c}{WC}  &
\multicolumn{2}{c}{WN}  \\
\multicolumn{1}{c}{} &
\multicolumn{4}{c}{(1150 - 11000 \AA)} & \multicolumn{4}{c}{(3150 - 11000
	\AA)} \\
\multicolumn{1}{c}{} &
\multicolumn{1}{c}{mean} & \multicolumn{1}{c}{max} &
\multicolumn{1}{c}{mean} & \multicolumn{1}{c}{max} &
\multicolumn{1}{c}{mean} & \multicolumn{1}{c}{max} &
\multicolumn{1}{c}{mean} & \multicolumn{1}{c}{max} \\
\multicolumn{1}{c}{} &
\multicolumn{1}{c}{(dex)} & \multicolumn{1}{c}{(dex)} &
\multicolumn{1}{c}{(dex)} & \multicolumn{1}{c}{(dex)} &
\multicolumn{1}{c}{(dex)} & \multicolumn{1}{c}{(dex)} &
\multicolumn{1}{c}{(dex)} & \multicolumn{1}{c}{(dex)} \\
\hline
\multicolumn{9}{c}{`Perfect' spectra}        \\
 \Tstar & 0.006 &  0.021 &  0.005 &  0.013 & 
	  0.010 &  {\bf 0.084} &  0.005 & 0.013 \\
 \dotM  & 0.008 &  0.021 &  0.008 &  0.020 & 
	  0.008 &  0.027 &  0.009 & 0.023 \\
 \logL  & 0.011 &  0.029 &  0.007 &  0.020 & 
	  0.012 &  0.035 &  0.007 & 0.023 \\
 \Vwind & 0.006 &  0.020 &  0.009 &  0.021 & 
	  0.007 &  0.023 &  0.008 & 0.021 \\
\hline
\end{tabular}
\ec
\tablecomments{0.86\textwidth}{All the columns and labels are as in 
Table~\ref{tab:test}. The second row marks corresponding results for 
the spectral ranges considered in the tests.
The values given in bold are those that are beyond the boundary accuracy
(0.05 dex) for the derived parameter.
There is 1 (one) value in total that is beyond the mentioned
boundary (1 for \TstarE), i.e. in $\sim 1$\% of the WN parameters: 
note that it is in the case when no UV data are used.
}
\end{table}

\section{Conclusions}
\label{sec:conclusions}

 We have developed a novel technique that enables direct fitting of observed spectra. To validate this technique, we have calculated 4D grids of model spectra for early spectral-class massive stars, specifically Wolf-Rayet stars (WC and WN spectral classes) and Blue Supergiants (BSGs) with low metallicity, similar to that of the Small Magellanic Cloud (SMC).

These model grids served as a testing ground  for developing and testing  our technique, as well as for estimating the expected numerical uncertainties. 
We demonstrated that this approach achieves numerical precision not exceeding 0.05 dex when estimating essential stellar parameters -- such as effective temperature, mass-loss rate, luminosity, and terminal wind velocity. This was confirmed through rigorous testing on designated `test' models. Even when Gaussian noise was added to the synthetic spectra, the mean absolute deviation consistently remained well below 0.05 dex for objects exhibiting both weak and strong stellar winds.
It is essential to note that the actual accuracy of the derived physical parameters depends not only on the fitting approach but also on how well the simplifying assumptions in the models and underlying theory reflect real conditions in stellar winds. However, without such (numerical) uncertainty estimates, even the most sophisticated methods would be unreliable in deriving and interpreting stellar parameters for physical analysis. Our results are encouraging and provide strong support for our goal of deriving fundamental stellar parameters from spectral analysis with the highest attainable accuracy.

 It is essential to note that the accuracy  of the derived parameters depends on the spectral range and the inclusion of ultraviolet (UV) spectral range  contributes to improved parameter derivations, particularly for objects with weak winds. This underscores the importance of  selecting the spectral range for comprehensive and accurate stellar parameter determination.

We explored the influence of unaccounted factors, such as variations in metal abundances, the chosen wind acceleration law, and the degree of clumping, on the precision of the derived parameters.
Notably, we found that variations in abundances predominantly affected the derived values of the mass-loss rate, particularly evident in scenarios with weak winds represented by our SMC grid. Interestingly, for weak winds, the derived values of effective temperature and luminosity remained largely unaffected by changes in abundances.

Furthermore, our tests revealed the significant influence of wind acceleration law on the accuracy of determined stellar parameters, with a noticeable decrease in precision, particularly up to 0.2 dex for objects with weak winds. In contrast, the application of different degree of clumping demonstrated good parameter precision (less than 0.05 dex) for objects with strong winds, but resulted in decreased precision  up to 0.2 dex for objects with weak winds.
This dichotomy underscores the interplay between the degree of clumping and the strength of stellar wind, emphasizing the need for careful consideration of these factors in the pursuit of precise stellar parameter determinations.
Consequently, our results affirm that the clumping scaling law is effective for objects characterized by strong winds, whereas its application to objects with weak winds results in a significant decline in precision, up to 0.2 dex.

	This suggests that the real accuracy of stellar parameters derived from any fitting technique using theoretical models will strongly depend on the systematic errors arising from the simplifying assumptions inherent in the theoretical models. 
	Specifically, a critical inference from our investigation is that the precise information about the wind velocity law is required  to ensure reliable determination of stellar parameters. Consequently, we plan to expand the parameter space 
	by including models  with diverse values of $\beta$. 
	This is necessary in order to enhance the accuracy of stellar wind parameter determinations.

On the technical side, we see at least two ways to achieve better accuracy in derived stellar parameters: (a) build `denser' model grids; (b) improve the numerical approximation of model spectra.

In practical terms, denser model grids with smaller parameter steps can enhance the accuracy of derived stellar parameters within the 4D grid fitting framework. However, this approach demands significant computational resources and time. The latter is due to the complexity of the grid-building process, which is not easily automated.

The need for numerical approximation arises because stellar atmosphere codes cannot produce detailed spectra of all chemical elements `instantaneously'. While it would be ideal to avoid numerical approximations, it is currently not feasible. With this respect, the accuracy of the derived stellar parameters will improve if we manage to find a `better' numerical approximation than those used in Section~\ref{subsec: gridmodelling} of this study.

Finally, we note that when analysing observed spectra of massive stars
there might be two major issues as illustrated by the numerical
experiments in this study. First, the wind of a real object may have
velocity profile different from that adopted in our 4D grids
($\beta$-law with $\beta = 1$). Second, its chemical composition may 
deviate from the abundance set used to calculate these grids. And, the
`combined' effect of both issues is hard to forsee. While the first
issue could be handled by adding a new `dimension' to the grids (no
matter whether `globally' or `locally', as discussed above), we do not 
think it is feasible to adopt the same approach for handling (estimating) 
the chemical composition from fitting the observed spectrum of a massive
star. The reason is that it is not possible to define a `global' parameter 
which could describe the variety of changes in the spectrum due to 
different chemical elements. Alternatively, adding more `dimensions' to
the grids even only for the most abundant elements is hardly to be
considered the way to go. Therefore, this `combined' effect needs be 
addressed in detail and the corresponding solution will likely adopt 
some iteration-procedure that must be carefully tested: a complex task 
that we plan to consider in a follow-up study.

\section{Data availability}
The calculated 4D grids (792 WC spectra, 432 WN spectra and 600 SMC
spectra) as well as their future updates (if any) 
will be available from this link:
\url{https://drive.google.com/drive/u/1/folders/16gLnF2Z3abES8CPUOxI2dKBw8j58MW6_}.
The model spectra are in the range 1000 - 11000~\AA~ with sampling of 1~\AA.

\begin{acknowledgements}
This research is based on observations made with the International
Ultraviolet Explorer (IUE), obtained
from the MAST data archive at the Space Telescope Science Institute,
which is operated by the Association of Universities for Research in
Astronomy, Inc., under NASA contract NAS 5–26555.
This research has made use of the NASA's Astrophysics Data System, and
the SIMBAD astronomical data base, operated by CDS at Strasbourg,
France.
\corr{We thank the anonymous referee for her/his valuable comments and suggestions, which
helped to improve this manuscript.}
\end{acknowledgements}
%

\bibliographystyle{raa}
\bibliography{ref}

\appendix

\section {4D-grid fitting of the UV-optical spectrum of WR~23}
\label{app:wr23}

	The accuracy of the derived stellar parameters of a studied object
	generally depends on the quality of its observed spectra. We recall that
	the 4D fitting procedure works with spectra in absolute flux units. Therefore, 
	the photometric absolute uncertainties associated with the observed spectrum will
	have an impact on the derived stellar parameters. Additionally, the spectral
	resolution (and sampling) should not be very `crude', because that will
	result in inaccurate emission line profiles which in turn would
	deteriorate the quality of the derived stellar wind velocity.
	
	The distance to the studied object, along with its associated uncertainties, 
	is another physical  quantity that is important for obtaining a reliable stellar parameters.
	 Specifically, it plays a key role in estimating stellar luminosity, 
	 which subsequently affects the derived values of both the mass-loss rate and stellar temperature. 
         This relationship arises , because the continuum is
	primarily determined by the stellar luminosity. Thus, changing the
	distance, which leads to changes in luminosity, causes changes in the ionization structure
	 of the stellar wind.
	To maintain consistency in the observed spectrum,  adjustments are necessary. 
	This involves `back-adjusting' the ionization structure by modifying the mass-loss rate and  stellar temperature. 
	The latter is particularly significant as it determines the quantity of photoionizing photons emitted.
	
	The focus of this study is not to derive stellar parameters of massive stars,
	 however, in this appendix we provide an example
	how the 4D grid-fitting procedure can be applied to real observations. 
	Therefore, in the light of the preceding discussion, WR~23 was selected. 
	WR~23 is categorized as a Wolf-Rayet star of the WC (carbon-rich) subtype 
	(e.g., the Galactic Wolf-Rayet
	Catalogue; \citealt{wr_catalogue}\footnote{Galactic Wolf
		Rayet Catalogue;
		\url{http://pacrowther.staff.shef.ac.uk/WRcat/index.php}}) at a Gaia
	distance of $2.3\pm0.1$~kpc \citep{crowther23}.
	
	The optical spectrum of WR~23 was taken from the STELIB spectroscopic
	library that has an intermediate spectral resolution of $\leq 3$\AA, 
	sampling of 1\AA, and overall absolute photometric uncertainty of 3\%
	\citep{leborgne03}.
	
	We expanded the spectral range by making use of the UV spectrum of WR~23
	from the International Ultraviolet Explorer (IUE): the data are
	taken from the MAST archive (Mikulski Archive for Space
	Telescopes\footnote{\url{https://archive.stsci.edu/}}). The IUE spectrum has 
	a sampling of 1.68~\AA~ at wavelengths $\lambda < 1979$~\AA~ and 
	2.67~\AA~ at wavelengths $\lambda > 1979$~\AA, respectively.
	
	Our fitting procedure of observed spectra consists of the following
	steps.
	
	First, we prepare the 4D grid of model spectra for the corresponding
	spectral binning (sampling) of the observed spectrum at hand. 
	
	Then, we derive the stellar parameters using the approach described in
	Section~\ref{subsec: gridmodelling} for the nominal distance of 2.3 kpc
	to WR~23. Also, we derive the stellar parameters for the cases of upper 
	(2.4 kpc) and lower (2.2 kpc) limits to the distance.

\begin{table*}
	\caption{4D grid-fitting results for WR~23.}
	\label{tab:wr23}
	\bc
	\begin{tabular}{lcc}
		\hline\hline
		\multicolumn{1}{c}{Parameter} &
		\multicolumn{1}{c}{UV$+$Optical}  &
		\multicolumn{1}{c}{Optical} \\
		\hline
		\Tstar (kK) & 93.86 [93.28; 97.97] &  99.05 [95.58; 101.78]  \\
		\dotM (\dotMyr)  & -4.559 [-4.595; -4.510] & -4.577 [-4.614; -4.545]
		\\
		\logL (\sunlum)  & 5.435 [5.360; 5.485] & 5.369 [5.320; 5.433]  \\
		\Vwind (\kms) & 2298 [2293; 2318] & 2251 [2240; 2259]  \\
        E(B-V) (mag) & 0.590 [0.584; 0.593] & 0.547 [0.542; 0.553]  \\
		\hline
	\end{tabular}
	\ec
\tablecomments{0.86\textwidth}{
Labels `UV$+$Optical' and `Optical' denote results from fitting the
entire WR~23 spectrum or only its optical part.
The derived fit parameters 
(stellar temperature, mass-loss rate, luminosity, stellar wind velocity,
reddening) with the associated uncertainties, i.e. the
range of parameter value given in parentheses (see Fig.~\ref{fig:wr23}).}
\end{table*}

\begin{figure*}
	\begin{center}
		\includegraphics[width=2.8in, height=2.0in]{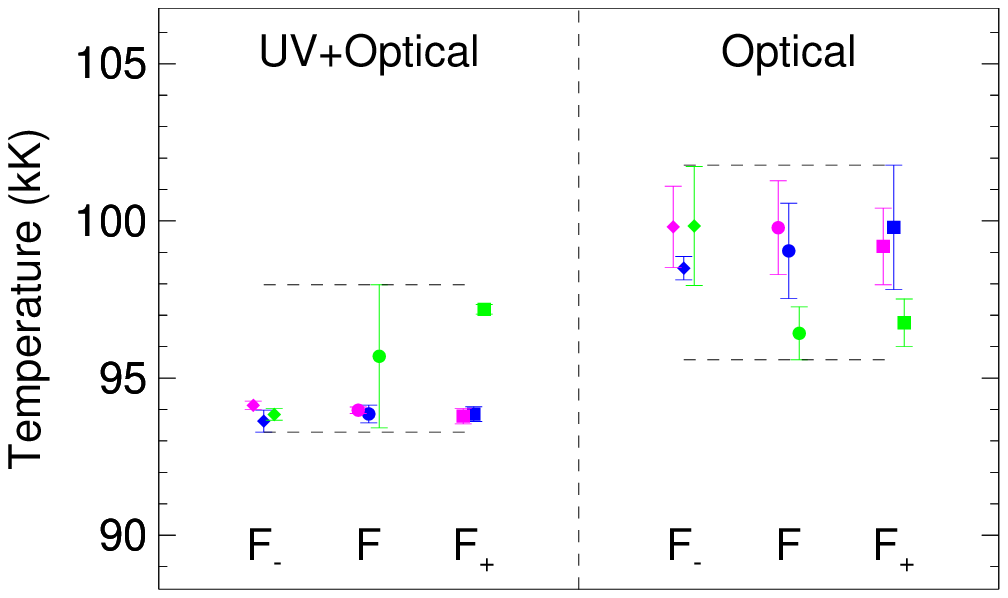}
		\includegraphics[width=2.8in, height=2.0in]{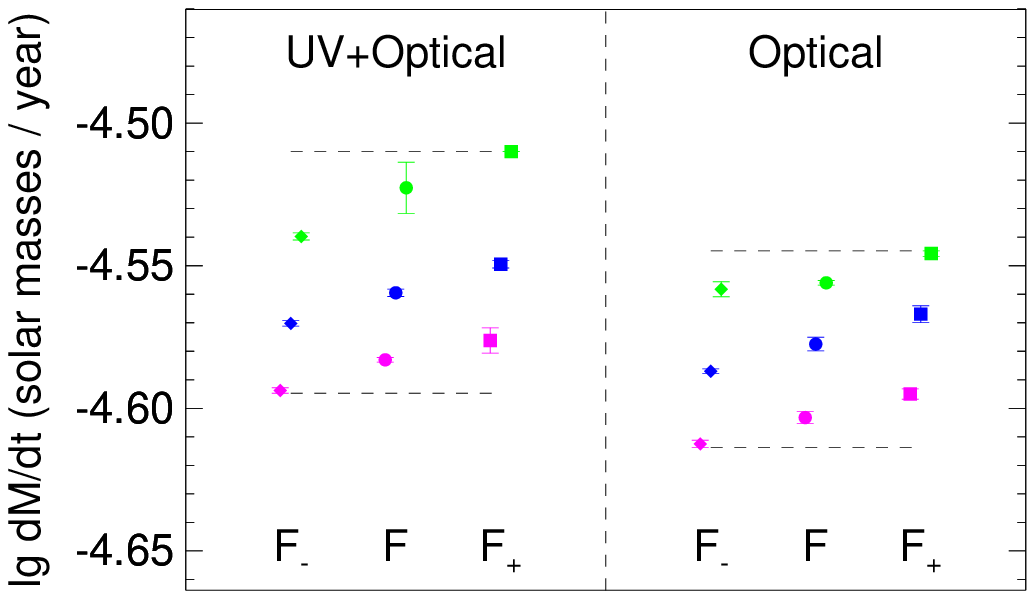}
		\includegraphics[width=2.8in, height=2.0in]{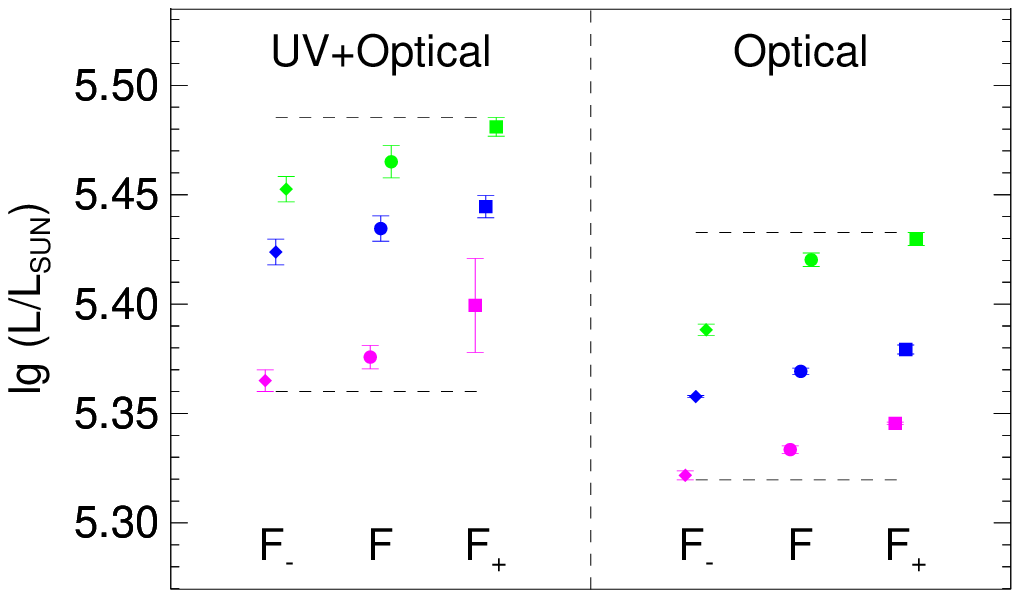}
		\includegraphics[width=2.8in, height=2.0in]{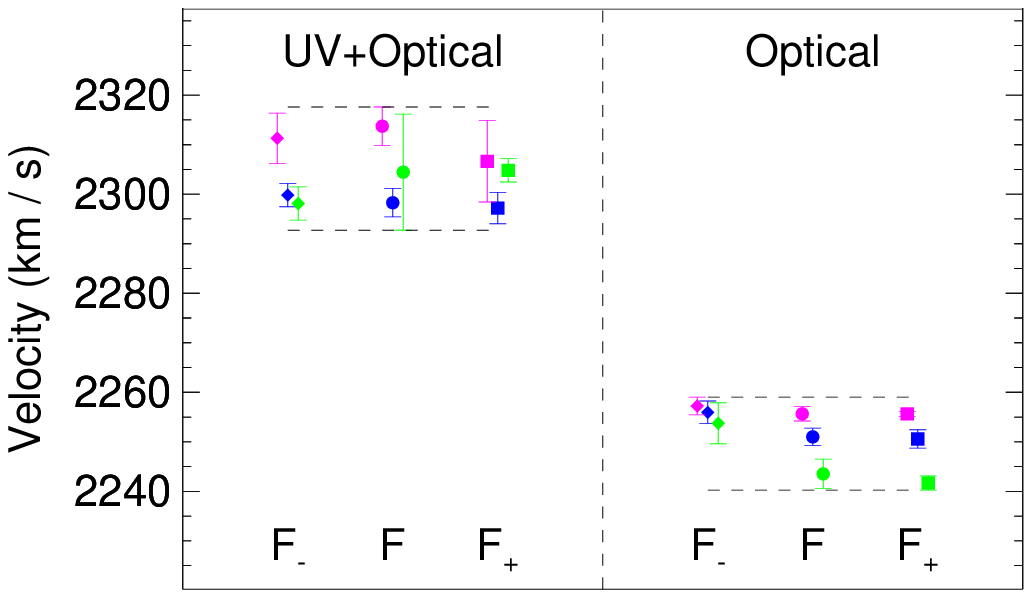}
	\end{center}
	\caption{
		Derived stellar parameters for WR~23: stellar temperature
		(\TstarE), mass-loss rate (\dotME), luminosity (\logLE) and terminal
		wind velocity (\VwindE).
		The results are color-coded as follows: values for the nominal distance to WR~23
		 of 2.3 kpc are shown in blue, those for the lower limit on distance (2.2 kpc) are in magenta,
		  and those for the upper limit on distance (2.4 kpc) are in green. 
		The results for the nominal spectra are marked by the `F' symbol, while
		those corresponding to the typical absolute flux uncertainty of $\pm
		3$\% are marked by the 'F$_+$' and 'F$_-$' symbols, respectively.
		Error bars present the uncertainties due to the adopted numerical
		technique (4D grid-fitting): if not visible, the error bars are smaller
		than the size of the plotted symbol.
		The horizontal dashed lines (in black) present the accumulated
		errors of each parameter due to the distance and absolute flux
		uncertainties (see Table~\ref{tab:wr23}).
		Labels `UV$+$Optical' and `Optical' denote results from fitting the
		entire WR~23 spectrum or only its optical part.
	}
	\label{fig:wr23}
\end{figure*}

\begin{figure*}
	\begin{center}
 	\includegraphics[width=2.8in, height=2.0in]{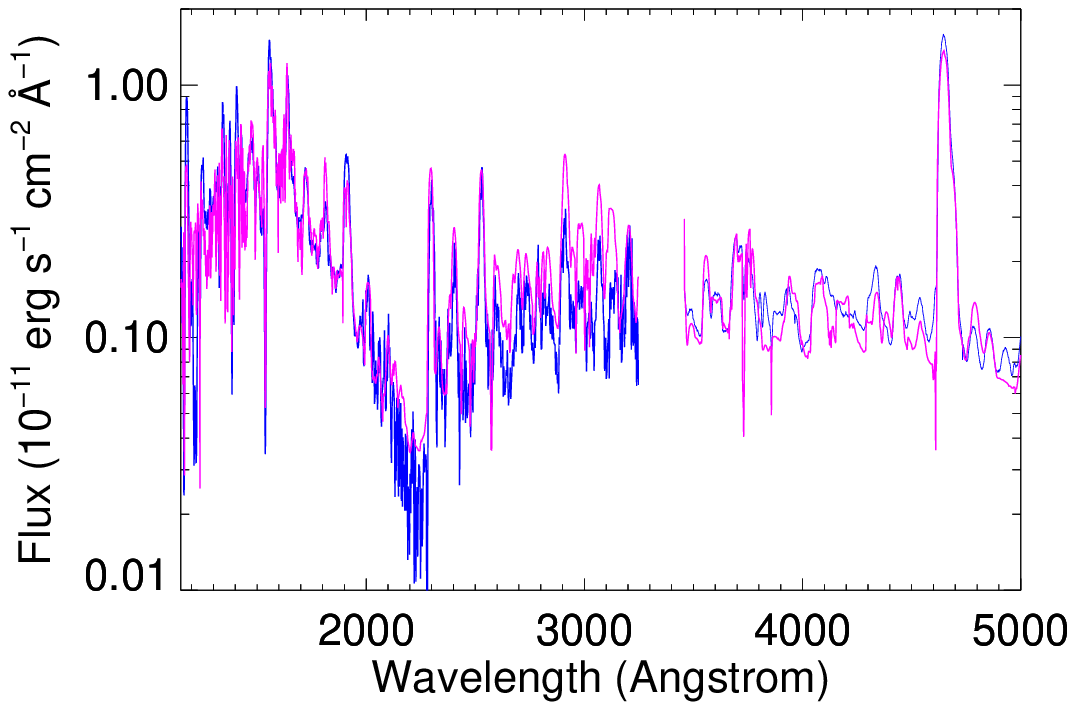}
 	\includegraphics[width=2.8in, height=2.0in]{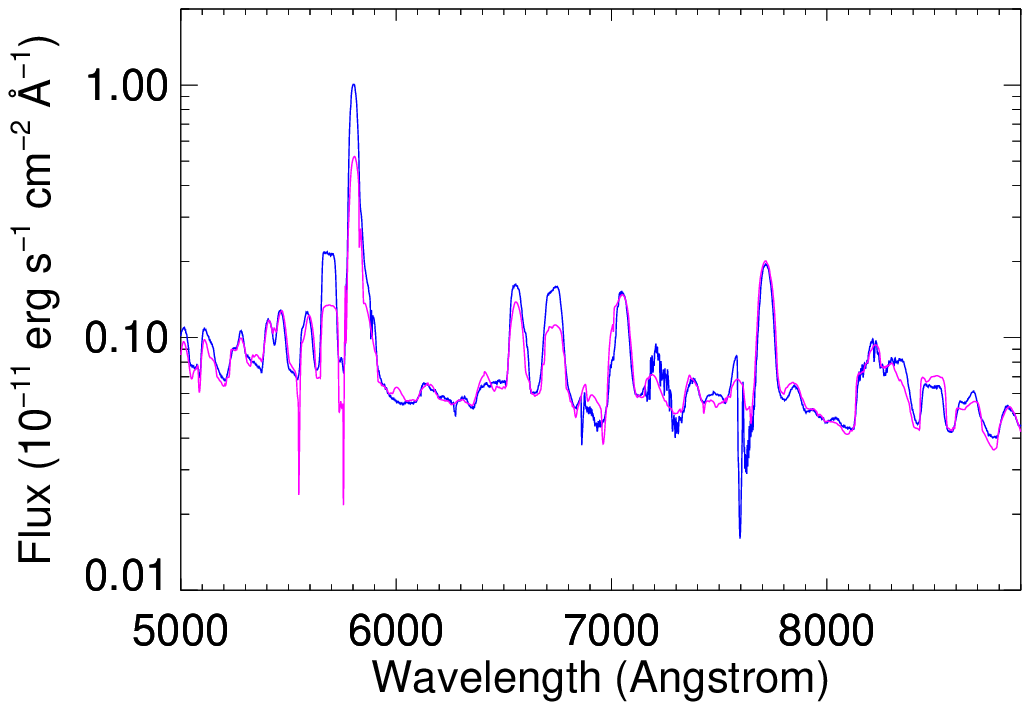}
 	\includegraphics[width=2.8in, height=2.0in]{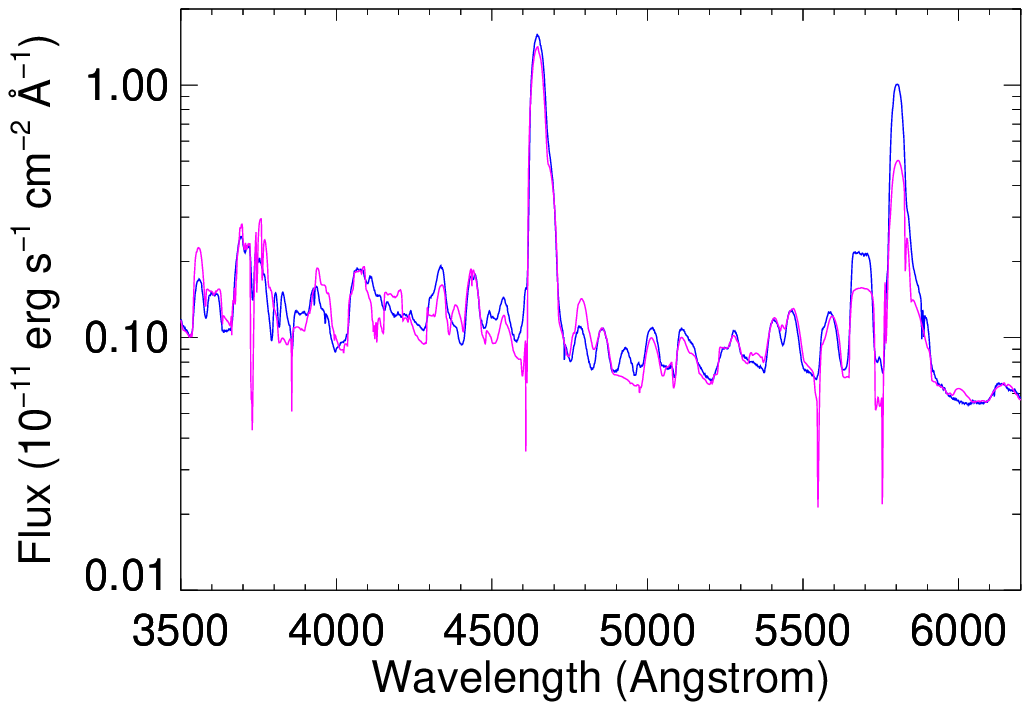}
 	\includegraphics[width=2.8in, height=2.0in]{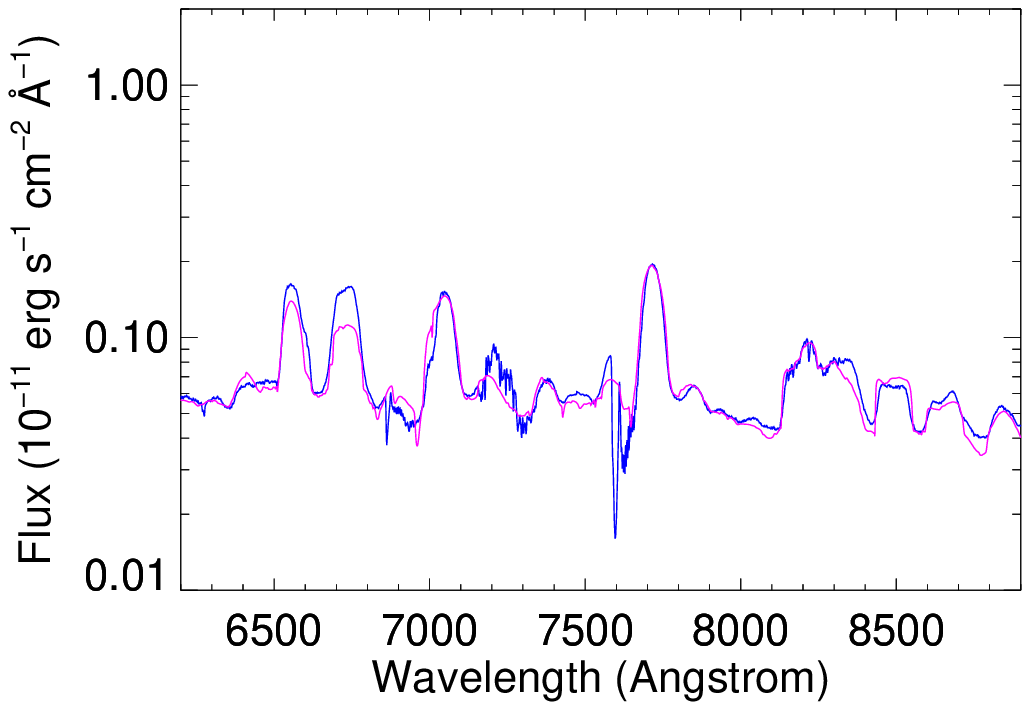}
		
	\end{center}
	\caption{
		The observed spectrum of WR~23 (in blue )
		overlaid with the best-fit model spectrum (in magenta) for the
		cases of nominal distance and nominal absolute flux calibration.
		The best-fit model spectra are calculated with \cmfgen using the 
		mean model parameter values as given in Table~\ref{tab:wr23}.
                Results from fitting the entire WR~23 spectrum or only its 
		optical part (denoted `UV$+$Optical' and `Optical' 
		in Table~\ref{tab:wr23}) are shown in the first and 
		second raw, respectively.
                We note the `gap'(at $\sim 3500$\AA) between the UV (from IUE) 
                and optical (from STELIB spectroscopic library) parts of the 
		WR~23 spectrum.
	}
	\label{fig:wr23a}
\end{figure*}
	
	Finally, we repeat these fits for the three values of the distance as
	for each case we perform the 4D fitting twice to take into account the
	absolute photometric uncertainties: i.e., considering the flux to be by
	3\% higher or lower than the nominal flux.
	
	In all the fits, we adopt the Galactic extinction curve of 
	\citet{fitzpatrick99} with R$_V = 3.1$.
	
	\corr{The derived stellar parameters and their associated uncertainties are
	given in Figs.~\ref{fig:wr23}, ~\ref{fig:wr23a}  and Table~\ref{tab:wr23}.}
	Note that the highest contribution to uncertainties of the derived
	parameters is due to the errors on distance to WR~23. Because luminosity
	scales as the square of the distance, the highest impact is on the derived value
	of stellar luminosity and the uncertainties on the absolute flux then
	accumulate: i.e., they add extra uncertainty. But in general, the
	uncertainties from distance and absolute flux are somehow
	`redistributed' between the derived values of luminosity, mass-loss rate
	and stellar temperature. It is worth nothing that the wind velocity remains 
	unaffected by this redistribution, likely because the wind velocity is mostly `confined' by the
	emission-line profiles.

Note that if any of the stellar parameters we are fitting for is already
constrained from other studies, we can incorporate that prior information
by  keeping the parameter fixed at its known value and then proceed to fit the remaining parameters.
Nevertheless, among the four fundamental stellar parameters, only the 
 terminal wind velocity can be constrained reliably from observations.
	
	Interestingly, the derived terminal velocity from the 4D
grid-fitting, {\it using no a priori information about its value}, 
	is in good correspondence with the value of 2280\kms from the classical
	analysis of IUE spectra of massive stars by \citet{prinja90} 
	and of 2342\kms by \citet{niedzielski02}.  This is 
	valid both for the fits to the entire spectrum (UV-optical) of WR~23 
	and only to its optical spectrum (Table~\ref{tab:wr23}).
	
	In general, the results of fitting the UV-optical and only the optical 
	spectrum of WR~23 are in acceptable correspondence between each other
	(see also Section~\ref{sec:tests}). 
	We attribute it to the presence of many strong emission lines, that help 
	constrain better the physical parameters of the studied massive stars. 
	Nevertheless, it is our understanding that the former (results from 
	fitting the entire spectrum) should be considered more reliable.

Finally, we note that reddening curves with R$_V \neq 3.1$ can be used 
in our 4D-grid fits. But, this can be adopted only if corresponding
information is available from other studies.

\section{Tests with different abundances, wind acceleration law and 
volume filling factor}
\label{app}

In the figures below, we provide the results from the 4D-grid modelling
(fitting) of test spectra (`perfect' observed spectra) for the objects
considered in this work (WC, WN and SMC). These tests are aimed at 
checking the corresponding effects if chemical composition (abundances), 
stellar wind acceleration ($\beta$-law) and clumping (volume filling 
factor) in the observed objects differ from those used to build our 
4D spectral grids of model spectra.

\begin{figure*}
\begin{center}
\includegraphics[width=2.8in, height=2.0in]{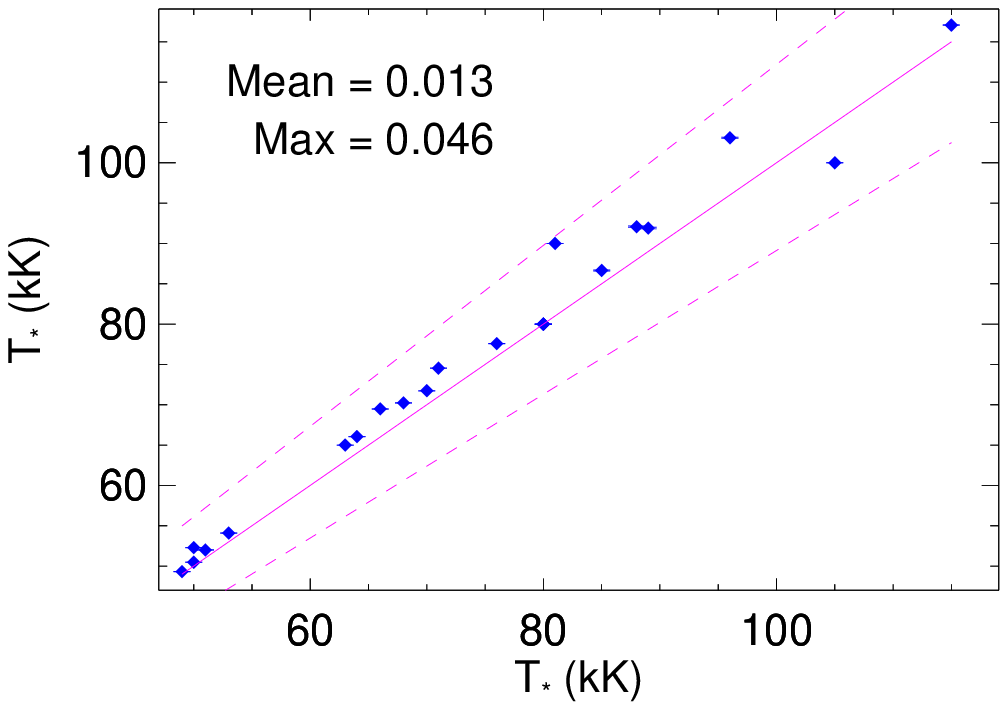}
\includegraphics[width=2.8in, height=2.0in]{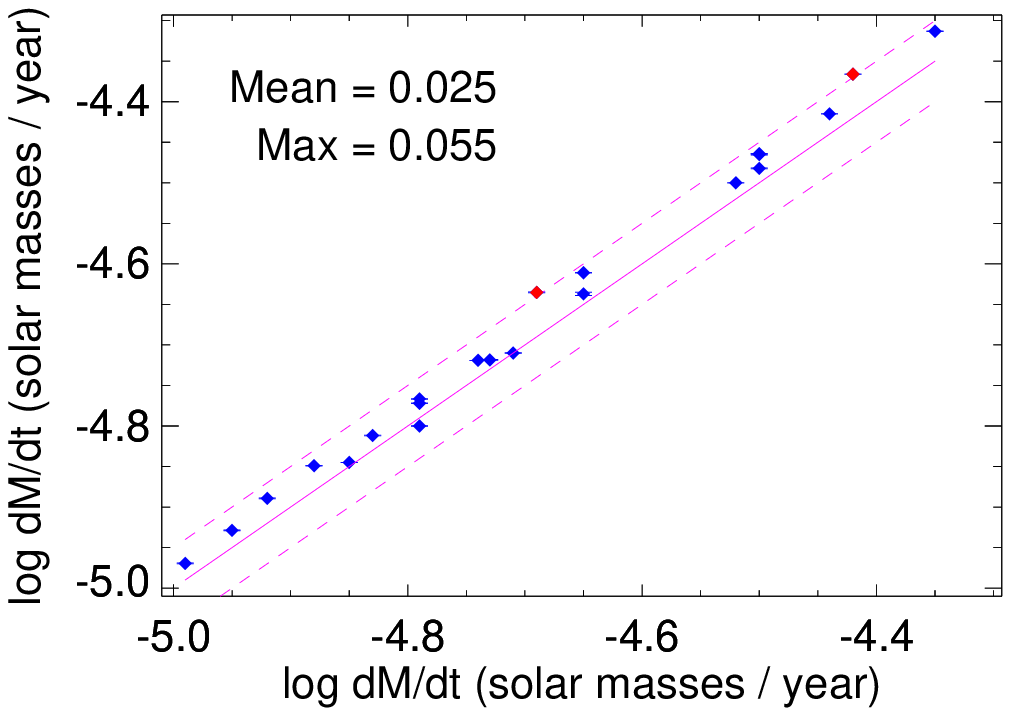}
\includegraphics[width=2.8in, height=2.0in]{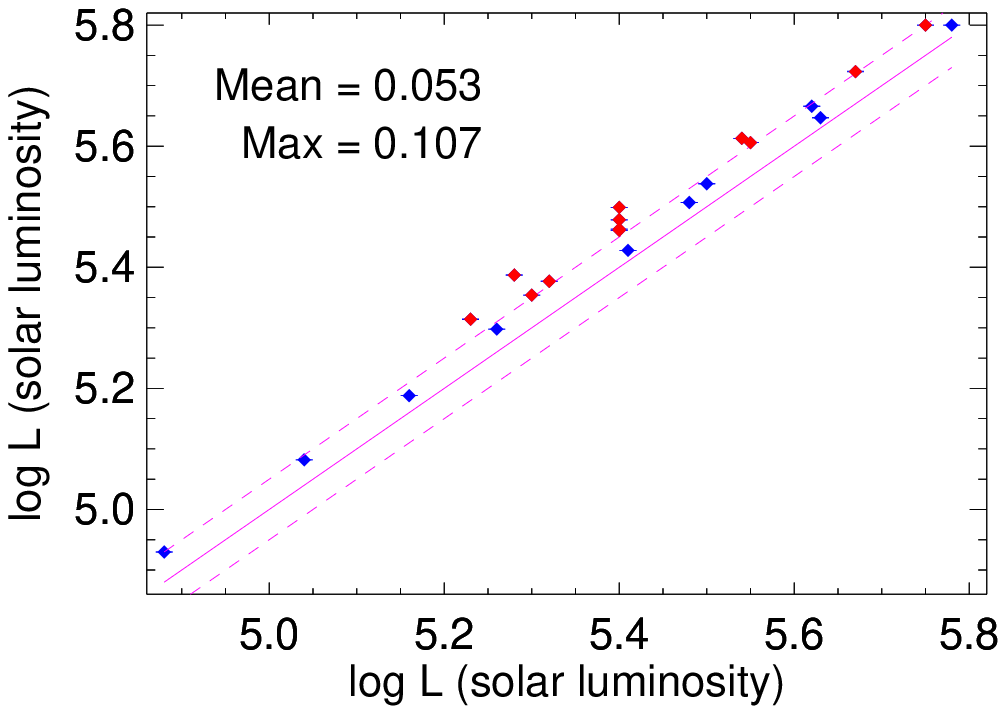}
\includegraphics[width=2.8in, height=2.0in]{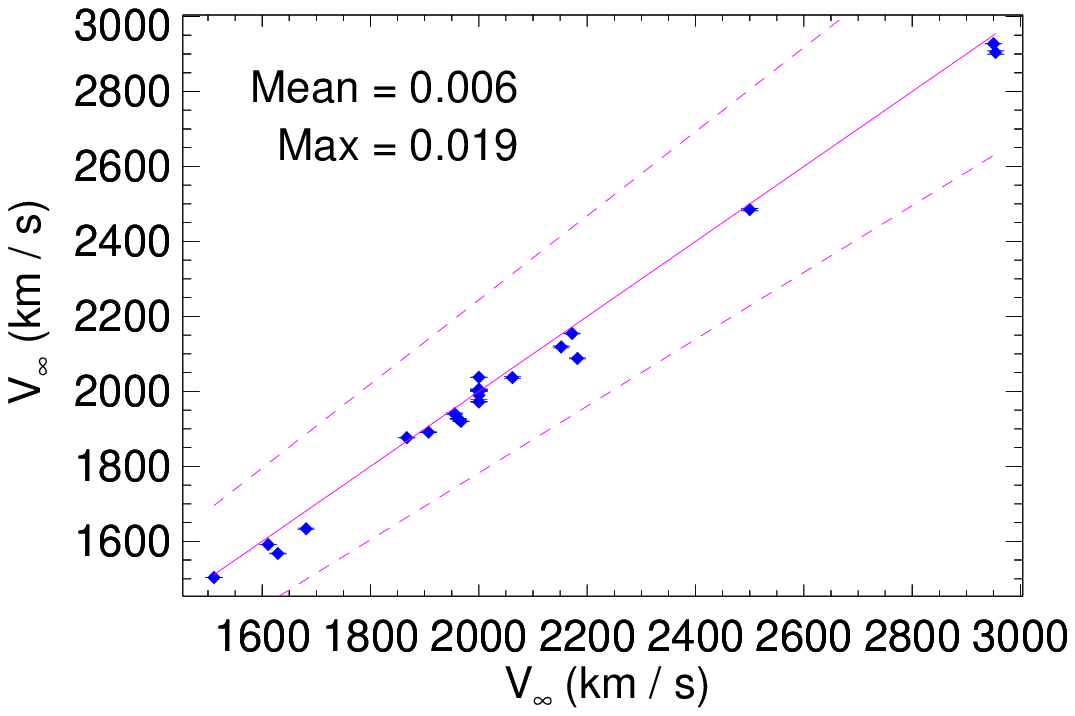}

\end{center}
\caption{
Test results for the `Add.' set of WC abundances (`perfect' observed spectra): 
effective temperature, mass-loss
rate, luminosity and terminal wind velocity. Lines in magenta colour:
(a) the solid line represents the perfect correspondence between the 
input model parameters and their values derived from the grid-modelling; 
(b) the two dashed lines represent the boundary for the $\pm 0.05$~dex 
absolute deviation from the expected perfect correspondence 
\corr{which corresponds to approximately a $\pm 12$ percent difference in linear terms.} 
The symbols shown in red colour are those that are beyond the boundary accuracy of
0.05 dex for the derived parameters. Mean and
Max labels denote the mean absolute deviation for the test sample and
its maximum value.
Error bars are the standard deviation of the four minimizations adopted.
}
\label{fig:WC_abu}
\end{figure*}

\begin{figure*}
\begin{center}
\includegraphics[width=2.8in, height=2.0in]{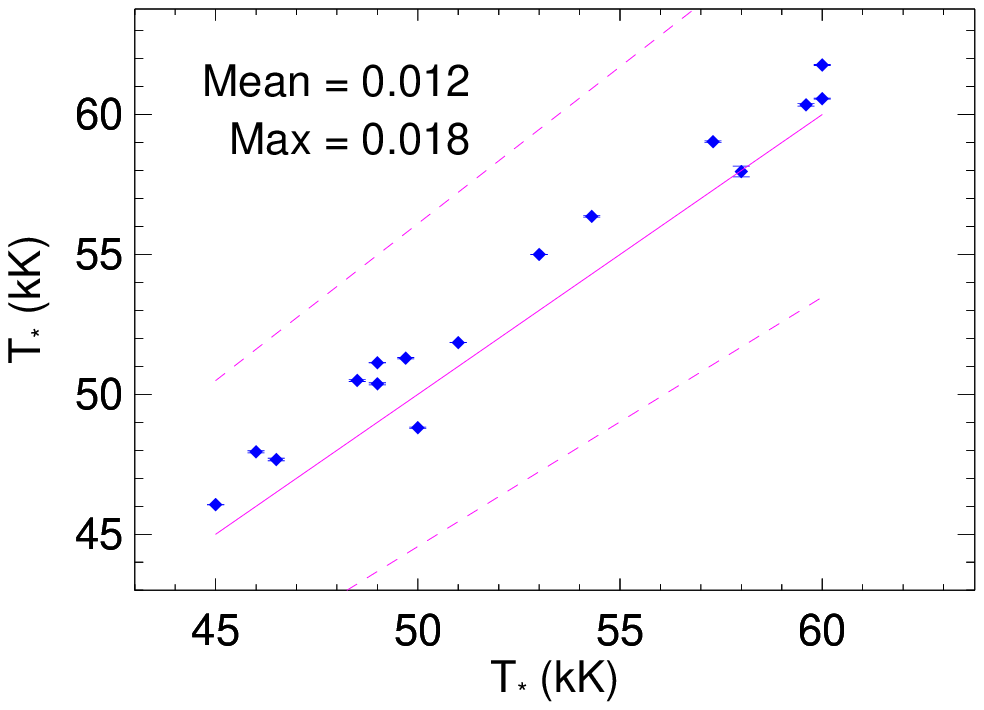}
\includegraphics[width=2.8in, height=2.0in]{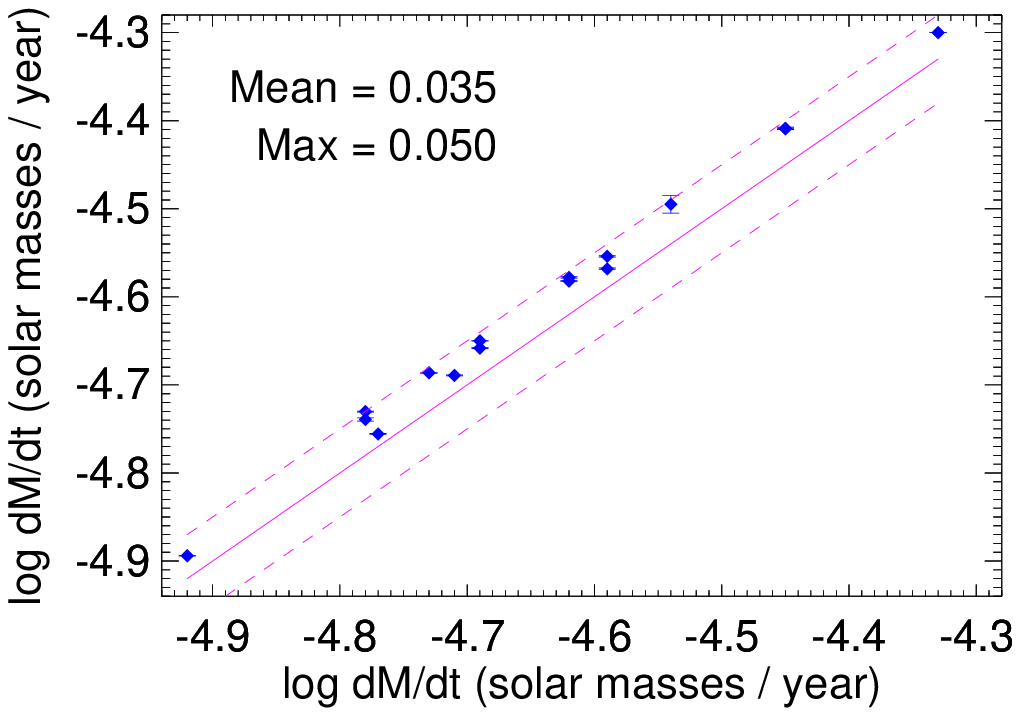}
\includegraphics[width=2.8in, height=2.0in]{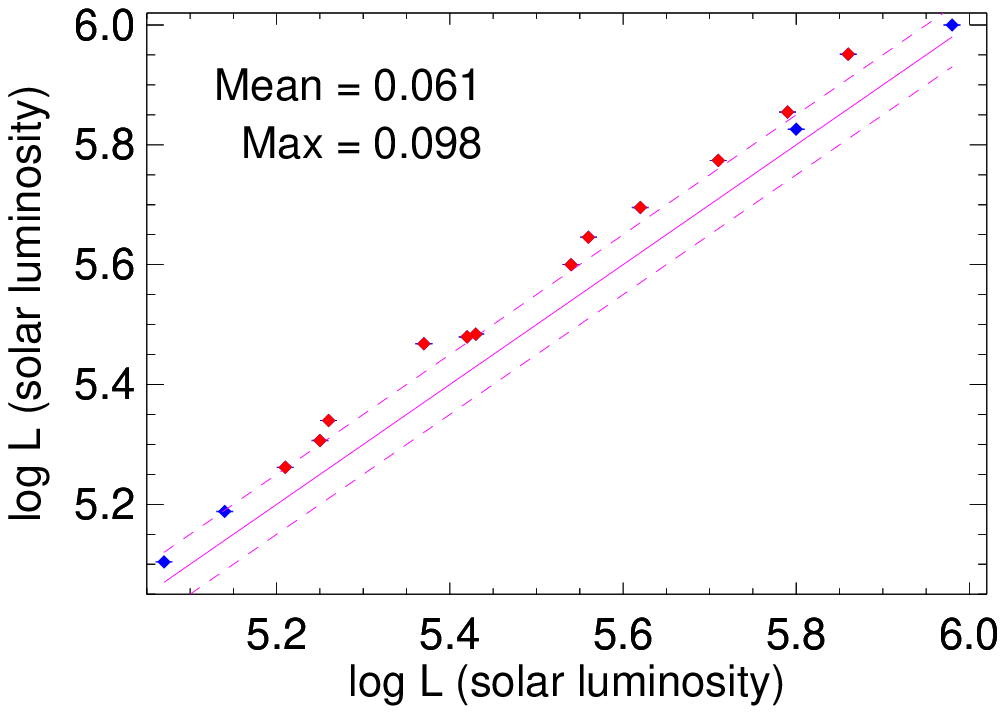}
\includegraphics[width=2.8in, height=2.0in]{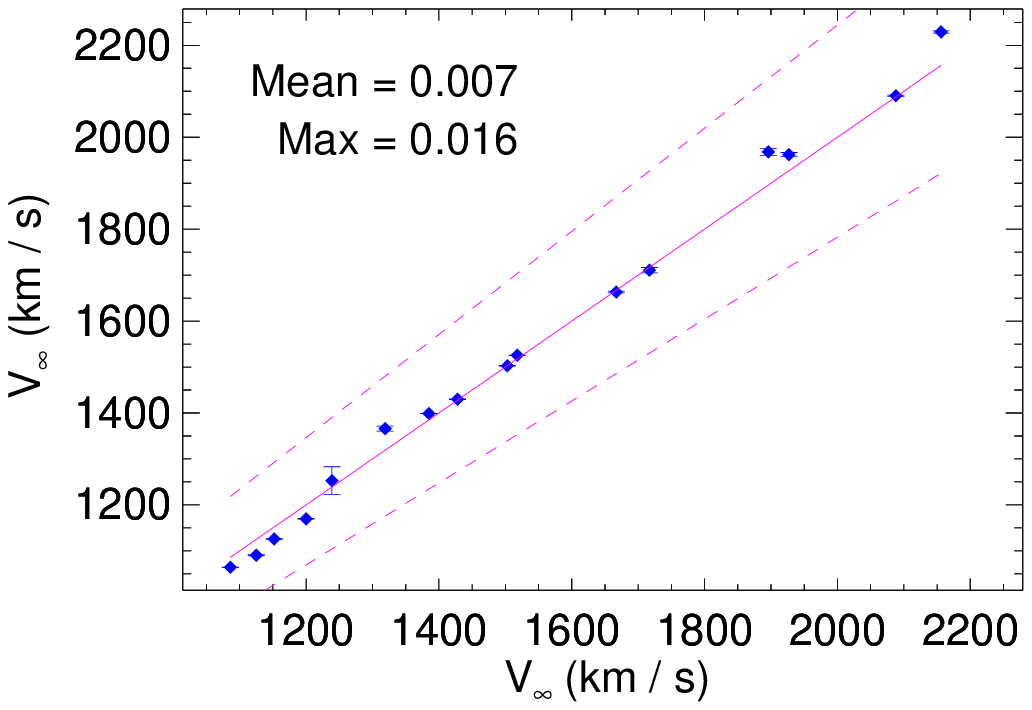}

\end{center}
\caption{The same as Fig.~\ref{fig:WC_abu} but for the `Add.' set of WN
abundances (`perfect' observed spectra).
}
\label{fig:WN_abu}
\end{figure*}

\begin{figure*}
\begin{center}

\includegraphics[width=2.8in, height=2.0in]{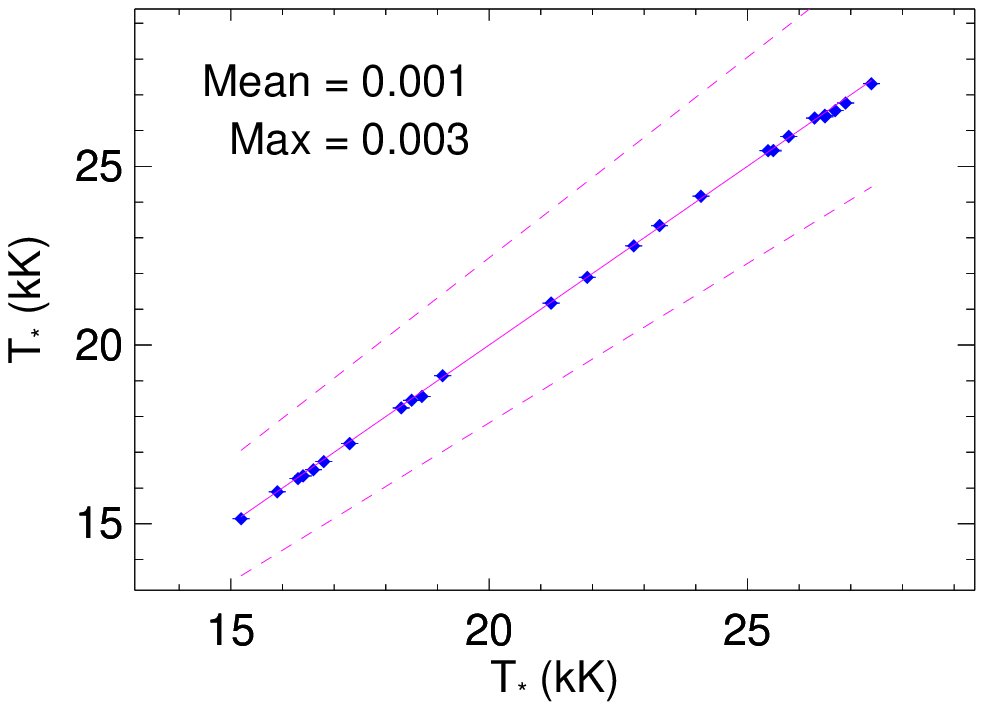}
\includegraphics[width=2.8in, height=2.0in]{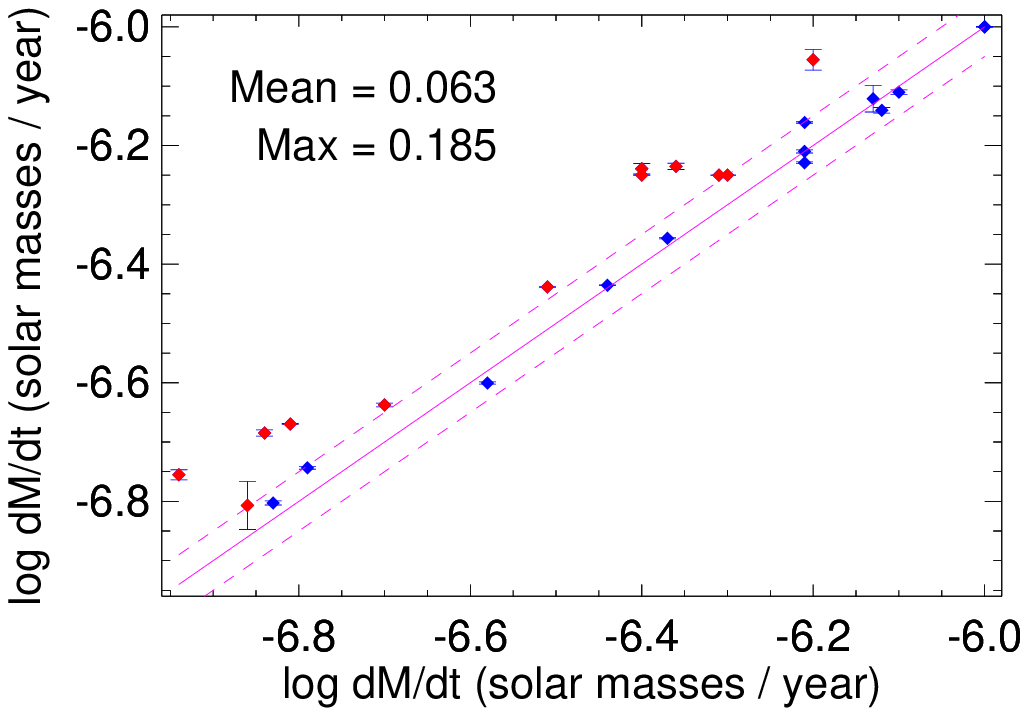}
\includegraphics[width=2.8in, height=2.0in]{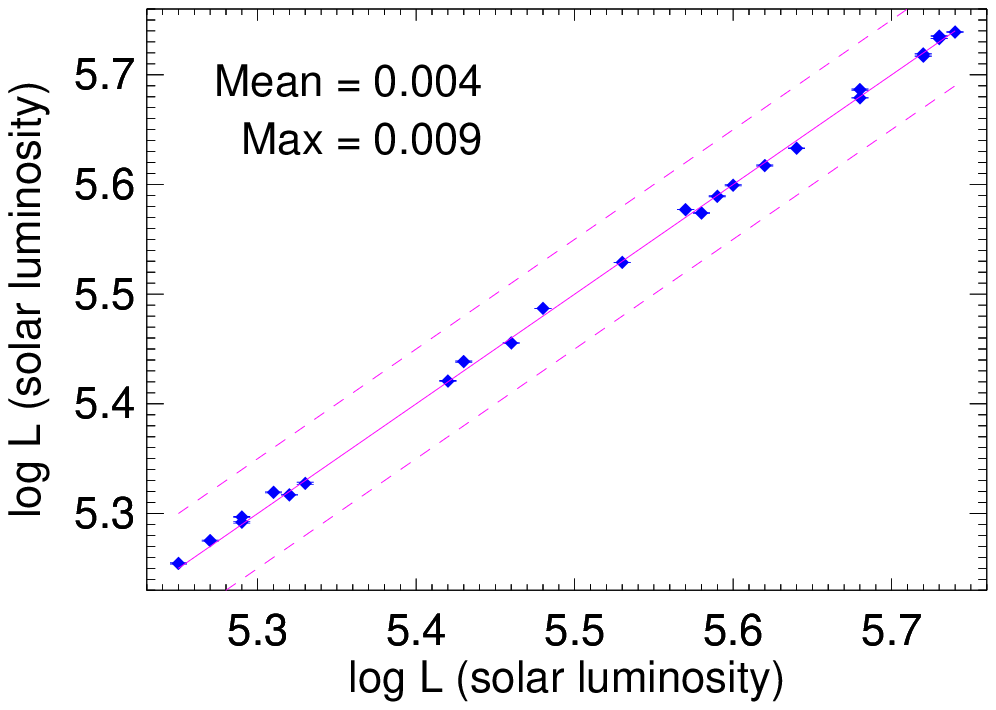}
\includegraphics[width=2.8in, height=2.0in]{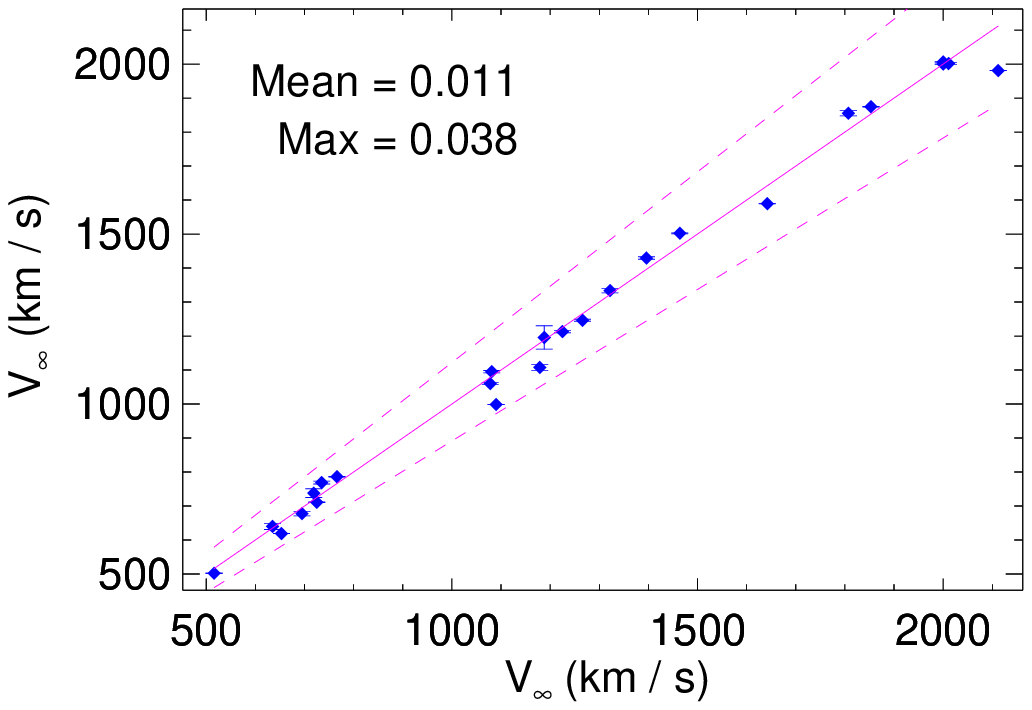}
 
\end{center}
\caption{The same as Fig.~\ref{fig:WC_abu} but for 
the `Add.' set of SMC abundances (`perfect' observed spectra).
}
\label{fig:SMC_abu}
\end{figure*}


\begin{figure*}
\begin{center}
\includegraphics[width=2.8in, height=2.0in]{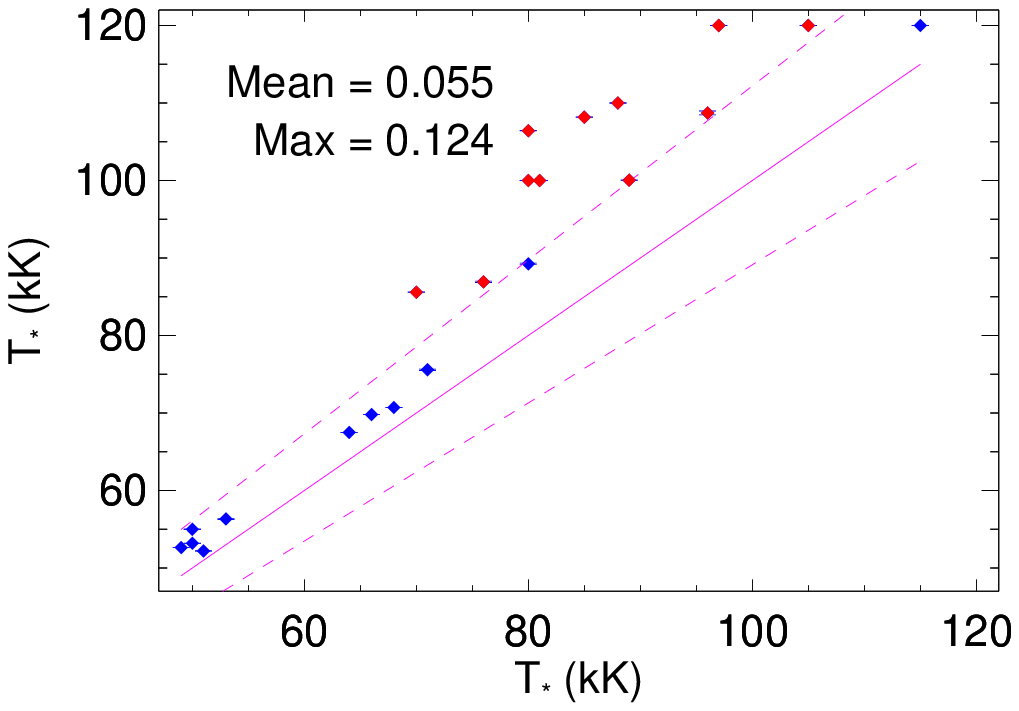}
\includegraphics[width=2.8in, height=2.0in]{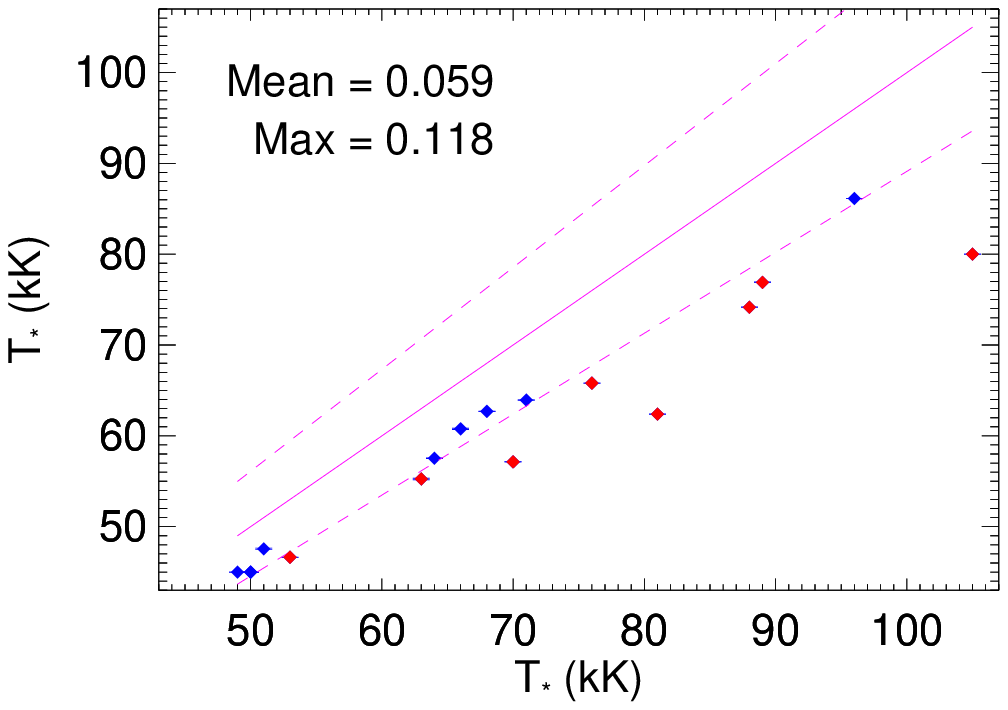}
\includegraphics[width=2.8in, height=2.0in]{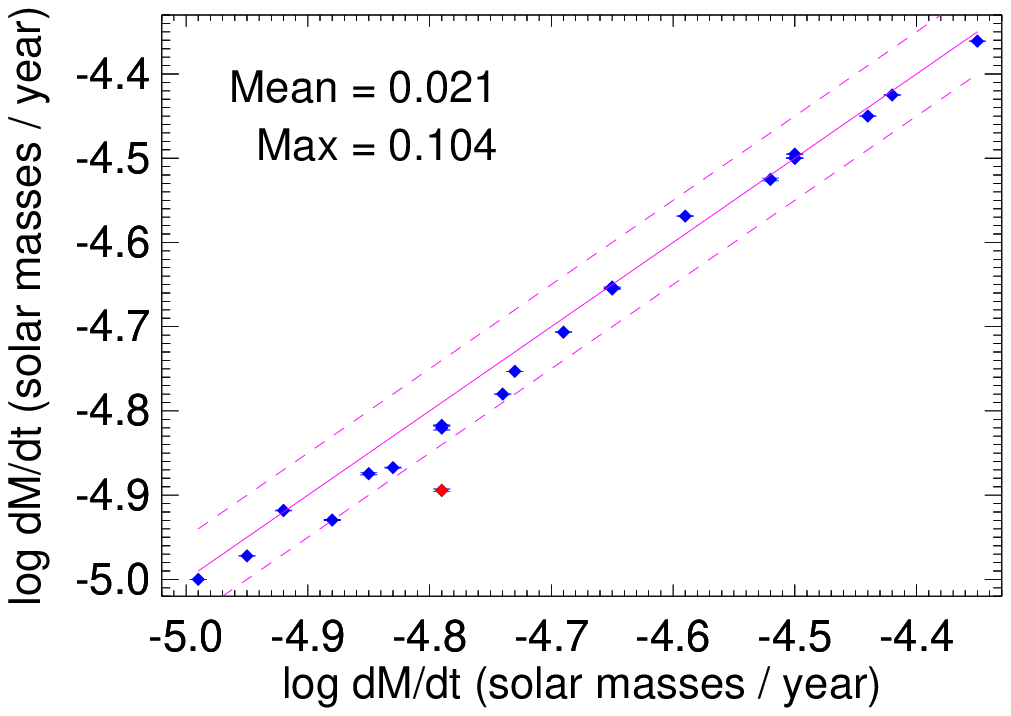}
\includegraphics[width=2.8in, height=2.0in]{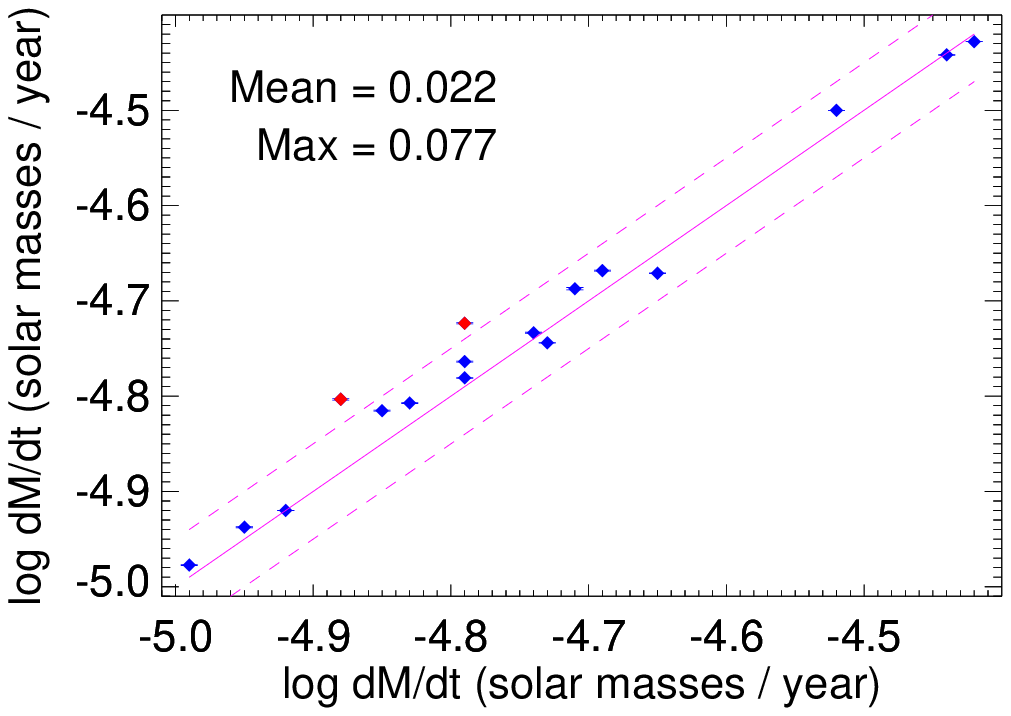}
\includegraphics[width=2.8in, height=2.0in]{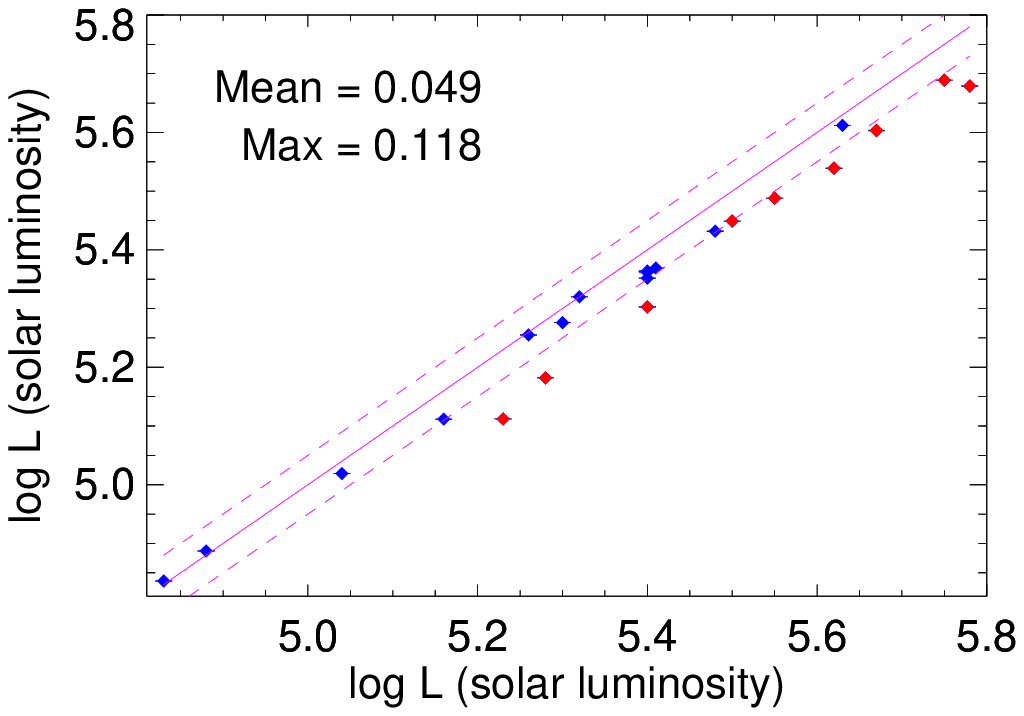}
\includegraphics[width=2.8in, height=2.0in]{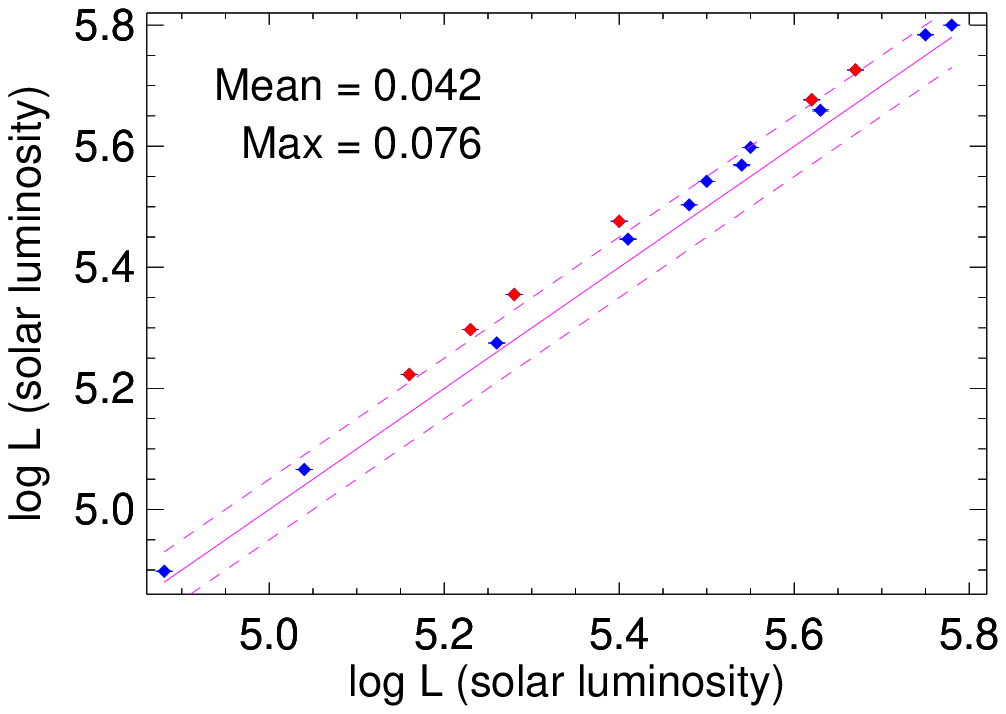}
\includegraphics[width=2.8in, height=2.0in]{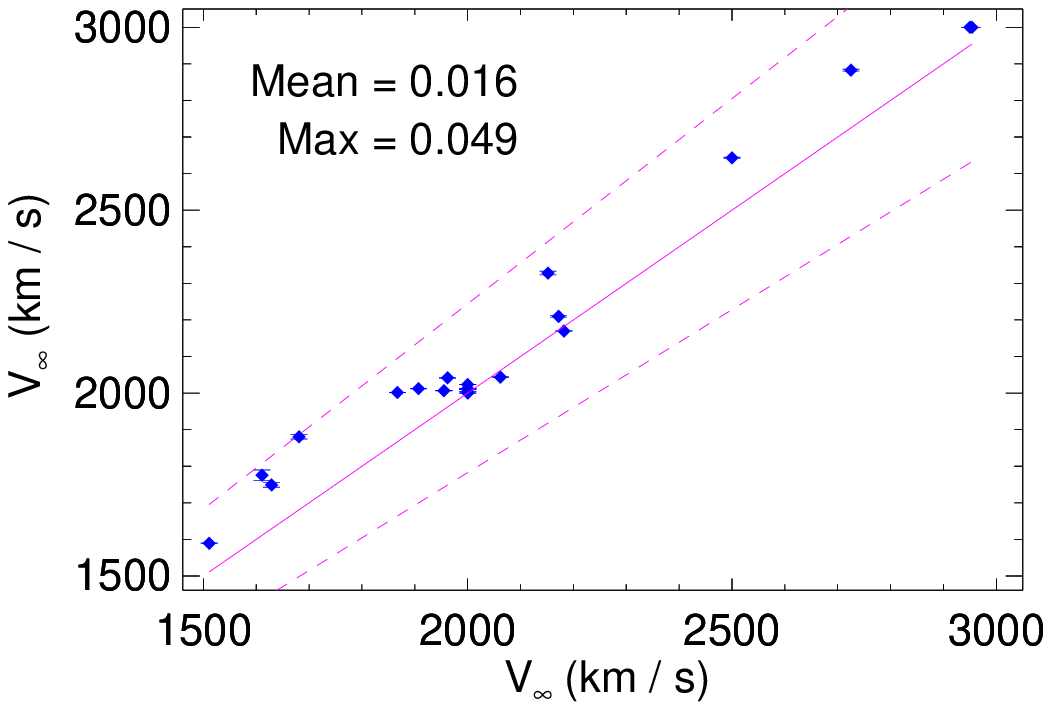}
\includegraphics[width=2.8in, height=2.0in]{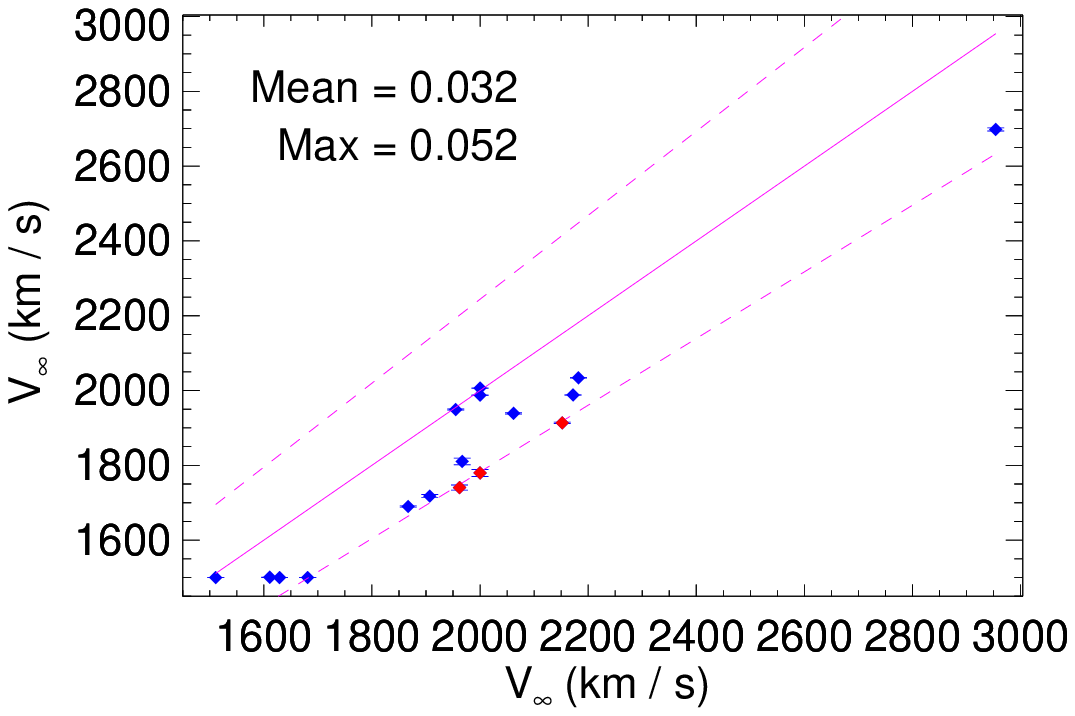}

\end{center}
\caption{The same as Fig.~\ref{fig:WC_abu} but for the WC grid and
$\beta$-law with $\beta = 0.5$ (left panels) and 
$\beta = 2$ (right panels) (`perfect' observed spectra).
}
\label{fig:WC_beta}
\end{figure*}

\begin{figure*}
\begin{center}
\includegraphics[width=2.8in, height=2.0in]{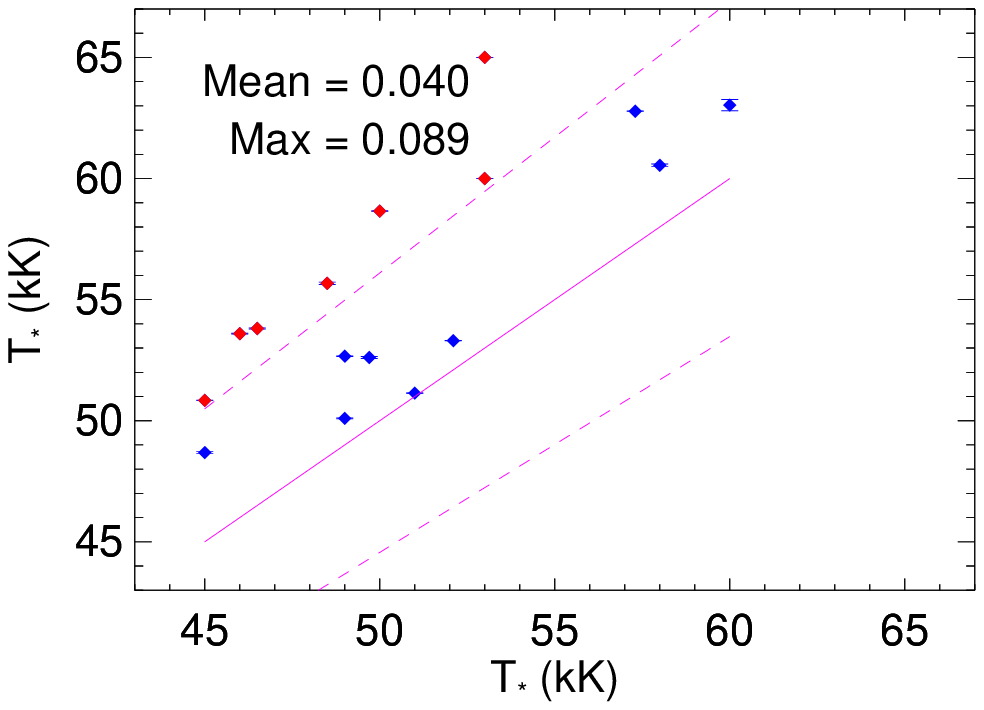}
\includegraphics[width=2.8in, height=2.0in]{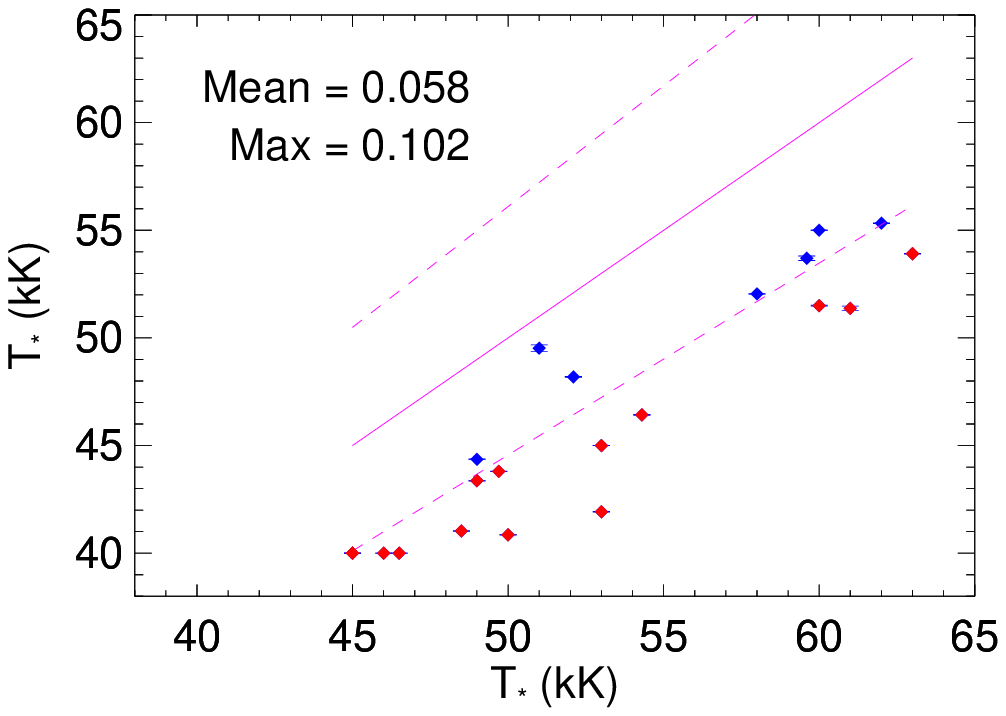}
\includegraphics[width=2.8in, height=2.0in]{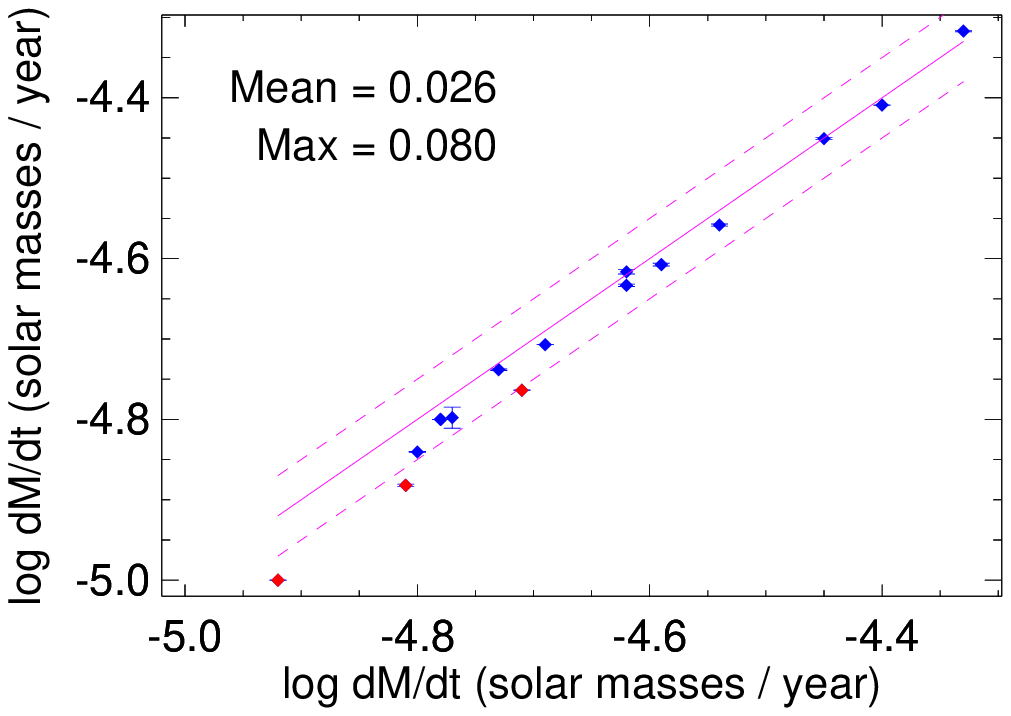}
\includegraphics[width=2.8in, height=2.0in]{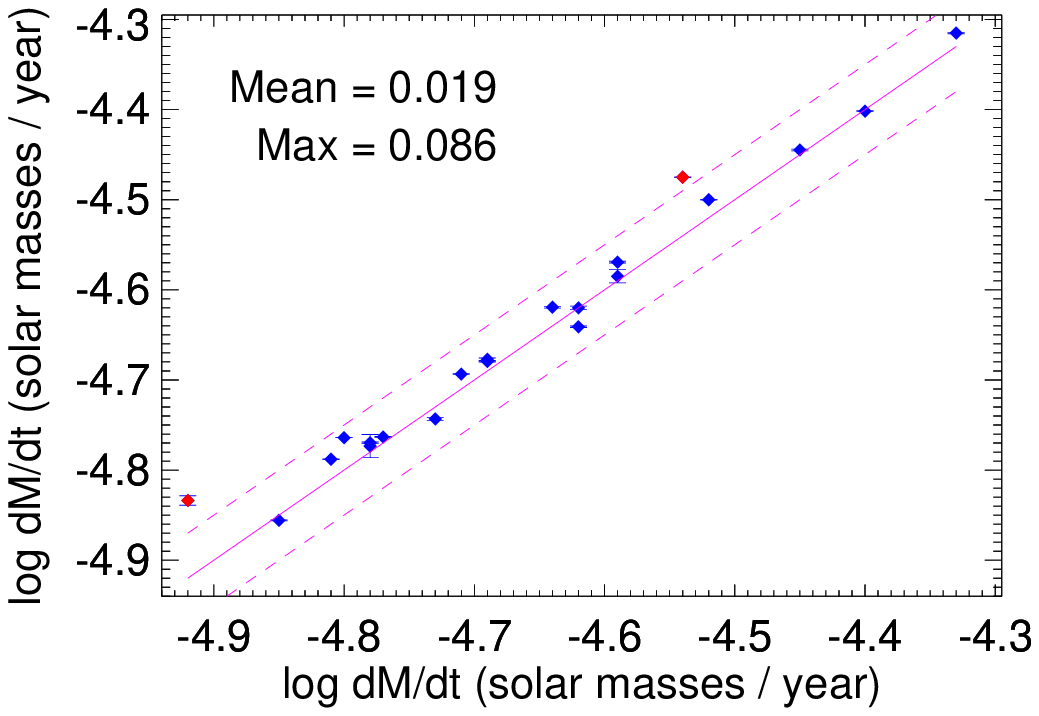}
\includegraphics[width=2.8in, height=2.0in]{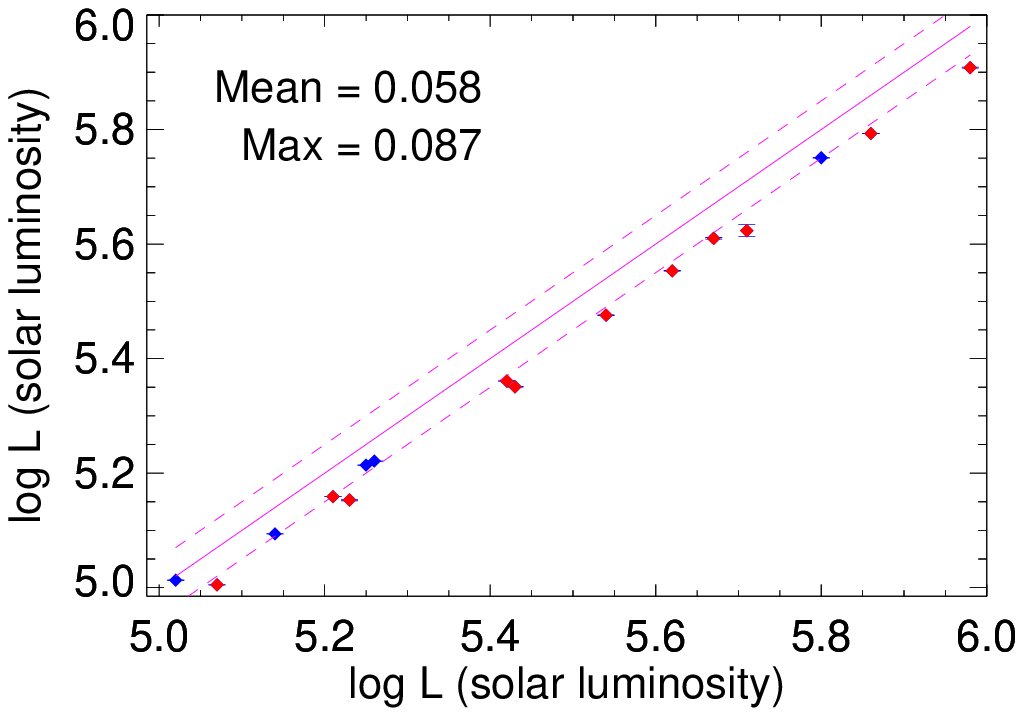}
\includegraphics[width=2.8in, height=2.0in]{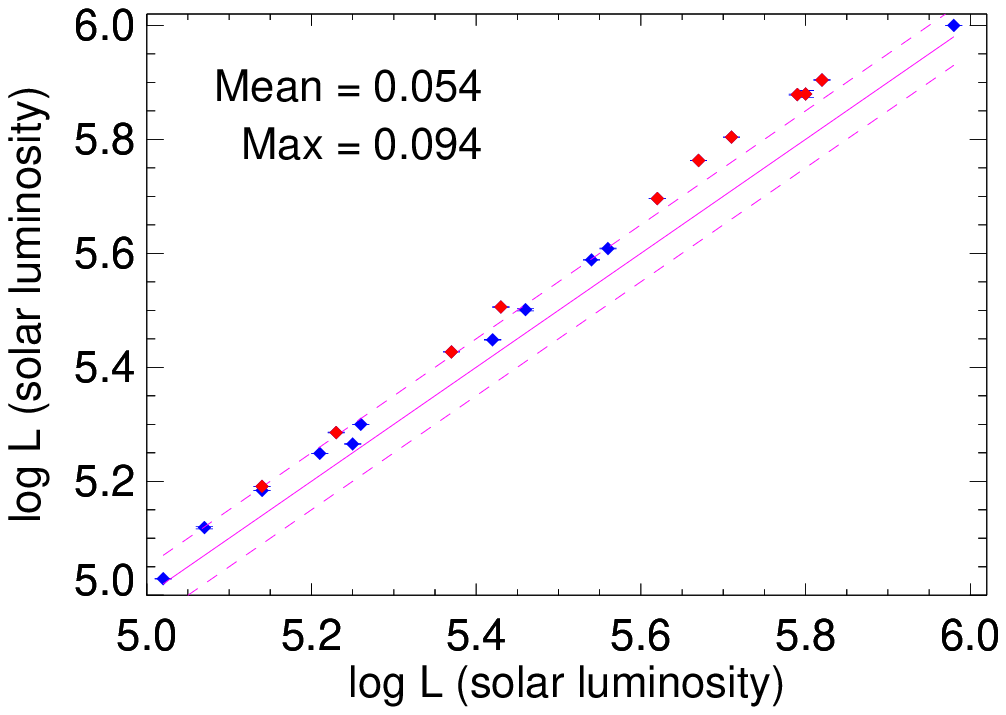}
\includegraphics[width=2.8in, height=2.0in]{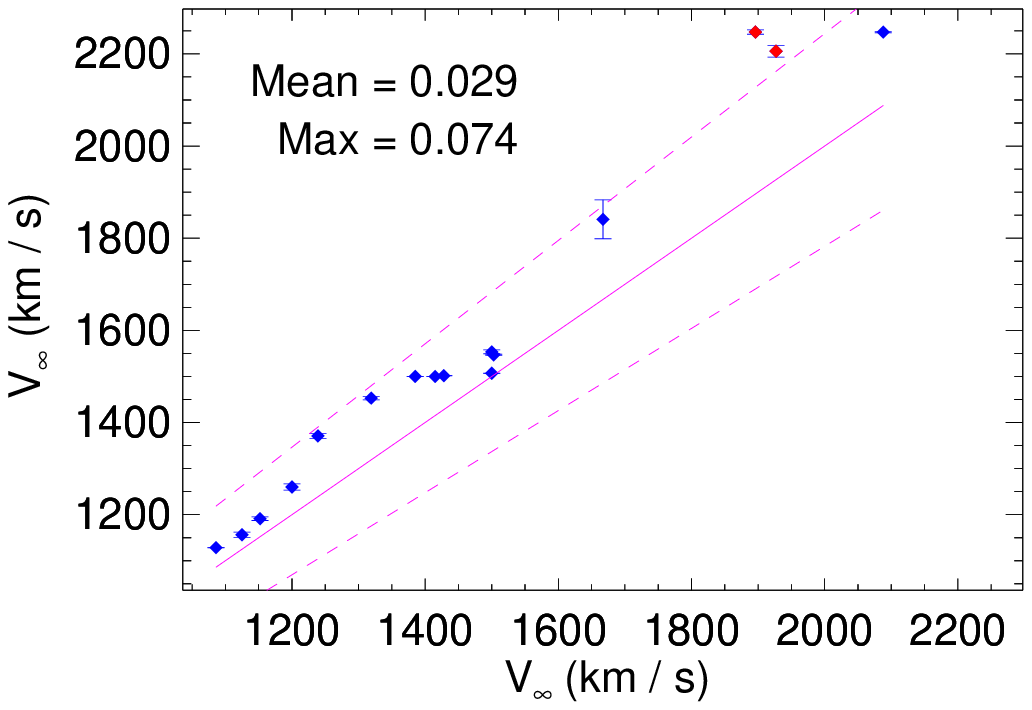}
\includegraphics[width=2.8in, height=2.0in]{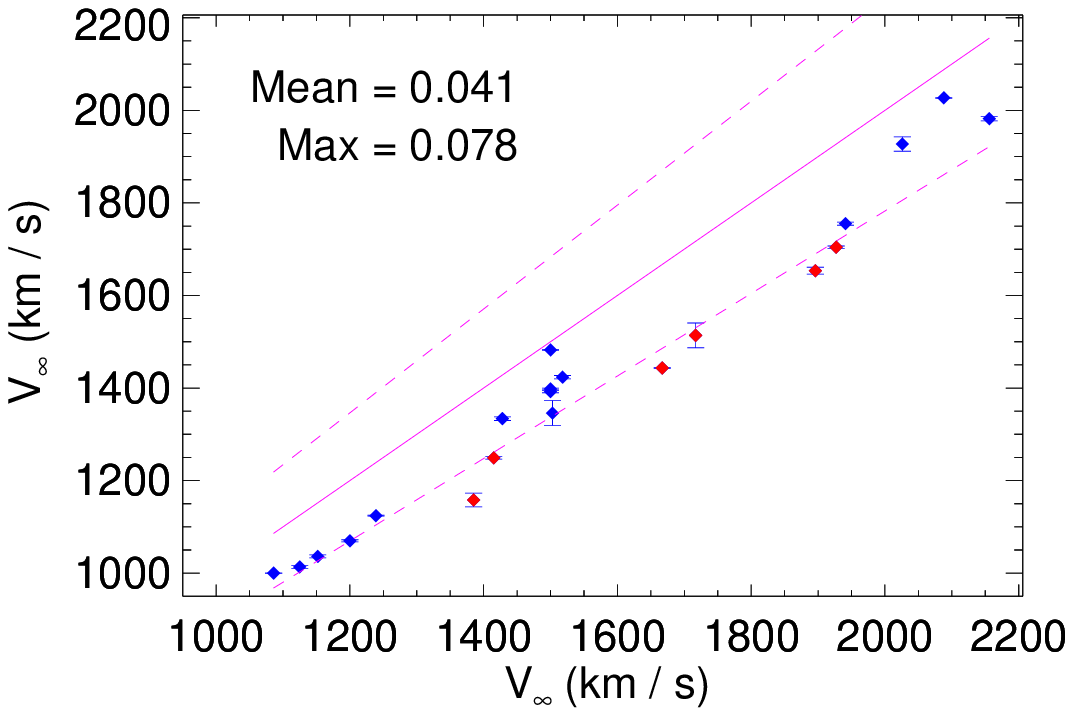}

\end{center}
\caption{The same as Fig.~\ref{fig:WC_abu} but for the WN grid and 
$\beta$-law with $\beta = 0.5$ (left panels) and
$\beta = 2$ (right panels) (`perfect' observed spectra).
}
\label{fig:WN_beta}
\end{figure*}

\begin{figure*}
\begin{center}
\includegraphics[width=2.8in, height=2.0in]{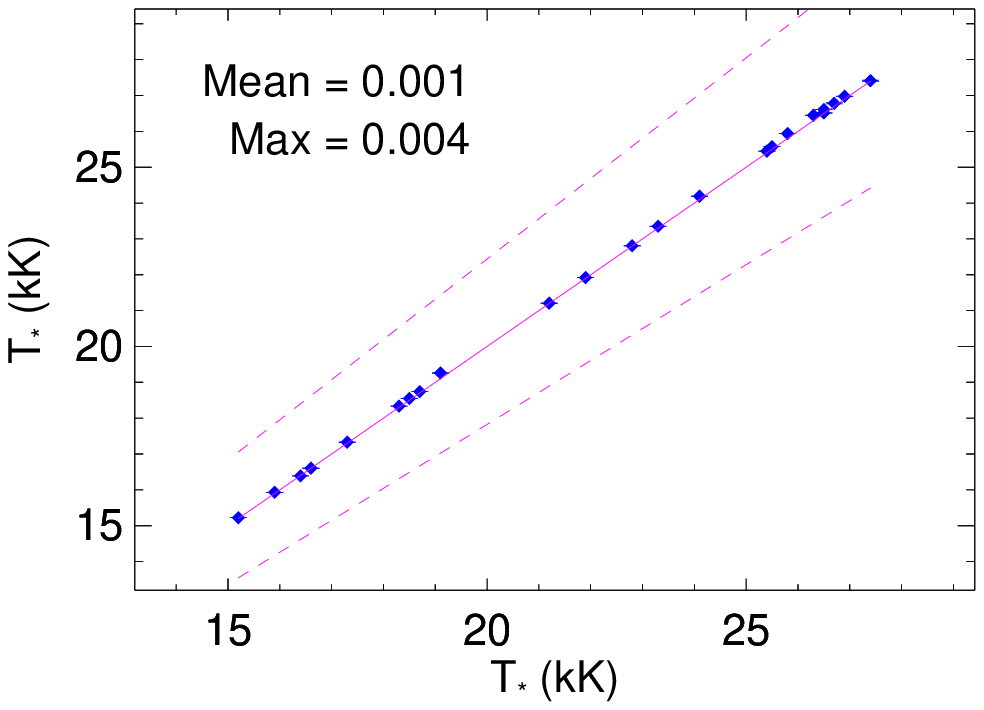}
\includegraphics[width=2.8in, height=2.0in]{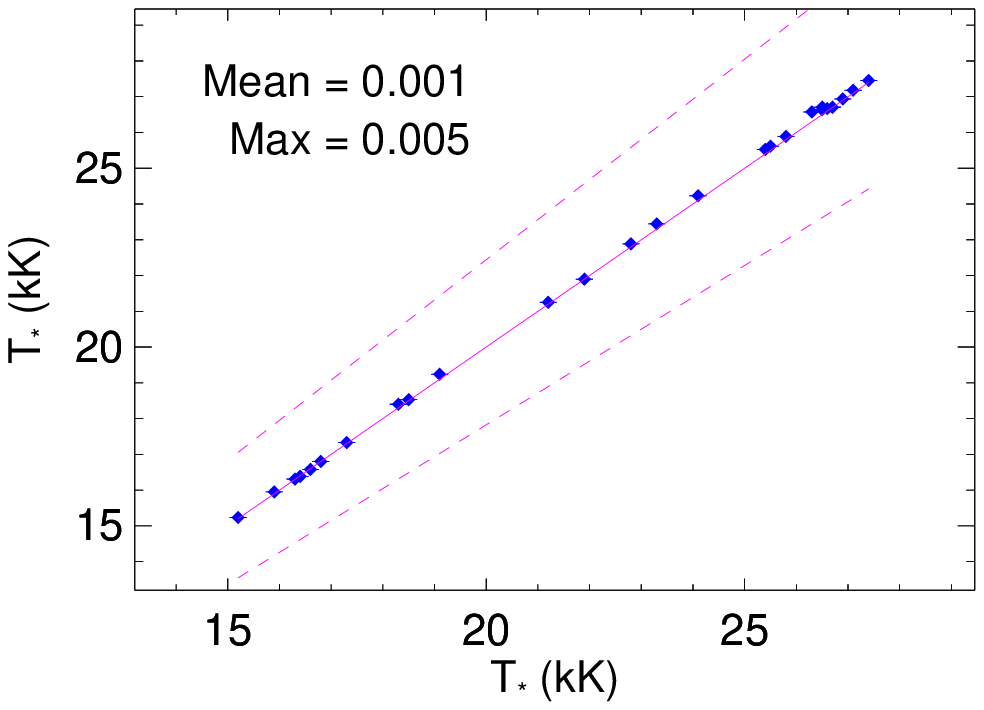}
\includegraphics[width=2.8in, height=2.0in]{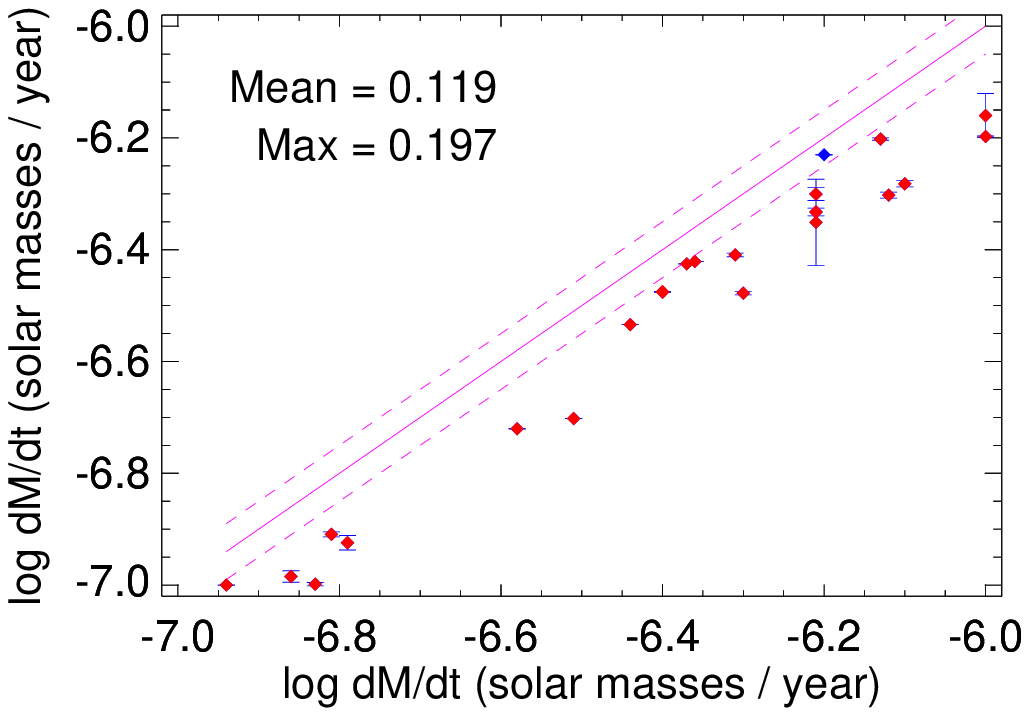}
\includegraphics[width=2.8in, height=2.0in]{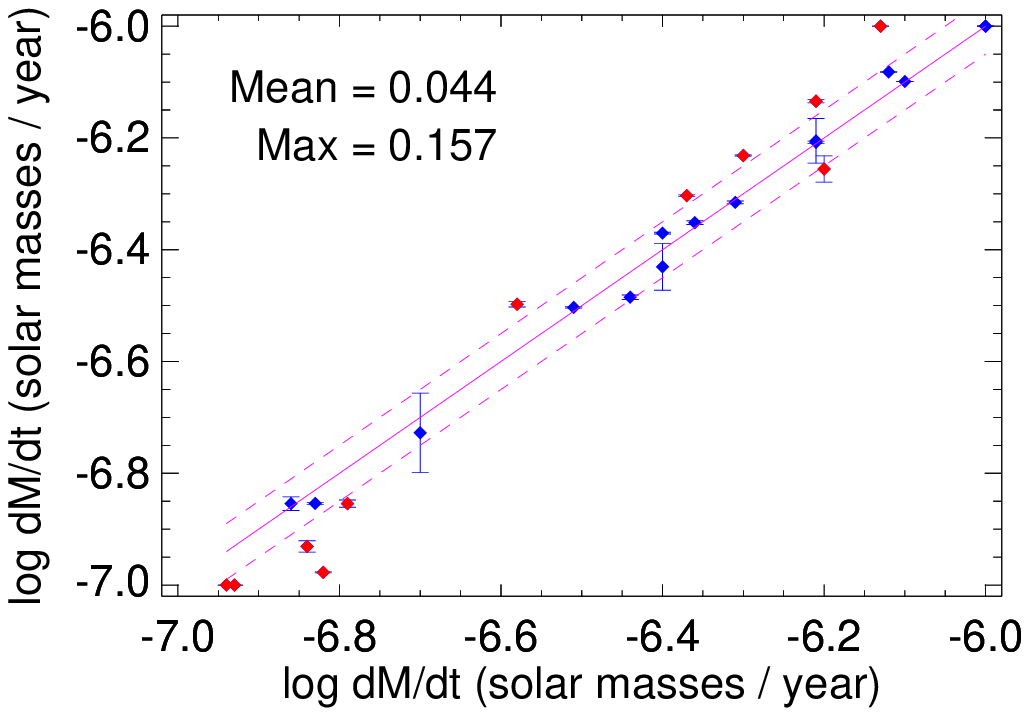}
\includegraphics[width=2.8in, height=2.0in]{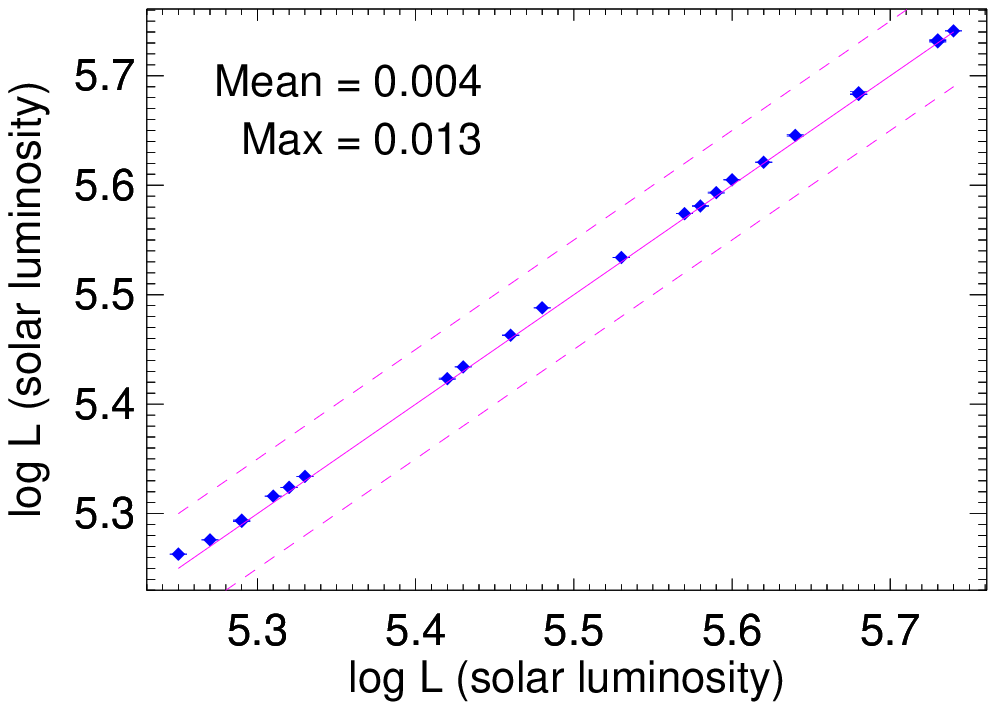}
\includegraphics[width=2.8in, height=2.0in]{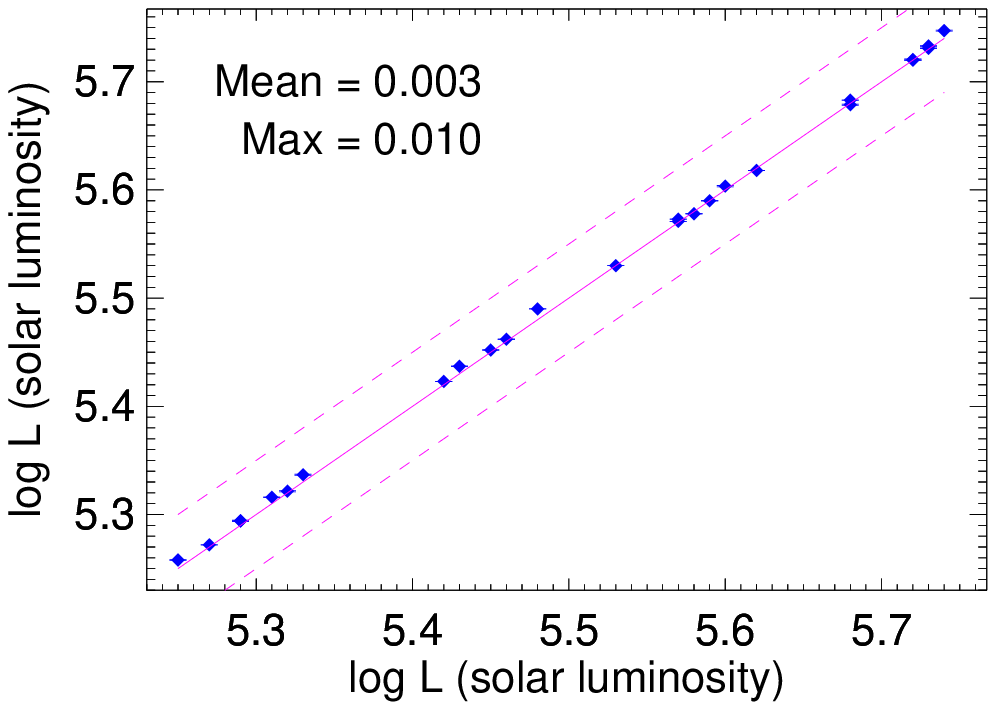}
\includegraphics[width=2.8in, height=2.0in]{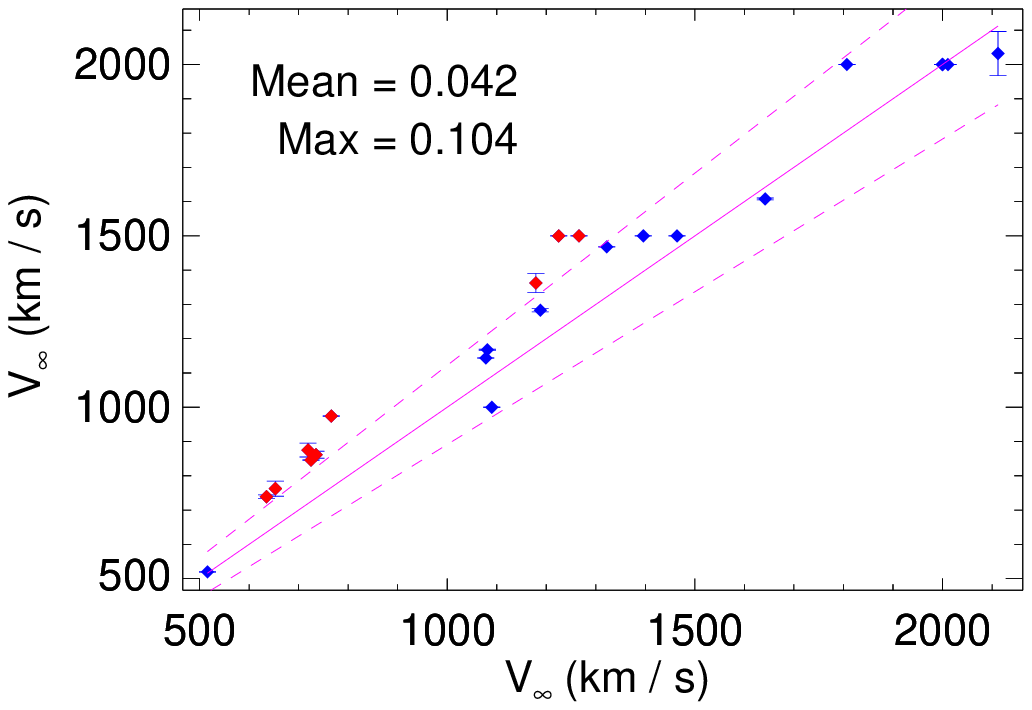}
\includegraphics[width=2.8in, height=2.0in]{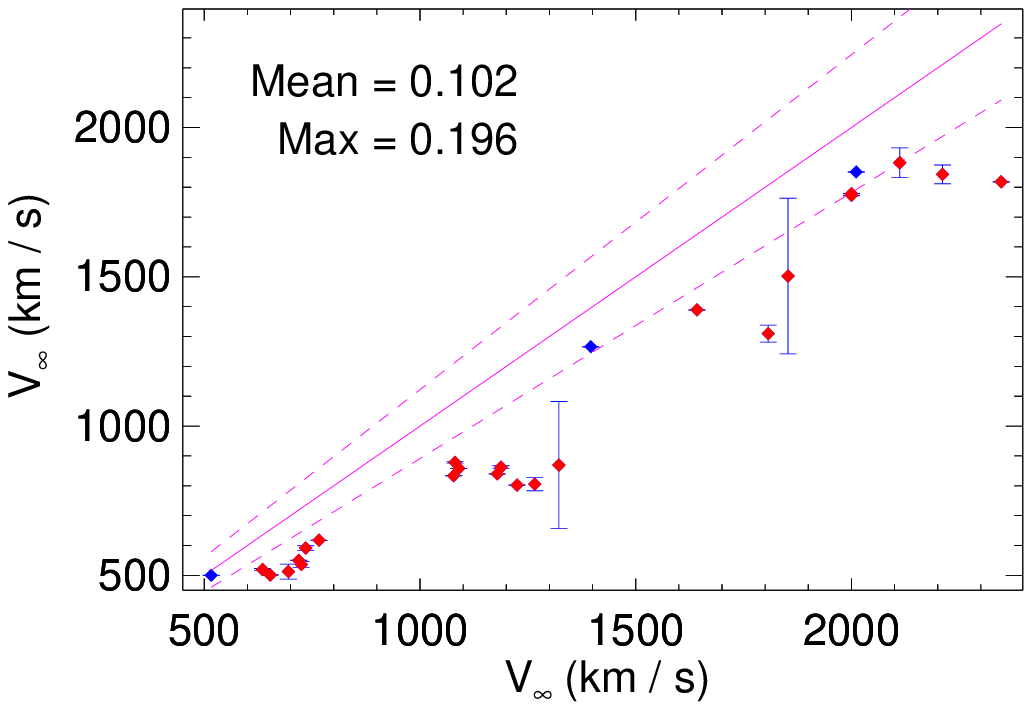}

\end{center}
\caption{The same as Fig.~\ref{fig:WC_abu} but for the SMC grid and
$\beta$-law with $\beta = 0.5$ (left panels) and
$\beta = 2$ (right panels) (`perfect' observed spectra).
}
\label{fig:SMC_beta}
\end{figure*}


\begin{figure*}
\begin{center}
\includegraphics[width=2.8in, height=2.0in]{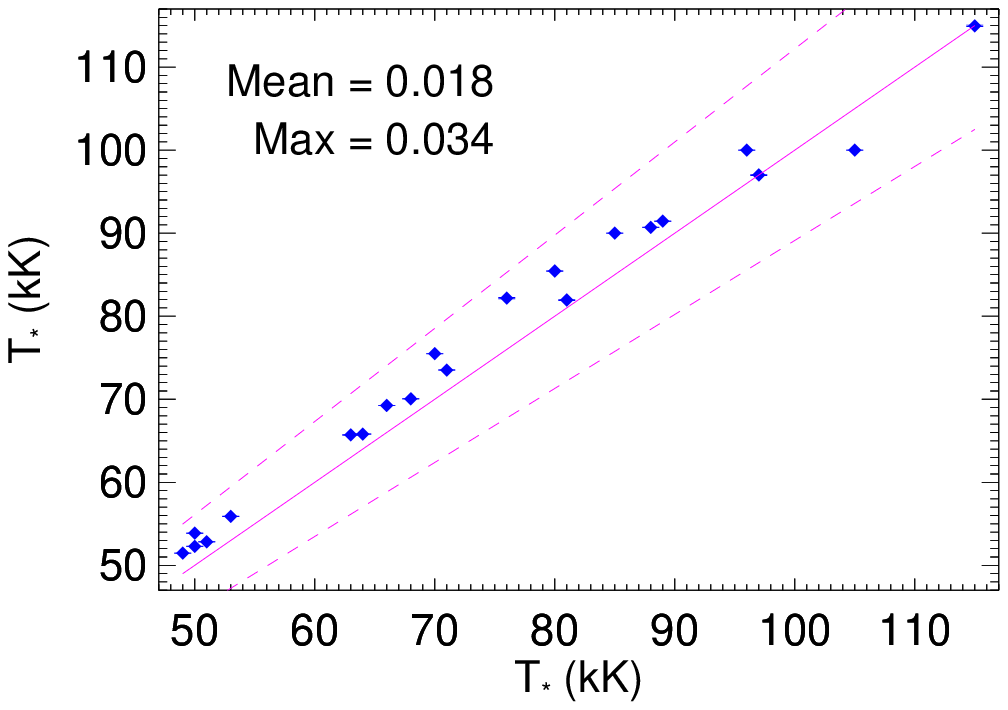}
\includegraphics[width=2.8in, height=2.0in]{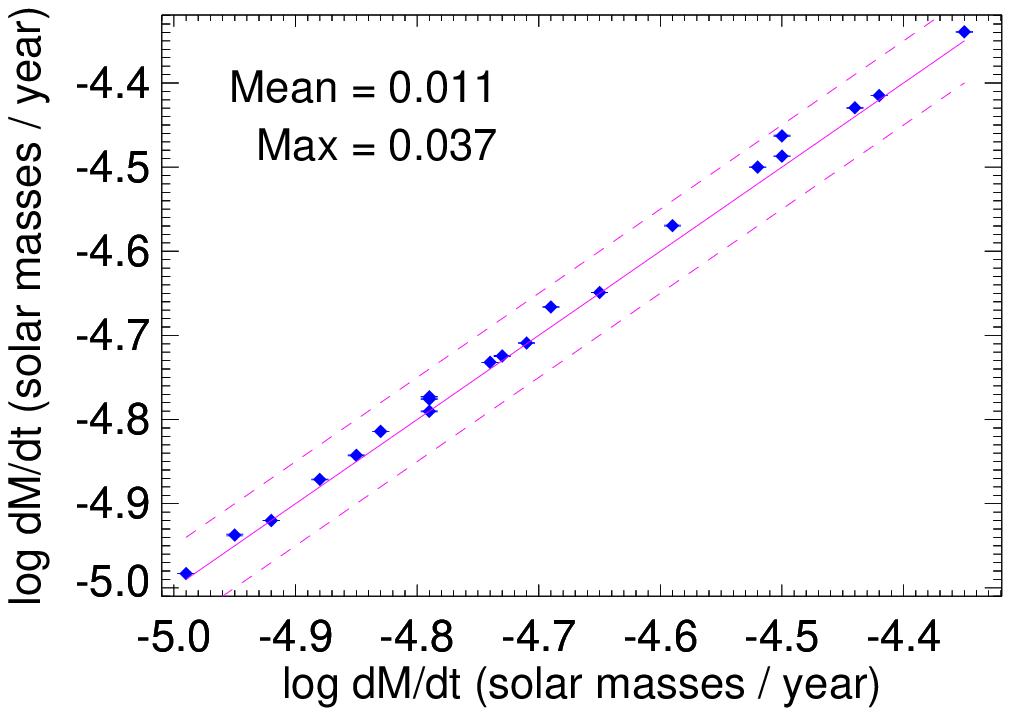}
\includegraphics[width=2.8in, height=2.0in]{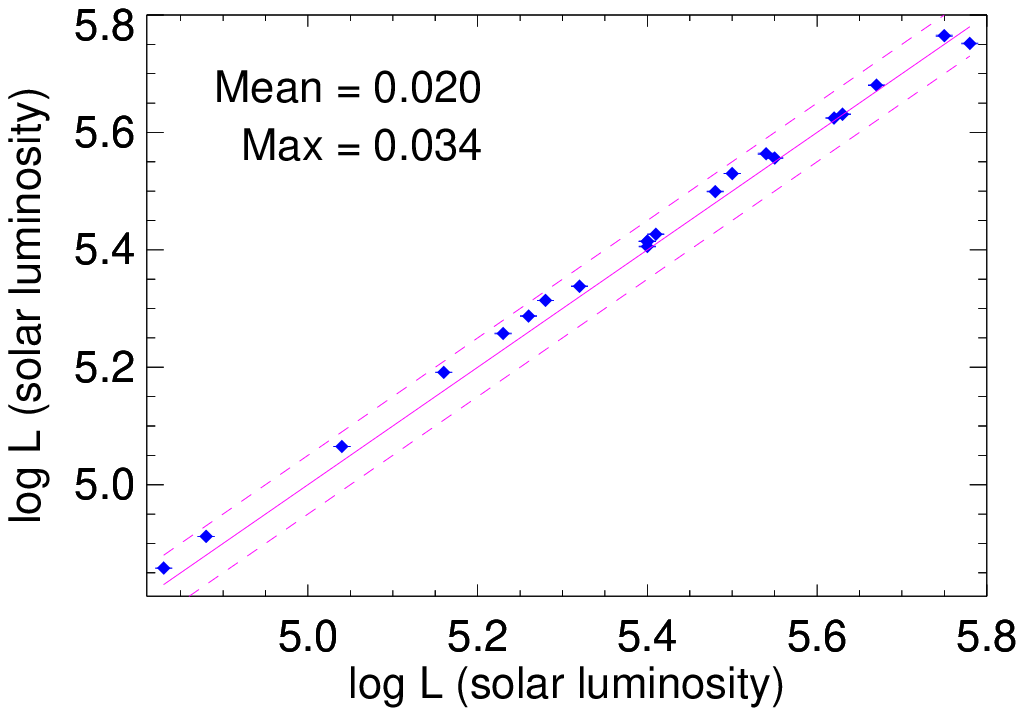}
\includegraphics[width=2.8in, height=2.0in]{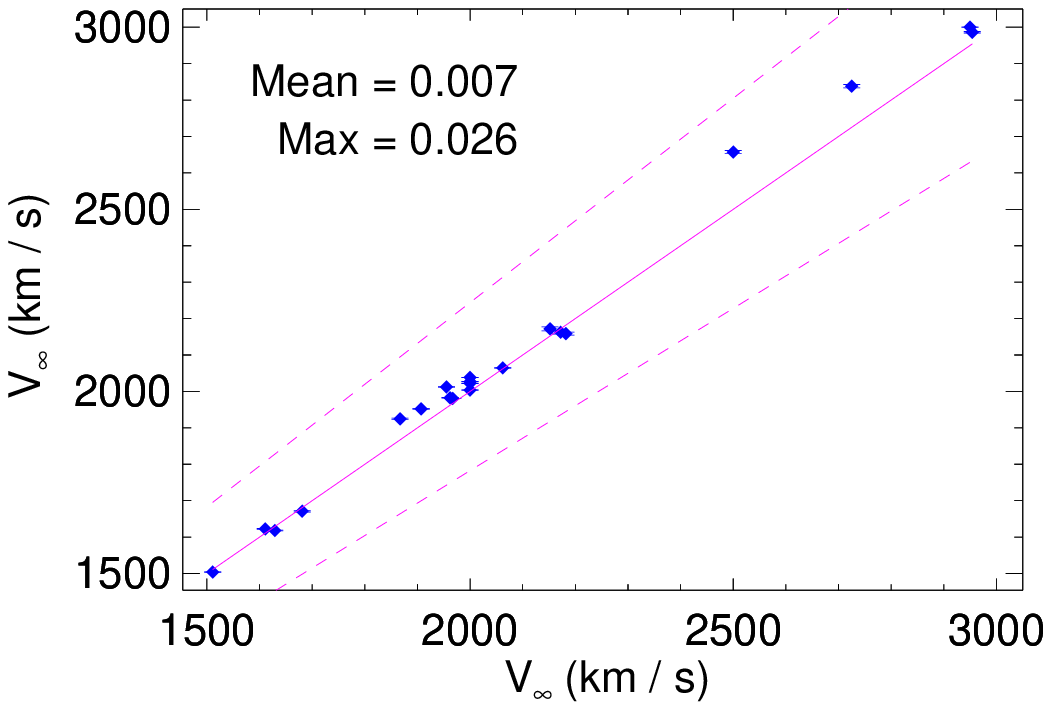}

\end{center}
\caption{The same as Fig.~\ref{fig:WC_abu} but for the WC grid and
volume filling factor $f_{\infty} = 0.25$ (`perfect' observed spectra).
}
\label{fig:WC_clump}
\end{figure*}

\begin{figure*}
\begin{center}
\includegraphics[width=2.8in, height=2.0in]{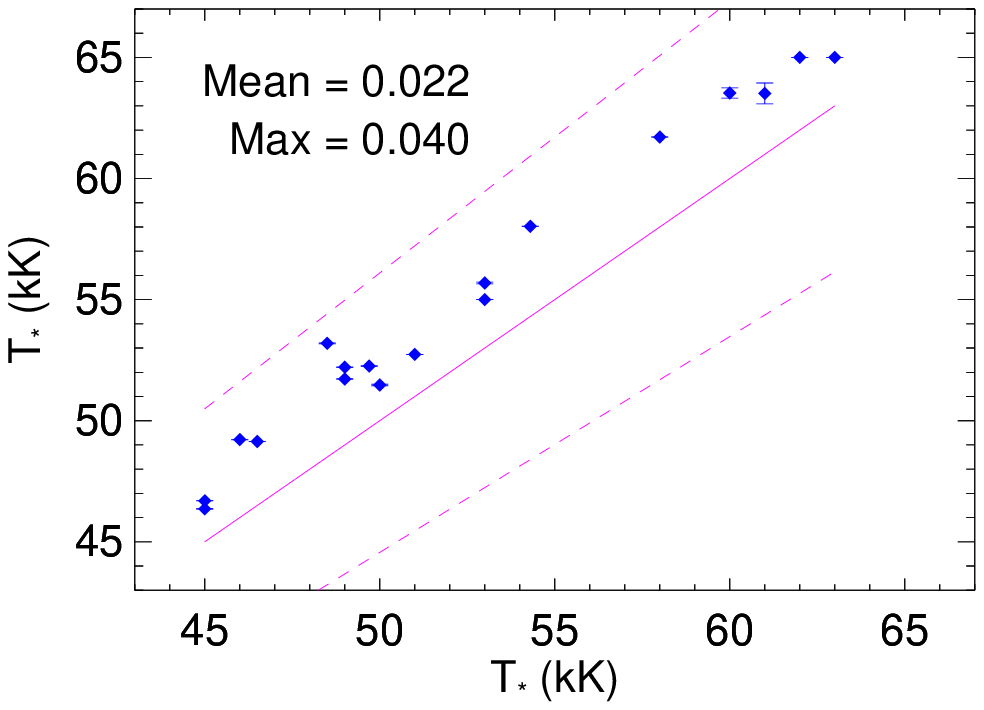}
\includegraphics[width=2.8in, height=2.0in]{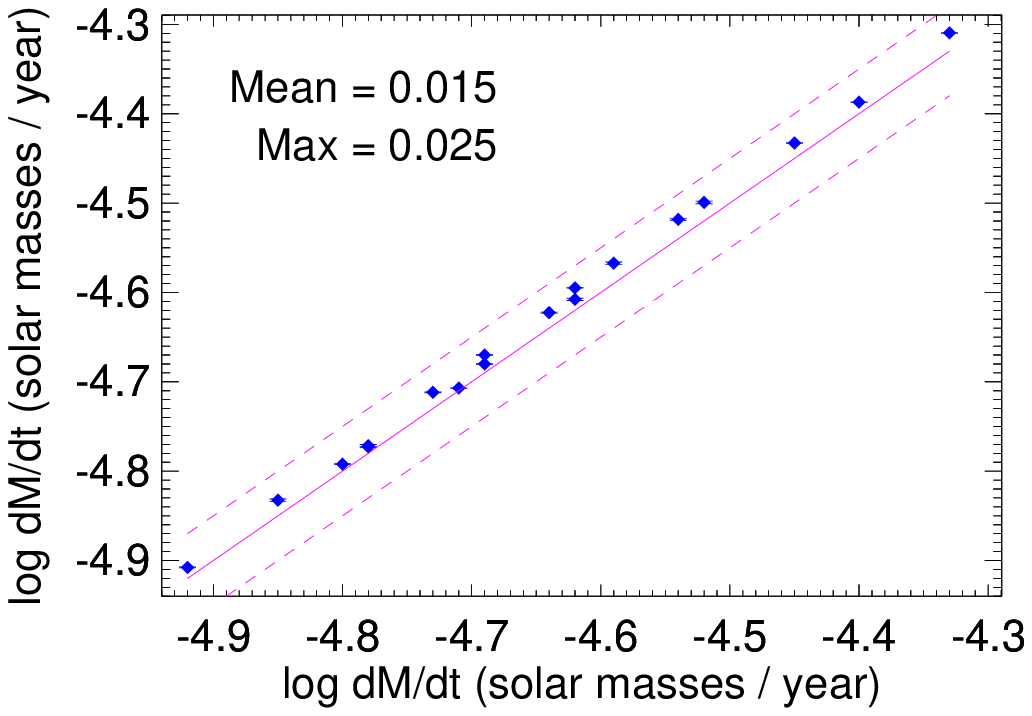}
\includegraphics[width=2.8in, height=2.0in]{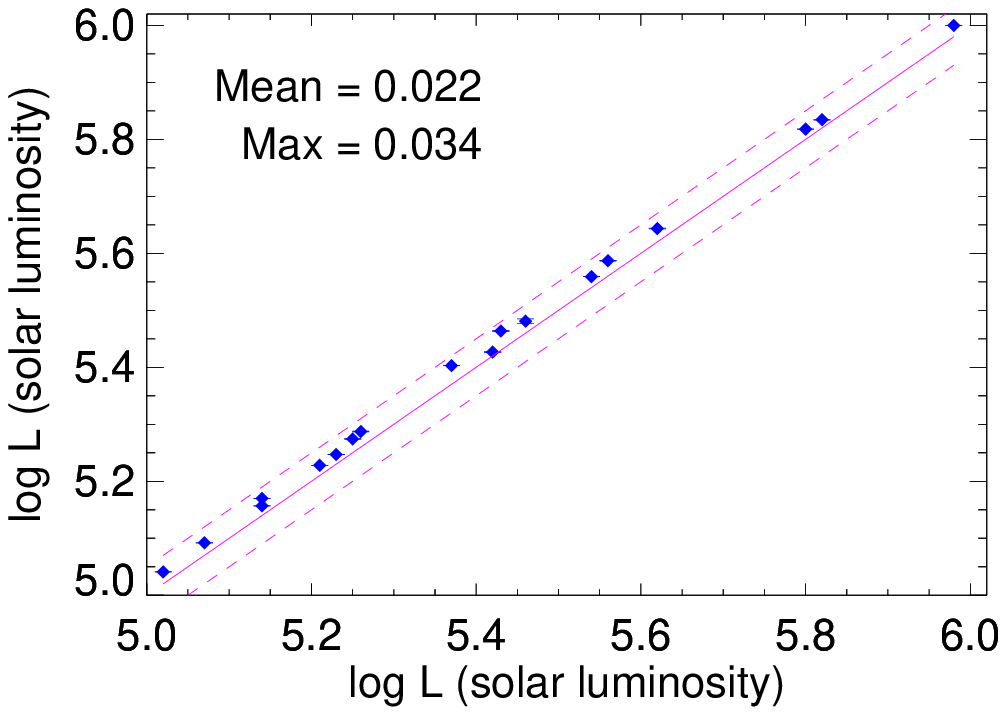}
\includegraphics[width=2.8in, height=2.0in]{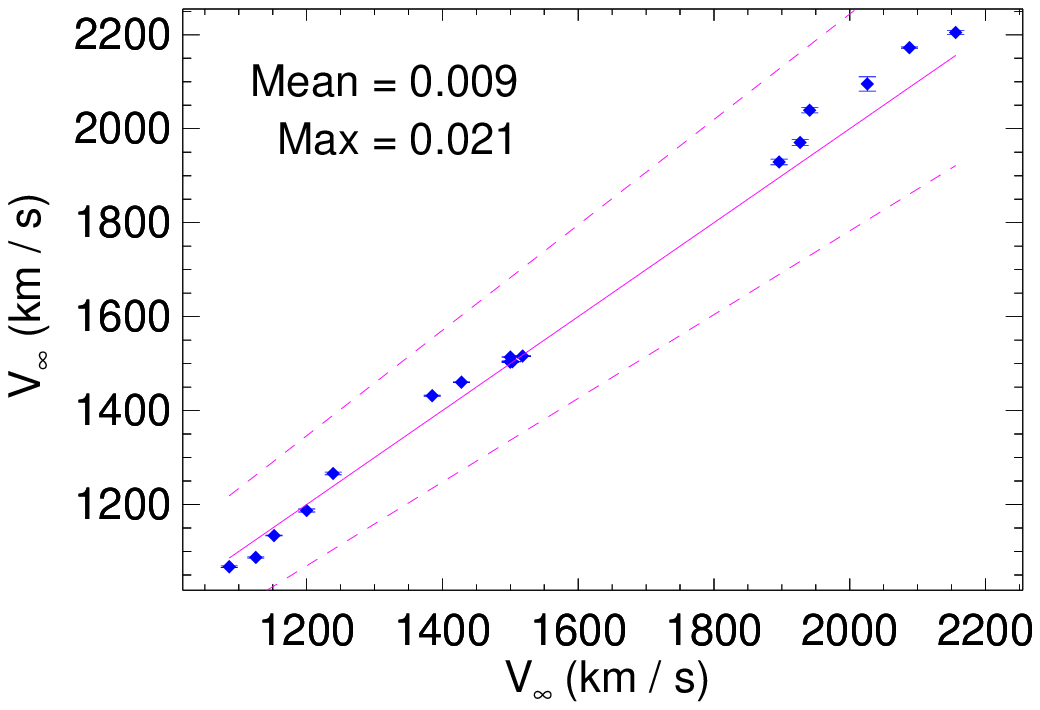}

\end{center}
\caption{The same as Fig.~\ref{fig:WC_abu} but for the WN grid and 
volume filling factor $f_{\infty} = 0.25$ (`perfect' observed spectra).
}
\label{fig:WN_clump}
\end{figure*}

\begin{figure*}
\begin{center}
\includegraphics[width=2.8in, height=2.0in]{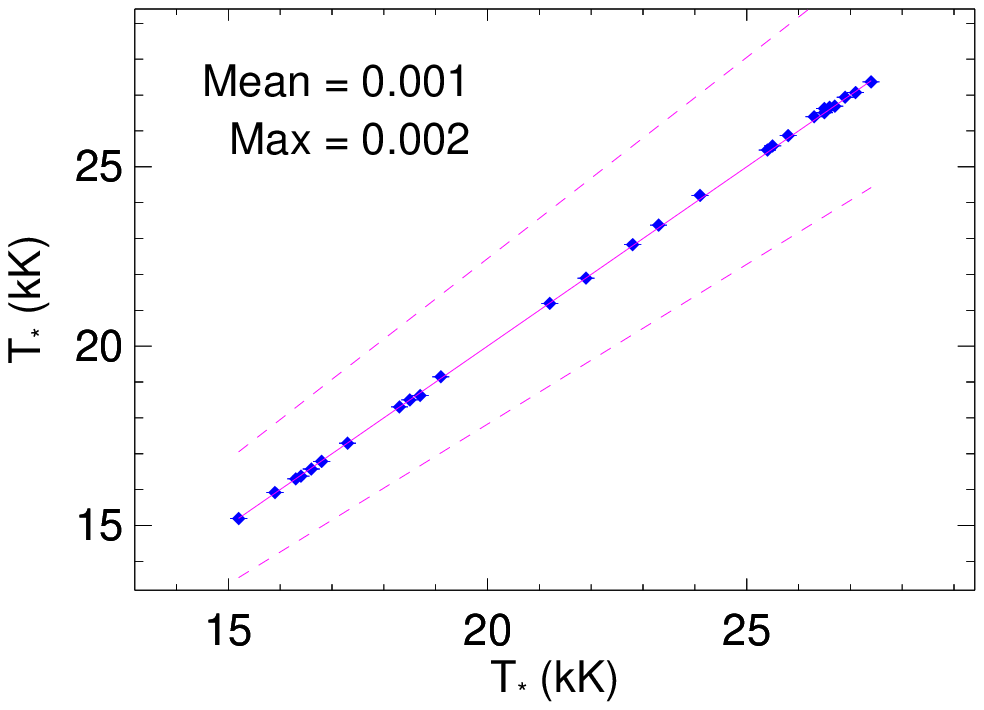}
\includegraphics[width=2.8in, height=2.0in]{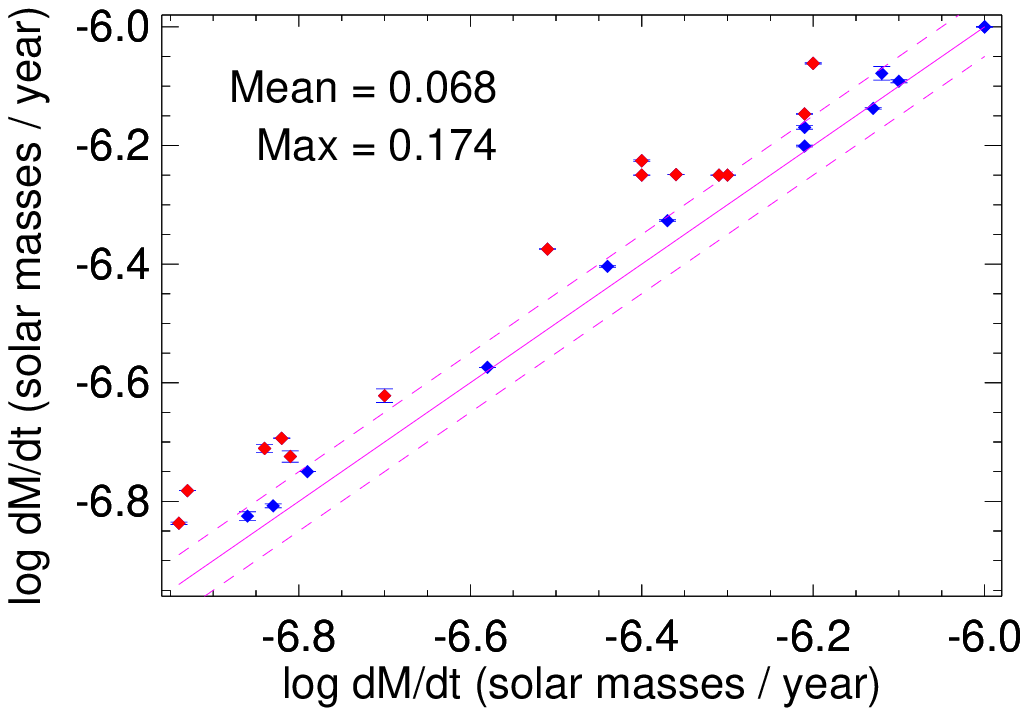}
\includegraphics[width=2.8in, height=2.0in]{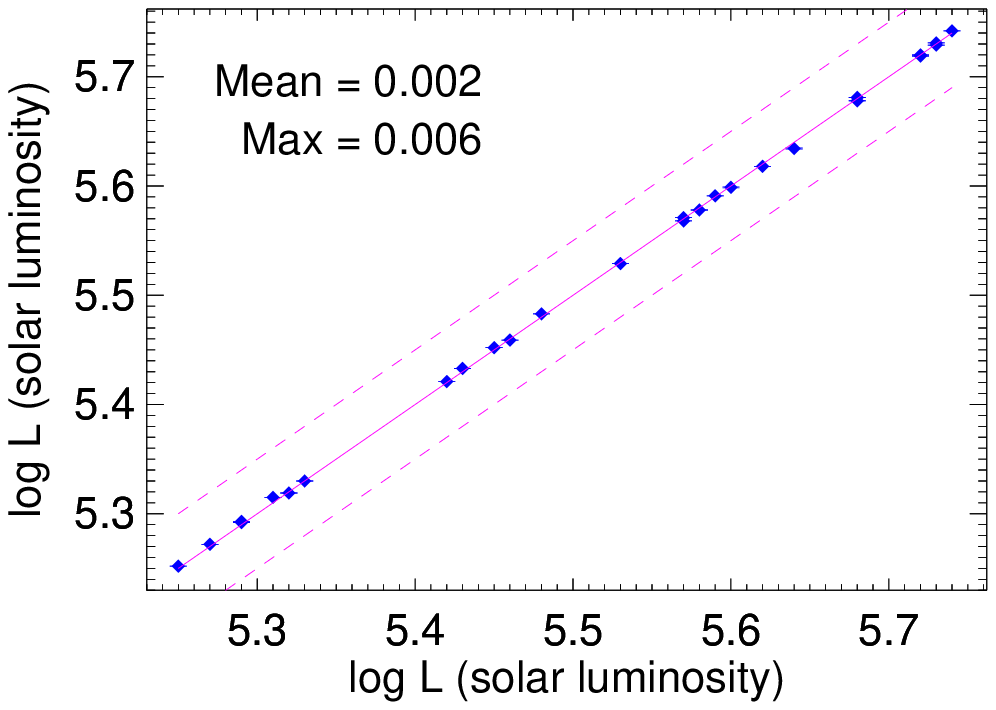}
\includegraphics[width=2.8in, height=2.0in]{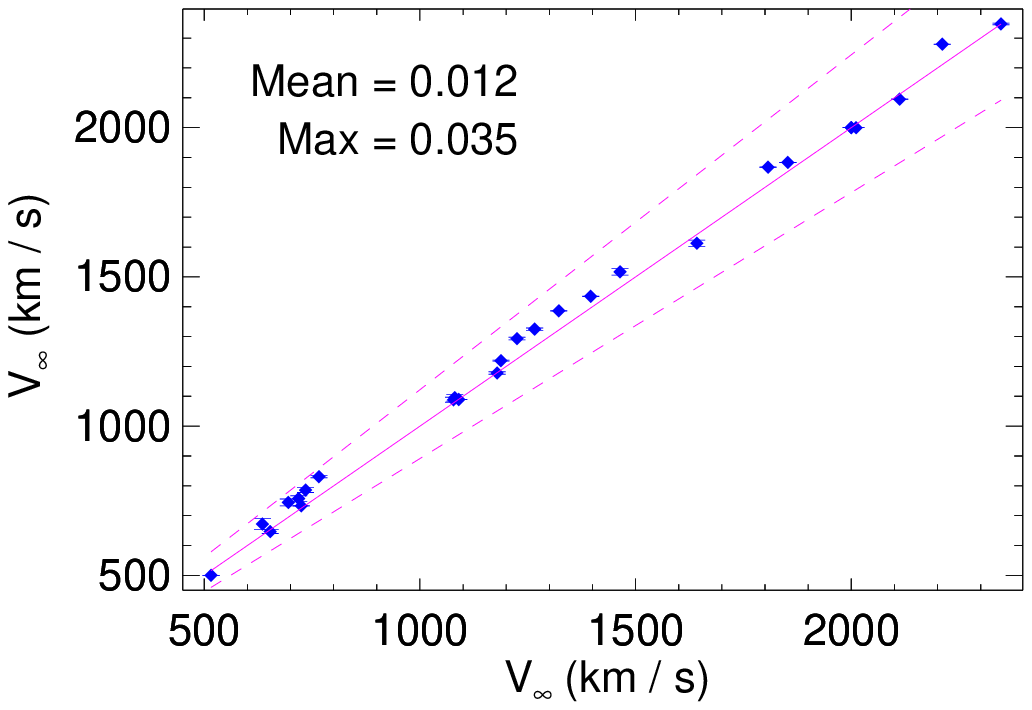}

\end{center}
\caption{The same as Fig.~\ref{fig:WC_abu} but for the SMC grid and
volume filling factor $f_{\infty} = 0.25$ (`perfect' observed spectra).
}
\label{fig:SMC_clump}
\end{figure*}

\label{lastpage}

\end{document}